\documentclass[fleqn,usenatbib]{mnras}
\pdfoutput=1

\usepackage{newtxtext,newtxmath}

\usepackage[T1]{fontenc}
\usepackage{ae,aecompl}

\usepackage{setspace}
\usepackage[switch]{lineno} 
\usepackage[final]{pdfpages}
\usepackage{makecell}
\usepackage{url}
\usepackage[utf8]{inputenc}
\usepackage{amsfonts}
\usepackage[english]{babel}
\usepackage{longtable}
\usepackage{afterpage}
\usepackage{booktabs}
\usepackage{mathtools}

\usepackage{graphicx}	
\usepackage{amsmath}	
\usepackage{amssymb}	

\newcommand{\swift}{{\it Swift~\/}}

\newcommand{\fermi}{{\it Fermi~\/}}

\def\H0{{\rm ~km~s^{-1}~Mpc^{-1}}}

\def\cf{{\rm cf.~\/}}
\def\la{\mathrel{\hbox{\rlap{\hbox{\lower4pt\hbox{$\sim$}}}{\raise2pt\hbox{$<$}}}}}
\def\ga{\mathrel{\hbox{\rlap{\hbox{\lower4pt\hbox{$\sim$}}}{\raise2pt\hbox{$>$}}}}}
\def\d25{D$_{25}$}

\def\deg{\hbox{$^\circ$}}

\newcommand{\ie}{{\rm i.e.}}
\newcommand{\eg}{{\rm e.g.}}

\newcommand{\ed}{\varepsilon_{\mathrm{d}}}
\newcommand{\eb}{\varepsilon_{\mathrm{b}}}
\newcommand{\ee}{\varepsilon_{\mathrm{e}}}
\newcommand{\rd}{r_{\mathrm{d}}}
\newcommand{\rph}{r_{\mathrm{ph}}}
\newcommand{\Lum}{L_{0,52}}
\newcommand{\Epz}{E_{\mathrm{p,z}}}

\title[Testing a model for subphotospheric dissipation in GRBs]{Testing a model for subphotospheric dissipation in GRBs: fits to \textit{Fermi} data constrain the dissipation scenario}
\author[B. Ahlgren et al.]{Bj\"{o}rn Ahlgren$^{1}$\thanks{E-mail:
bjornah@kth.se}, Josefin Larsson$^{1}$, Erik Ahlberg$^{1}$, Christoffer Lundman$^{1}$,
\newauthor Felix Ryde$^{1}$, Asaf Pe'er$^{2,3}$
\\
$^{1}$KTH Royal Institute of Technology, Department of Physics, and the Oscar Klein Centre, AlbaNova, SE-106 91 Stockholm, Sweden \\
$^{2}$Physics Department, University College Cork, Cork, Ireland
$^{3}$Department of Physics, Bar-Ilan University, Ramat-Gan 52900, Israel}

\date{Accepted 2019 January 09. Received 2019 January 09; in original form 2018 January 24}

\pubyear{2019}
\hypersetup{draft}
\begin{document}
\label{firstpage}
\pagerange{\pageref{firstpage}--\pageref{lastpage}}
\maketitle

\begin{abstract}
It has been suggested that the prompt emission in gamma-ray bursts (GRBs) could be described by radiation from the photosphere in a hot fireball. Such models must be tested by directly fitting them to data. In this work we use data from the \textit{Fermi Gamma-ray Space Telescope} and consider a specific photospheric model, in which kinetic energy of a low-magnetisation outflow is dissipated locally by internal shocks below the photosphere. 
We construct a table model with a physically motivated parameter space and fit it to time-resolved spectra of the 36 brightest Fermi GRBs with known redshift. We find that about two thirds of the examined spectra cannot be described by the model, as it typically under-predicts the observed flux. However, since the sample is strongly biased towards bright GRBs, we argue that this fraction will be significantly lowered when considering the full population. From the successful fits we find that the model can reproduce the full range of spectral slopes present in the sample. For these cases we also find that the dissipation consistently occurs at a radius of $\sim 10^{12}$ cm and that only a few percent efficiency is required. Furthermore, we find a positive correlation between the fireball luminosity and the Lorentz factor. Such a correlation has been previously reported by independent methods. We conclude that if GRB spectra are due to photospheric emission, the dissipation cannot only be the specific scenario we consider here.
\end{abstract}




\section{Introduction}
Prompt GRB emission mechanisms are still a matter of debate, and remain an unsolved problem in GRB physics. The principal approach to finding the origin of GRB prompt emission has been to fit spectra with empirical functions and use the best-fitting parameters to constrain different emission scenarios. For a considerable time the empirical Band function \citep{1993ApJ...413..281B}, a smoothly broken power law, has been the predominant way of fitting these spectra. The Band function often provides good fits to the prompt emission spectra and the fits have commonly been interpreted in terms of synchrotron emission \citep{1996ApJ...466..768T,Briggs:1999tg,2009Sci...323.1688A,2016ApJ...816...72Z}. However, it has been shown that synchrotron radiation cannot consistently produce the full distribution of spectral shapes we observe when fitting with the Band function, including the low-energy slope \citep{Preeceetal_1998A_ApJ} and the spectral width of the peak \citep{2015MNRAS.447.3150A,2015A&A...583A.129Y}. In addition, the Band function does not represent a physical scenario, and in order to obtain physical information we must instead ultimately use a physically motivated model and fit to data (see also \citealt{2014ApJ...784...17B,2015MNRAS.451.1511B}).

Observations of GRBs with spectra close to blackbodies (BBs) \citep{2004ApJ...614..827R, Ghirlandaetal_2013A_MNRAS,2015ApJ...800L..34L}, as well as spectra described by a BB and an additional component \citep{Rydeetal_2010A_ApJ, 2011ApJ...727L..33G, 2012ApJ...757L..31A, Burgess_2014A_ApJ,2016ApJ...833..139A}, suggest that at least some bursts have prompt emission of a photospheric origin. Conversely, for most GRBs, the simplest models of photospheric emission cannot provide an adequate description of the observed spectra. In particular, the low-energy spectral slope of most bursts is significantly softer than a Planck function. 

It has been shown that more realistic photospheric models will result in spectra that are broader than a Planck function. The broadening can be a result of, \eg , subphotospheric dissipation \citep{2005ApJ...628..847R, 2006ApJ...642..995P,2006A&A...457..763G,2010MNRAS.407.1033B,2011ApJ...738...77V,2015ApJ...802..132C}, or geometric effects \citep{2008ApJ...682..463P,Lundmanetal_2013A_MNRAS}.
Subphotospheric dissipation has previously been invoked to interpret the `top-hat' spectral shape of GRB~110920 \citep{2015MNRAS.450.1651I} and the broadening of the initially very narrow spectrum of GRB~090902B \citep{2011MNRAS.415.3693R}. Though subphotospheric dissipation has been used to describe several bursts, the different model scenarios must be examined further, particularly using larger data samples and proper techniques when fitting models to data.

In our previous work \citep{2015MNRAS.454L..31A}, A15 from here on, we fitted a model for subphotospheric dissipation to prompt GRB emission data from the \textit{Fermi Gamma-ray Space Telescope}. This was the first time such a model was fitted to data and the study served as an important proof-of-concept. 
Specifically, we fitted GRB~090618, which can also be described by a Band function with typical parameters, and GRB~100724B, which has previously been fitted with a composite Band+BB model. The results showed that subphotospheric dissipation is a viable description of both bursts and justified further investigations. 

In this work we build on the work presented in A15 by increasing the model parameter space and by including approximate adiabatic cooling. We have also significantly expanded our sample, now analysing 36 bursts, and have performed Monte Carlo simulations to quantify some of the model's inherent uncertainties. We also consider parameter correlations and physical implications.

The main goal of the work is to test the scenario of localised dissipation by internal shocks below the photosphere in a low-magnetisation outflow. We emphasise that the constraints found in the paper do not apply to photospheric emission in general or other dissipation scenarios. We confine our study to the case where almost all the dissipated energy goes into the electrons and explore the effects of three free parameters, as described in section~\ref{tablemodel}.
The second main objective is to present the continued development of our table model, which we plan to make publicly available.

The paper is organised as follows: section 2 describes the physical scenario and the table model implementation. In section 3 we describe our data sample, binning and fitting procedure. Section 4 presents the results and section 5 is the discussion. Section 6 is summary and conclusions. 
Finally, in appendices \ref{appendix:A}-\ref{appendix:E:burstsumtable} we present more technical details about the model and analysis.

\section{The Model}

\subsection{Physical scenario}
\label{section:Themodel}
In this paper we consider the same physical model as in A15. It is a model based on the fireball model, specifically a hot fireball, using a kinetic code by \cite{2005ApJ...628..857P}. We described this model in A15, but for completeness we present an overview here as well. Any deviations from the model as stated in A15 will be explicitly pointed out. Note that all quantities in this section are in the observer frame unless otherwise stated.

In the fireball model (see \eg ~\citealt{2015AdAst2015E..22P} for a recent review), an isotropic equivalent luminosity, $L_{0} = L_{0,52}10^{52}$~erg~s$^{-1}$, is emitted from the central engine in the form of baryons, electrons, $B$-fields, and photons. The outflow starts at the initial jet radius, $r_0$, where the bulk Lorentz factor is $\Gamma \sim 1$. The photons give rise to a radiative pressure which accelerates the outflow into highly relativistic collimated jets. The outflow is accelerated until the saturation radius, $r_{\mathrm{s}} = \Gamma r_0$. At this point the outflow is in thermal equilibrium and coasts with constant $\Gamma$. The baryons, mainly protons, carry the bulk of the kinetic energy of the outflow. In our model we define the dissipation radius as $r_{\mathrm{d}} = r_{\mathrm{ph}}/\tau = L \sigma_{\mathrm{T}} / 4\pi \tau \Gamma^3  c^3 m_{\mathrm{p}}$, where $\tau$ is the optical depth due to the baryonic electrons (\ie ~the electrons which are associated with the protons of the outflow). At this radius we dissipate a fraction $\varepsilon_{\mathrm{d}}$ of the kinetic energy of the protons, out of which the fraction $\varepsilon_{\mathrm{e}}$ goes to electrons and a fraction $\varepsilon_{\mathrm{b}}$ is used to generate magnetic fields.

The dissipation creates a non-equilibrium situation between the photon bath and the heated electron population, where they interact through Compton and inverse Compton scattering, as well as pair production and pair annihilation. We also treat synchrotron radiation and synchrotron self-absorption. It is assumed that electrons are injected as a Maxwellian distribution with a dimensionless temperature, $\theta = kT/m_\mathrm{e}c^2 \propto \ed \ee$, in the range of 5.5 - 220. Both newly injected electrons and the existing population will interact with the photons. Before the dissipation essentially all energy is carried by the baryons in the jet. After the dissipation the dissipated energy has been transferred to the photons, but the baryons still carry the majority of the energy.

Possible dissipation mechanisms in GRB jets include magnetic reconnection \citep{1994MNRAS.270..480T, 2002A&A...391.1141D, 2005A&A...430....1G, 2006NJPh....8..119L}, hadronic collisions \citep{2010MNRAS.407.1033B} or internal shocks \citep{1997ApJ...490...92K,1998MNRAS.296..275D,2005ApJ...628..847R}. 
In our current implementation we are working with the internal shocks paradigm by assuming that $r_{\mathrm{d}} = \Gamma^2 r_{0}$. We also assume that the dissipation is a continuous process localised between $r_{\mathrm{d}}$ and $2r_{\mathrm{d}}$. Note that this is different from the scenario with continuous subphotospheric dissipation throughout the jet employed in \eg ~\cite{2016ApJ...831..175V}. 

A caveat is that large a value of $\varepsilon_{\mathrm{d}}$ would remove the majority of the kinetic energy of the outflow and would inevitably decelerate it. This in turn may lead to a re-acceleration phase, which is not treated in our code. We therefore avoid the corresponding part of the parameter space. Since the energy is passed to either the electrons or the magnetic fields, this translates to $\varepsilon_{\mathrm{d}}\varepsilon_{\mathrm{e}} + \varepsilon_{\mathrm{d}}\varepsilon_{\mathrm{b}}$ being the constrained quantity. We have decided to require $\ed \leq 0.4$ to minimise this effect while still keeping a large amount of parameter space to work with. Even at this level of dissipation one could expect a second acceleration phase \citep{2014ApJ...792...42B}, but we expect this to have a limited impact on our results.

There are additional physical effects not accounted for in the code. Examples include spatial effects, such as the aforementioned geometric broadening, the jet opening angle, jet hydrodynamics (\eg ~\citealt{2009ApJ...700L..47L}) and the effects from a fuzzy photosphere \citep{2011ApJ...737...68B,2013ApJ...767..139B}. Finally, we do not at this stage consider neutrino emission or afterglow predictions.

\subsection{Table model}
\label{tablemodel}
The {\scriptsize XSPEC} table model implementation \citep{1999ascl.soft10005A} allows us to build a numerical model from simulated spectra. A table model consists of a grid of model spectra, corresponding to different parameter values. Interpolation between spectra is used to obtain spectral shapes between the grid points. Using our numerical code we perform a large number of simulations for different parameter values, spanning the grid
\begin{align*}
\Gamma &= 100,150,200,250,300,350,400,450,500\\
L_{0,52} &= 0.1, 0.5, 1, 5, 10, 50, 100, 200, 300 \\
\varepsilon_{\mathrm{d}} &= 0.01,0.025,0.05,0.75,0.1,0.15, 0.2,0.25, 0.3,0.35, 0.4
\end{align*}
which yields a total of 891 grid points. Note that other physical parameters often discussed in the context of subphotospheric dissipation, such as $\rd$ and $r_0$, are derived quantities according to the expressions defined in section~\ref{section:Themodel}.
Furthermore, the model obtains a normalisation and redshift as two additional parameters when it is implemented in {\scriptsize XSPEC}. These two parameters are kept constant during the fits, with the redshift given from observations and the normalisation set by the luminosity distance.

As in A15 we let $\eb = 10^{-6}$ and $\ee = 0.9$, in order to test the scenario where almost all the energy goes into electrons and where we have no significant contribution from synchrotron. We note that this leaves $\eb + \ee < 1$. However, this simply means that the remaining energy is never dissipated.
The main difference to the parameter space employed in A15 is that we now have set $\tau = 35$ instead of letting it be a free parameter. This represents a scenario with the dissipation occurring moderately deep in the outflow. There are several reasons for this choice. Firstly, testing shows that most fits 
prefer high values of $\tau$, with very few spectra being described by $\tau \sim$ a few. However, at $\tau \gtrsim 15$ this parameter turns out to be relatively insensitive. Changes in $\tau$ at these values usually result in small changes of the fit statistic, predicted spectral shape in the part of the GBM energy range where the statistics are best, and small changes in other model parameters. Additionally, $\tau$ cannot always be given a meaningful physical interpretation. Due to significant pair production, it does not directly correspond to the number of scatterings in the outflow, it also does not correspond directly to $\rd$. Finally, keeping $\tau$ fixed decreases degeneracies in the model.
Compared to the model in A15, we have also expanded $\ed$ to lower values, from 0.1 down to 0.01, and made all parameters more tightly spaced in the grid. These improvements from the previous model reduce uncertainties pertaining to interpolation in the grid (see discussion in appendix~\ref{appendix:B}).

With these modifications we have captured most of the physically relevant parameter space for the three free parameters of our model. The main exception is $\Gamma$, which could in principle lie outside the range 100-500, although most Lorentz factors are expected to lie within our parameter range, see \eg ~\cite{2018A&A...609A.112G}. The upper bound for $\ed$ is set by dynamical concerns, as discussed in section~\ref{section:Themodel}, and the lower bound by efficiency requirements. 
To a good approximation, the fireball luminosity, $\Lum$, relates to the observed isotropic equivalent luminosity as $L_{\mathrm{iso}} \sim \Lum \ed$. Hence, the maximum fireball luminosity is $\Lum < L_{\mathrm{iso,max}} / \ed$, for a maximal observed luminosity, $L_{\mathrm{iso,max}}$. \cite{2017ApJ...837..119A} find the limit $E_{\mathrm{iso}} < 3\times 10^{54}$ erg for observed isotropic energies. Assuming a typical duration for long GRBs of $\sim 50$ s, \citep{2016ApJS..223...28N}, we obtain the limits $\Lum <  600 $ and $\Lum \lesssim 100$, for the minimum $\ed = 0.01$ and a more typical $\ed = 0.05$, respectively. Thus, considering variations in $\ed$ and burst duration, the limit, $\Lum \leq 300$, we have set is reasonable, though not absolute. In section~\ref{subsection:discuss_badfits} we discuss the implications of our choices of parameter limits.

In A15 we referred to our table model as DREAM (Dissipation with Radiative Emission as A table Model). Since the goal is to make DREAM publicly available, we adopt a naming convention to help discriminate between and refer to the different versions of the table model as we develop it. The first number relates to the physical scenario considered when generating the model spectra, using the numerical code from \cite{2005ApJ...628..857P}. The physical scenario decides the parametrisation in the numerical code, without changing the underlying numerical scheme or the relevant physics. The second number denotes which grid has been used, including minor revisions such as post-processed treatment of adiabatic cooling. With this naming convention, the model of A15 becomes DREAM1.1, and the model herein is DREAM1.2. Throughout the paper, "model" is, unless otherwise stated, referring to the table model, DREAM1.2. "Physical scenario" is used to refer to the underlying physical model (subphotospheric dissipation by internal shocks with negligible magnetic fields). Any references to a "code" refers to the numerical code presented in \cite{2005ApJ...628..857P}. 

\begin{figure*}
\includegraphics[scale=0.32]{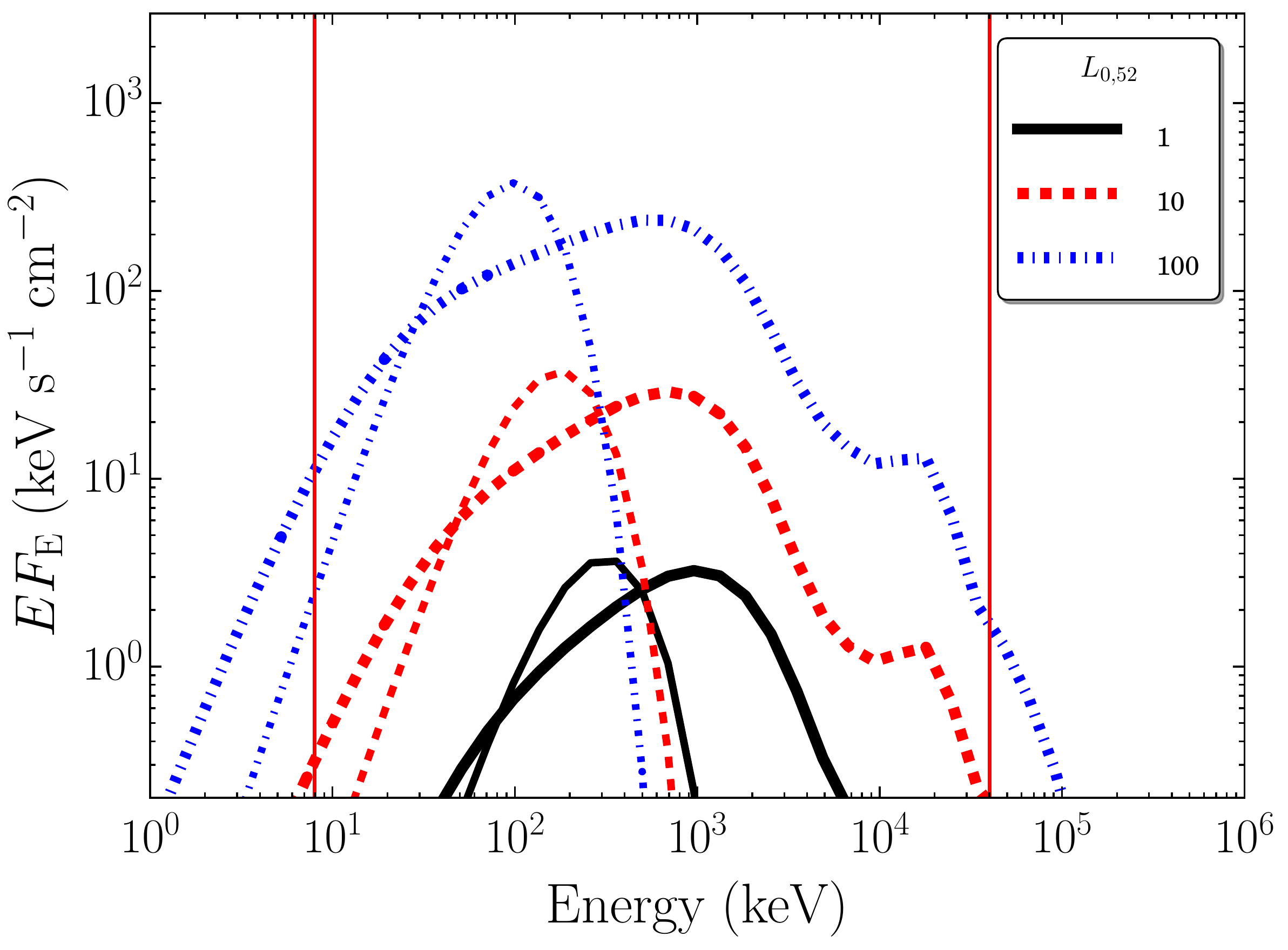}
\includegraphics[scale=0.32]{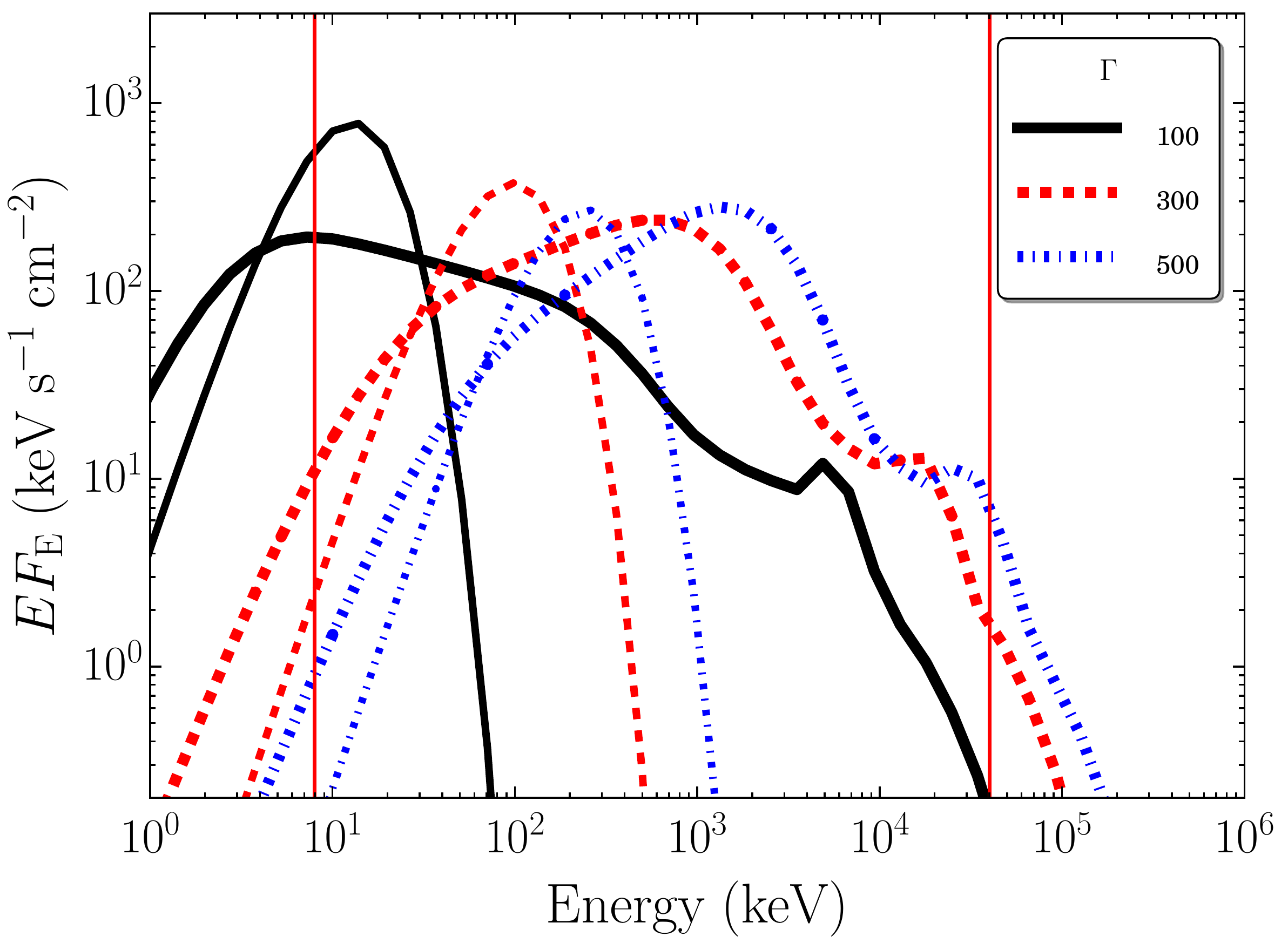}
\includegraphics[scale=0.32]{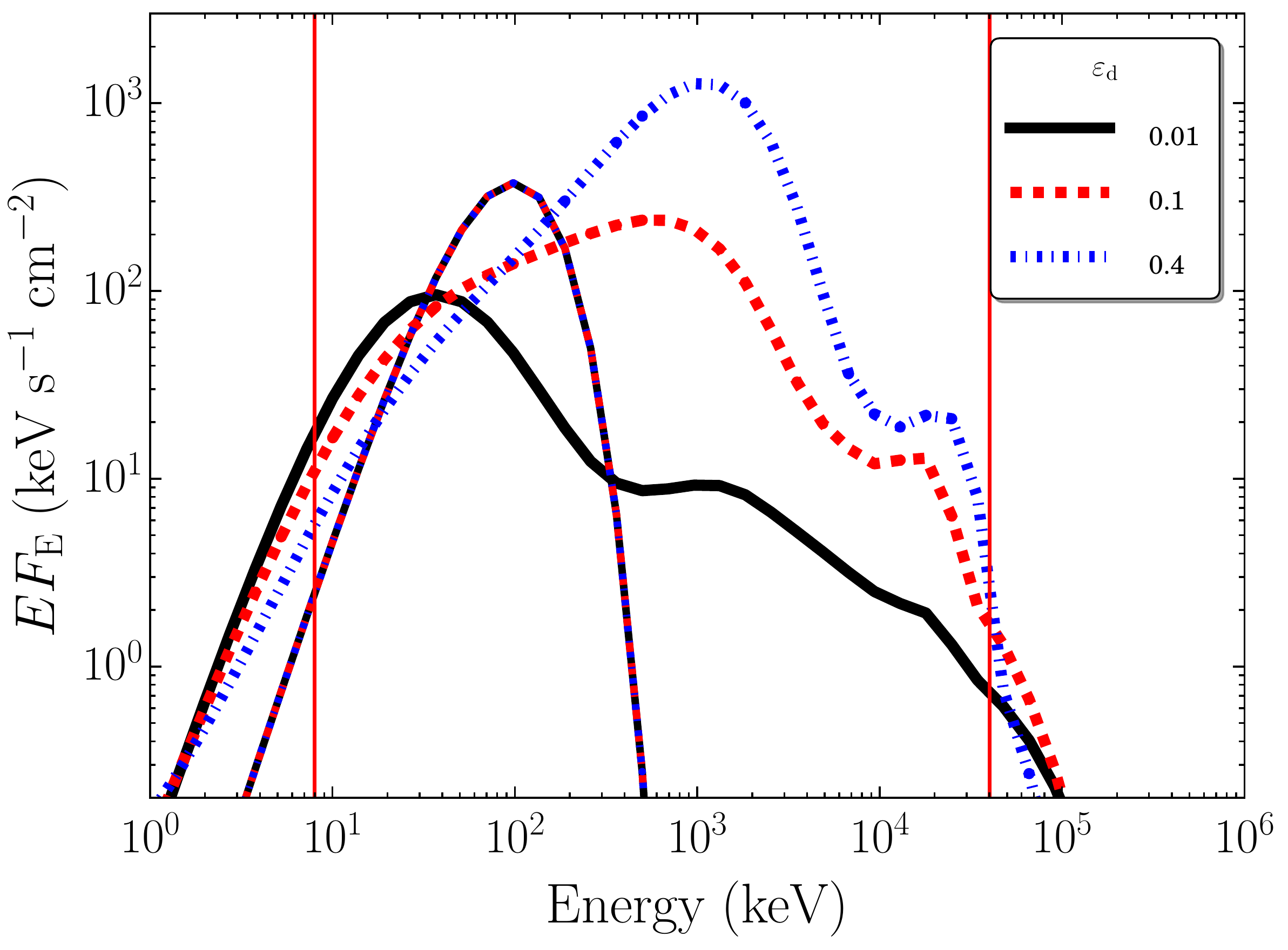}
\caption{Examples of spectra resulting from different free model parameters. The upper left, right, and bottom plot show spectra for varying $L_{0,52}$, $\Gamma$ and $\ed$, respectively. Additionally, the spectra have the following corresponding non-varying parameter values: 
$[\ed = 0.1$,$\Gamma = 300]$, 
$[\ed = 0.1$,$L_{0,52} = 100]$, 
$[L_{0,52} = 100$, $\Gamma = 300]$.
The thick lines represent the output spectrum from the code, from which we produce the table model. For illustrative purposes we also show the seed photon BB distribution, plotted in the same line style as the corresponding spectrum, but with thinner lines. The red, vertical lines show the \fermi GBM energy range, which is the fitted energy interval for the majority of GRBs. It should be noted that the plotted spectra include the adiabatic cooling factor $a$, introduced in section \ref{subsection:adcool}.}
\label{paramVarPlot}
\end{figure*}

In Fig.~\ref{paramVarPlot} we exemplify some spectral shapes as functions of the free parameters. As the dependency of spectral shape on our parameters is non-linear and we have a multi-dimensional parameter space, it is not possible to give an exact accounting for how the spectral shape changes with any single parameter. However, there are some clear features that are typical to certain parameters. Using Fig.~\ref{paramVarPlot} we give a qualitative overview of the main impact of our parameters on the spectral shapes. 

Generally, the spectrum will consist of a BB peak and an inverse Compton peak at higher energies. For a small amount of dissipation we expect the BB peak to dominate and to yield a single-peaked spectrum. When the dissipation increases the inverse Compton peak becomes more pronounced and we can obtain a single-peaked spectrum with the peak at high energies or, if the peaks are comparable in strength, we obtain a flat, `top-hat' shape, or a double peaked spectrum. 
For low values of $\varepsilon_{\mathrm{b}}$, as we consider here, the spectrum is dominated by inverse Compton scattering and the contributions from synchrotron radiation are negligible. Another feature is the pair production peak at $\sim \Gamma m_\mathrm{e} c^2 /(1+z)$.

The parameter with the clearest effect on the spectra is the fireball luminosity, $L_{0,52}$, which mainly sets the flux, since it is positively correlated to the number of photons in the spectrum and spans several orders of magnitude. Additionally, $L_{0,52}$ has the effect of setting the distance between the peaks in the spectrum caused by inverse Compton scattering. This is caused by the electron temperature being independent of $L_{0,52}$, while the temperature of the seed BB is inversely dependent on $\Lum$ \citep{2005ApJ...628..857P}. 

One of the major effects of $\Gamma$ is the Doppler shift, which sets the observed energies of the spectrum. Additionally, the initial BB temperature increases as a function of $\Gamma$, which sets the position and normalisation of the first peak in the spectrum. Increasing $\Gamma$ also yields a higher pair multiplicity, which is why we see the spectra with increasing $\Gamma$ in Fig.~\ref{paramVarPlot} get increasingly thermalised as $\Gamma$ increases. Note that this is different from the commonly assumed solution to the compactness problem, where an increasing $\Gamma$ suppresses photon-photon annihilation. Due to the assumption that $r_0 = \rd/\Gamma^2$, $\Gamma$ will alter quantities such as volume and electron number, resulting in a net effect of positive correlation between $\Gamma$ and pair multiplicity. Note that $\Gamma$ also contributes to setting the number of photons, but with an inverse dependence. However, due to its limited range and its other effects on the spectrum, it is the luminosity which works as the main driver of the predicted flux.

The last free parameter of our model, $\varepsilon_{\mathrm{d}}$, mainly changes the comoving electron temperature, $\theta$, \citep{2006ApJ...642..995P}. This sets the normalisation of the inverse Compton peak, with a higher $\ed$ resulting in a stronger peak. However, the position of the this peak is not sensitive to $\ed$ for much of the parameter space. This is because the first up-scattered photons may pair produce and yield a population of cooler electrons, usually outnumbering the baryonic electrons by a factor of several. The observed inverse Compton peak in the spectrum is a result of multiple scatterings with electrons of a wide energy distribution, making its position almost independent of the initial electron temperature.

\subsection{Additional scatterings and adiabatic cooling}
\label{subsection:adcool}
\begin{figure}
\includegraphics[scale=0.55]{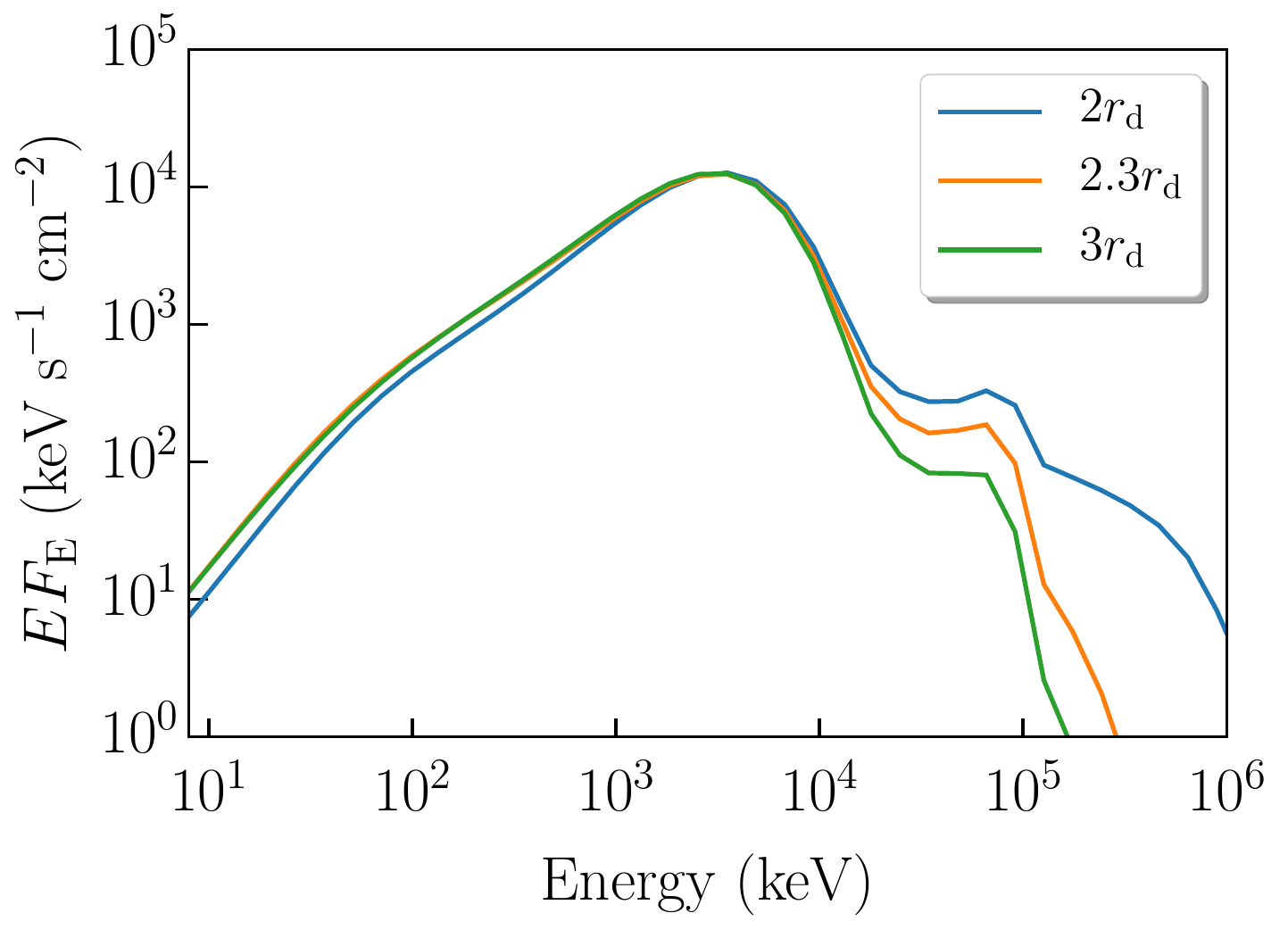}
\caption{Examples of the spectral shape for different stopping radii of the numerical code. Note that the dissipation stops at $2\rd$ for all spectra. The shape asymptotically approaches the spectrum at $\rph$, with most of the spectral evolution occurring early. These spectra were produced using the model parameters $\Lum = 300$, $\Gamma = 300$ and $\ed = 0.4$.}
\label{fig:adcool:diff_R_end}
\end{figure}

In A15 the numerical code was run only as long as there was dissipation, \ie ~between $\rd$ and $2\rd$. This takes care of the most important physics and relevant processes, but it does not include the last of the scatterings, between $2\rd$ and $\rph$. Running the code all the way to the photosphere from an optical depth of $\tau = 35$ is not feasible with the current set-up. However, it is sufficient to run the code a fraction of the way to the photosphere to capture most of the spectral changes occurring after the dissipation. In Fig.~\ref{fig:adcool:diff_R_end} we show examples of spectral shapes as a function of where the simulation ends. It is clear from the figure that there is little change already after $2.3\rd$ In this work we have chosen to let the code run between $\rd$ and $2.3\rd$. We note that the factor of 2.3 was found empirically to be a good trade-off between computational efficiency and capturing changes in the spectral shape.

As stated in A15, the code does not in its current form include adiabatic cooling. However, we would expect adiabatic cooling to yield a shift of the entire spectrum to lower energies., see \eg ~\cite{2004ApJ...613..448P}. This effect is caused by a continuous cooling of the photons as they do work on the outflow for as long as they scatter, \ie ~roughly as long as $r < r_{\mathrm{ph}}$ in our scenario. In a more realistic scenario we would also expect the photons to start decoupling before $\rph$, mainly leading to a broadening of the spectral shape \citep{2011ApJ...737...68B,2013ApJ...767..139B}. The photons continuously isotropise during collisions by converting internal photon energy to bulk kinetic energy of the outflow, and thus stay isotropic in the rest frame. As long as we have a thermal spectrum this results in a constant shift of the spectrum to lower energies. For non-thermal spectra, each scattering will also distort the spectral shape. We begin by considering only pure adiabatic cooling.
As it is a constant shift of the spectrum, this effect may be treated by analytic post-processing. We introduce an adiabatic cooling factor $a = 2(\tau_{\mathrm{d}})^{-2/3}$, where $\tau_{\mathrm{d}}$ is the optical depth at the end of the dissipation, with which we simply shift the spectrum. Hence the number of photons at a given energy, $N(E)$ is shifted such that $N(E) = N_{\mathrm{ad}} (E_{\mathrm{ad}})$, with $E = aE_{\mathrm{ad}}$ and where $E_{\mathrm{ad}}$ is the shifted energy. For more details on the adiabatic cooling factor, including its limitations, see \cite{2011ApJ...737...68B} and \cite{2013ApJ...767..139B}. The constant shift using $a$ is an approximation to taking into account all scatterings up to the photosphere. 

\begin{figure}
\includegraphics[scale=0.55]{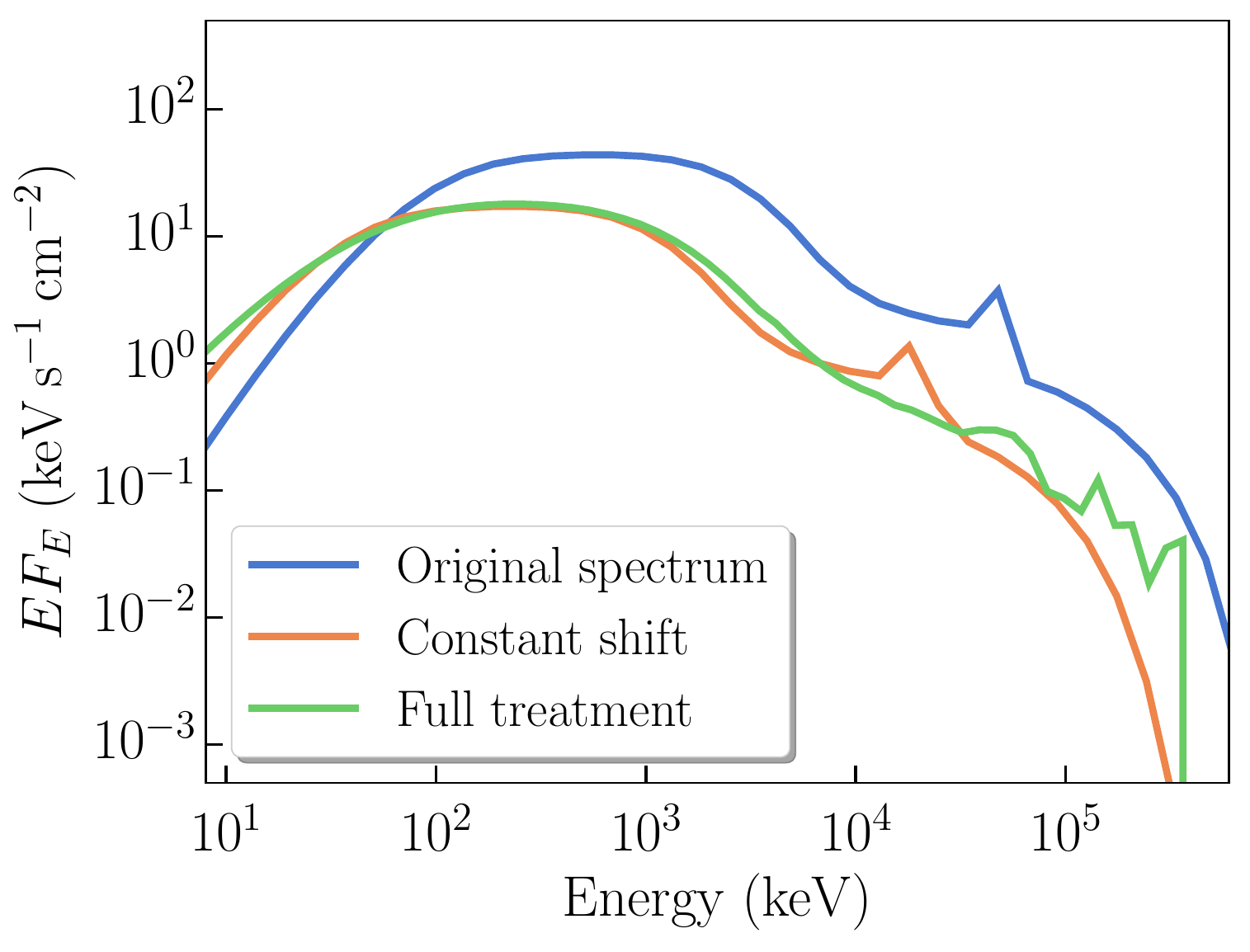}
\caption{Illustration of the effect of adiabatic cooling. The plot shows the original model spectrum with parameters $\Lum = 10$, $\Gamma = 250$ and $\ed = 0.05$ assuming a redshift $z=1$, together with the result of our approximation of a constant shift, as well as the full treatment of all scatterings and all geometric effects of a spherical outflow. The full post-process treatment is calculated using a numerical code from \protect\cite{2017arXiv170802633L}. The spiky behaviour at high energies in the curve of the full treatment is numerical noise and the small peak in the original and artificially cooled spectra is the pair annihilation peak.}
\label{fig:adcool:cool}
\end{figure}

In Fig.~\ref{fig:adcool:cool} we show an example of an original model spectrum, as calculated from our numerical code, compared with the spectrum after having applied the adiabatic cooling factor, as well as the spectrum after a full treatment of all scatterings, adiabatic cooling and geometric effects up to the photosphere. The full treatment was performed using a numerical code from \cite{2017arXiv170802633L}, which dynamically couples hydrodynamics to Monte Carlo radiative transfer inside a spherical outflow. The radiation can then be followed as it cools adiabatically, and self-consistently decouples from the flow at a fuzzy photosphere. This code can be used to study the spectral evolution after dissipation. However, the code is computationally very expensive, which prevents us from using it for all spectra. 
Fig.~\ref{fig:adcool:cool} illustrates that our approximation of a constant shift yields good agreement of the integrated model flux.
However, we note a small discrepancy between the spectrum with the constant shift approximation and the full treatment in the low energy slope $< 20$ keV. This effect is a result of the full treatment taking a fuzzy photosphere and geometric effects into account. We thus conclude that the approximate treatment of the adiabatic cooling is good, and that it represents a significant improvement over the original model spectrum, but that the inclusion of geometric effects would lead to a slightly softer low-energy slope.

\section{Fitting with the model}
\label{section:fitting_section}

\subsection{Data sample}
\label{subsection:fitting:datasection}
Our sample contains the brightest GRBs with known redshifts observed by \fermi before 2016-06-01. The known redshift\footnote{From \url{http://www.mpe.mpg.de/~jcg/grbgen.html.}} helps us fix the normalisation parameter of the model spectrum via the corresponding luminosity distance, instead of leaving it as a free parameter. This is important because we want to be able to test the model's ability to correctly predict the GRB flux. We chose a fluence cut of $>10^{-5}$ erg cm$^{-2}$, in order to allow us to perform a time-resolved analysis with good signal strength. The resulting sample contains 36 GRBs, listed in Table ~\ref{burstsample}.

For all GRBs we use Time Tagged Event (TTE) data from the Gamma-ray Burst Monitor (GBM) NaI and BGO detectors \citep{2009ApJ...702..791M}. We select detectors by the scheme outlined in \cite{2016yCat..22230028B}, choosing up to three NaI detectors with an angle towards the source lower than $60 \deg$ as well as the BGO detector with the lowest angle towards the source. The energy range fitted was 8--1000~keV (NaI) and 200~keV -- 40~MeV (BGO).

There are also LAT-LLE data available at high energies for nine bursts in the sample (GRB~080916C, GRB~090323A, GRB~090902B, GRB~090926A, GRB~110731A, GRB~130518A, GRB~141028A, GRB~150403A and GRB~160509A). The LLE data are a type of LAT data designed for the use with bright transients to bridge the energy range of the LAT and GBM \citep{2010arXiv1002.2617P}. The LLE data (pass7) were obtained using gtburst version v02-01-03p50 and \fermi Science Tools version v9r32p5. The fitted energy range is 30--100~MeV. As noted in section \ref{subsection:adcool}, it is in this energy range our model has the largest uncertainties. However, we note that the inclusion or exclusion of the LLE data has little to no effect on the results (see appendix~\ref{appendix:C:LLEdata} for a discussion on the impact of the LLE data on our analysis).

\begin{table}
\centering
\caption{GRB sample. The sample consists of bursts with fluence $>10^{-5}$ erg cm$^{-2}$ and with a known redshift. There are 36 bursts in total in the sample, which was collected before 2016-06-01.}
\label{burstsample}
\setlength\tabcolsep{6pt}
\begin{tabular}{llll} 
\hline \\[-9pt] GRB & T90  & Fluence  & redshift \\  & (s)& (erg cm$^{-2}$) & \\ \hline \\[-7pt]080916C  &  63.0  &  6.03e-05  &  4.35   
\\081121  &  42.0  &  1.53e-05  &  2.512  
\\081221  &  29.7  &  3.00e-05  &  2.26   
\\090323A  &  135.2  &  1.18e-04  &  3.57  
\\090424  &  14.1  &  4.63e-05  &  0.544 
\\090516  &  123.1  &  1.72e-05  &  4.109   
\\090618  &  112.4  &  2.68e-04  &  0.54  
\\090902B  &  19.3  &  2.22e-04  &  1.822  
\\090926A  &  13.8  &  1.47e-04  &  2.1062   
\\090926B  &  55.6  &  1.08e-05  &  1.24  
\\091003A  &  20.2  &  2.33e-05  &  0.8969  
\\091127  &   8.7  &  2.07e-05  &  0.49  
\\100414A  &  26.5  &  8.85e-05  &  1.368   
\\100728A  &  165.4  &  1.28e-04  &  1.567  
\\100814A  &  150.5  &  1.49e-05  &  1.44  
\\100906A  &  110.6  &  2.33e-05  &  1.727
\\110731A  &   7.5  &  2.29e-05  &  2.83  
\\111228A  &  99.8  &  1.81e-05  &  0.714  
\\120119A  &  55.3  &  3.87e-05  &  1.728  
\\120711A  &  44.0  &  1.94e-04  &  1.405  
\\120716A  &  237.1  &  1.44e-05  &  2.486   
\\130215A  &  143.7  &  1.86e-05  &  0.597  
\\130420A  &  105.0  &  1.16e-05  &  1.297  
\\130518A  &  48.6  &  9.46e-05  &  2.488 
\\130925A  &  215.6  &  8.48e-05  &  0.347 
\\131105A  &  112.6  &  2.38e-05  &  1.686  
\\140206A  &  27.3  &  1.55e-05  &  2.73 
\\140213A  &  18.6  &  2.12e-05  &  1.2076  
\\140423A  &  95.2  &  1.81e-05  &  3.26  
\\140512A  &  148.0  &  2.93e-05  &  0.725  
\\141028A  &  31.5  &  3.48e-05  &  2.33 
\\150314A  &  10.7  &  8.16e-05  &  1.758   
\\150403A  &  22.3  &  5.47e-05  &  2.06  
\\150821A  &  103.4  &  5.21e-05  &  0.755  
\\151027A  &  123.4  &  1.41e-05  &  0.81 
\\160509A  &  369.7  &  1.79e-04  &  1.17   \\
\hline
\end{tabular}
\end{table}

\subsection{Time-resolved spectral analysis}
\label{binning_section}
Our spectral analysis is time-resolved, using Bayesian blocks \citep{2013arXiv1304.2818S} as binning method\footnote{We use the python implementation v1.1.1.1 of the algorithm from the \textit{Fermi} Science Tools}. The algorithm is performed on unbinned TTE data, with the false alarm rate parameter $p_0 = 0.05$. We perform the binning using the NaI detector with the lowest angle of incidence. All signal-to-noise ratio (SNR) presented are also calculated from this detector. The background is fitted with polynomials. We account for repointing of the spacecraft and temporal evolution in the background by using rsp2 responses when available.

Bayesian blocks with an SNR cut is the most appropriate choice of binning given our model. The model does not include any time-evolution of the physical properties of the jet, meaning that any spectral evolution observed is assumed to be caused by central engine activity. Hence we assume constant central engine activity and no spectral evolution in each time bin we analyse. A constant central engine activity may be expected to give rise to a constant Poisson rate of photons in the light curve, which is what we recover with the Bayesian blocks scheme. However, this method may yield bins with a very low SNR. 
Thus, an SNR cut of $4$ was placed on the bins in accordance with what we found in the validation tests (see appendix~\ref{appendix:A}). We use an expression for the SNR as derived in \cite{2018ApJS..236...17V}, for a Poisson measurement and a Gaussian background. 
Previous work by \eg ~\cite{2016A&A...588A.135Y} also employs an SNR requirement when performing time-resolved spectral analysis. The large difference in the level of the SNR cut between our work and theirs is a result of the different expressions used for calculating the SNR. See also \cite{2014MNRAS.445.2589B} for a discussion about different binning techniques, and \cite{2015A&A...578A..80W} for general transient timing using Bayesian blocks.

\begin{figure}
\includegraphics[scale=0.58]{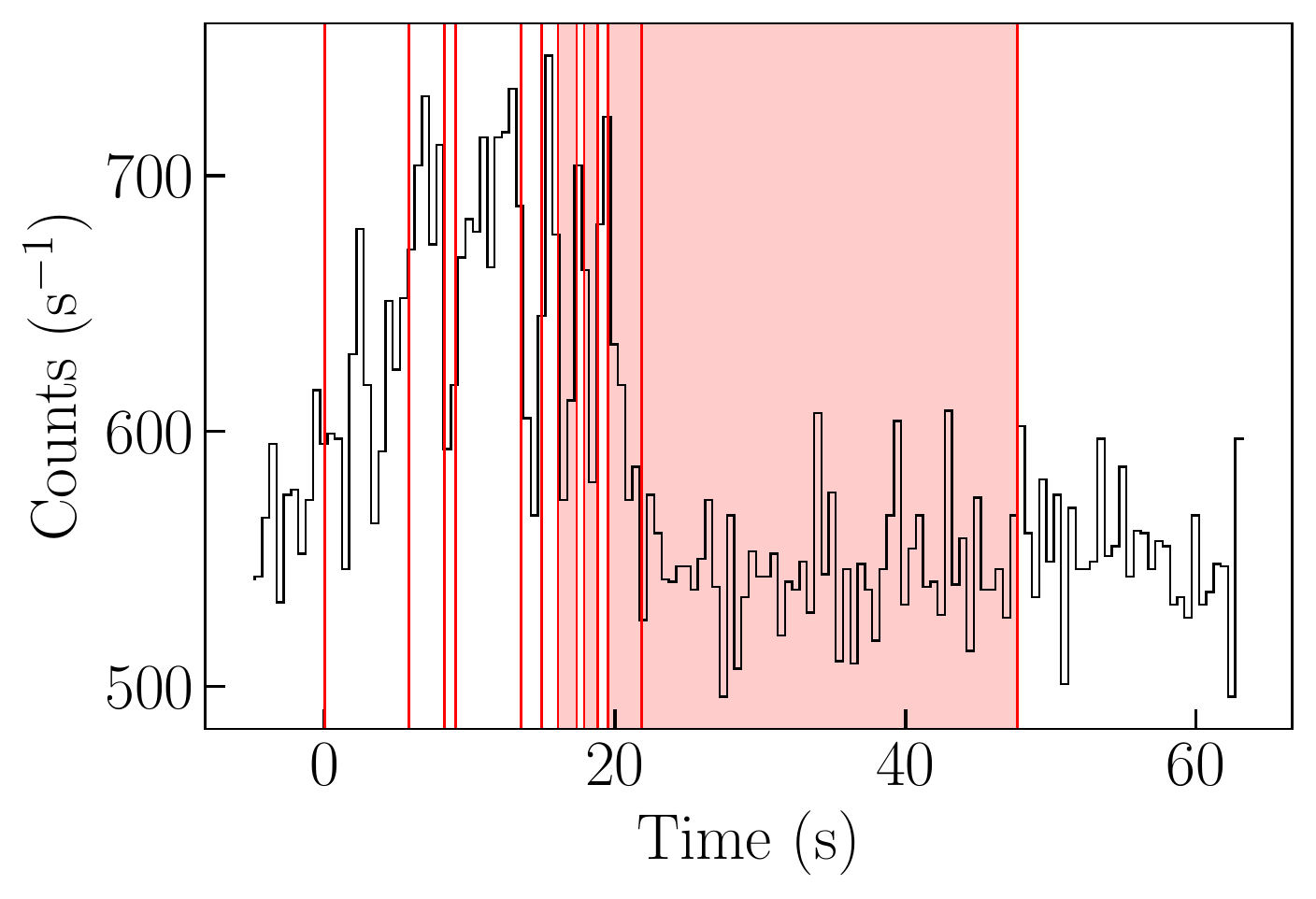}
\caption{Example of Bayesian blocks binning with SNR cuts applied. The burst in this example is GRB081121. The red vertical lines outline the time bins found using Bayesian blocks. Time bins 8, 10, 12 and 13 (corresponding to times $16.1-17.4$, $17.9-18.8$, $19.5-21.8$ and $21.8-47.7$ s) are excluded from the analysis due to the SNR cut at 4. These bins are shaded red. The light curve is produced with 0.3 s bins, using the NaI 11 detector, which is the same detector used to bin the data. The figure presents raw detector counts from all energy channels, with no background subtraction or off-axis effects accounted for.}
\label{binning_example}
\end{figure}

In Fig.~\ref{binning_example} we show an example of a light curve binned with Bayesian blocks, including 4 bins excluded at SNR$<4$. Our burst sample, defined in section \ref{subsection:fitting:datasection}, contains two bursts which yield no bins above the SNR threshold and are thus not analysed. The bursts which did not make the SNR cut are GRB~090516 and GRB~140423A. In total 145/779 bins are removed before analysis due to low SNR, giving us 634 time bins to analyse. 

The analysis is performed using {\scriptsize PyXspec}, a python implementation of HEASARC's {\scriptsize XSPEC} 12.8.1g, with pgstat statistics and with PHA (Pulse Height Analyser) data. We fit both our table model, DREAM1.2, as well the Band function, since the latter is used as a standard function in the field. We want to examine possible correlations between Band function parameters (\ie ~the low energy slope, $\alpha$, the peak energy, $E_{\mathrm{p}}$, and the high energy slope, $\beta$), and the model parameters of DREAM1.2. 

We note that the finite resolution of the table model grid results in additional uncertainties to those found from the regular fitting with Maximum Likelihood. Examining the points in the parameter space where we expect the largest resolution-related uncertainties, we find the median of the uncertainties to be less than $4-8$ per cent, depending on the parameter. $\Lum$ and $\ed$ have the largest uncertainties. This evaluation is presented in appendix~\ref{appendix:B}. 
These uncertainties are not propagated into further analyses.

We have tested for effective area corrections between the detectors and find that the results do not change significantly if we allow for this correction or not. The size of the errors on fitted parameters are also not systematically or significantly affected. We present the fit results obtained without effective area corrections.

We assume standard $\Lambda$-CDM cosmology with a Hubble constant of $H_0 = 67.3$, a cosmological constant, $\Omega_{\Lambda} = 0.685$ and a matter density $\Omega_{\mathrm{M}} = 0.315$, \citep{2014A&A...571A..16P}. All uncertainties are 1 $\sigma$ unless otherwise stated. Statistical uncertainties on parameter estimates are obtained using the  {\scriptsize XSPEC} \textit{error} command, where model parameters are varied until the fit statistic changes by the desired level.

\subsection{Goodness of fit}
\label{subsection:gof}
We wish to have a metric by which to judge if a fit should be rejected since we do not expect a fit of poor quality to give physically meaningful information. For this purpose we employ a Monte-Carlo (MC) parametric bootstrap method, using {\scriptsize XSPEC}. For each fit to real data we simulate 5000 spectra from the given best-fitting parameter values, using {\scriptsize XSPEC}'s built-in function \textit{fakeit}. The spectra are simulated using the response and background files of the original data. We then re-fit the model to the fake spectra and sample the fit statistic.

We perform a goodness of fit (GOF) test by comparing the pgstat value of the model fitted to the real data with the pgstat distribution obtained from the MC simulations. We choose to reject fits which have a fit statistic outside the $99.7$ per cent central confidence interval of the sampled distribution (corresponding to $3 \sigma$).
We performed tests with cuts corresponding to values between 1 and 5 $\sigma$, and found that it does not affect the overall shape of the distribution of parameter values in the accepted fits. This indicates that our results are qualitatively robust to the level of this cut.

\section{Results}
\label{section:Results}
After binning the data in Table~\ref{burstsample} and performing the SNR$>4$ cut we fit DREAM1.2 to the data. We discard all fits with a parameter on the boundary and perform our GOF test on the remaining fits. Out of 634 time bins with $\mathrm{SNR}>4$, 267 have no parameter on the boundary of the parameter space. 171 of these fits pass the GOF test and constitute our sample of accepted fits, which we analyse further. This corresponds to approximately 27 per cent of analysed spectra being accepted, with 10 bursts having at least 50 per cent accepted fits.
The rejected fits and their implications on our model assumptions are discussed in section~\ref{subsection:discuss_badfits}. See appendix~\ref{appendix:E:burstsumtable} for a table summarising the number of accepted and rejected time bins for each burst.

\subsection{Best-fitting parameters}
\label{subsec:Results:bestfitpars}
\begin{figure*}
\includegraphics[scale=0.6]{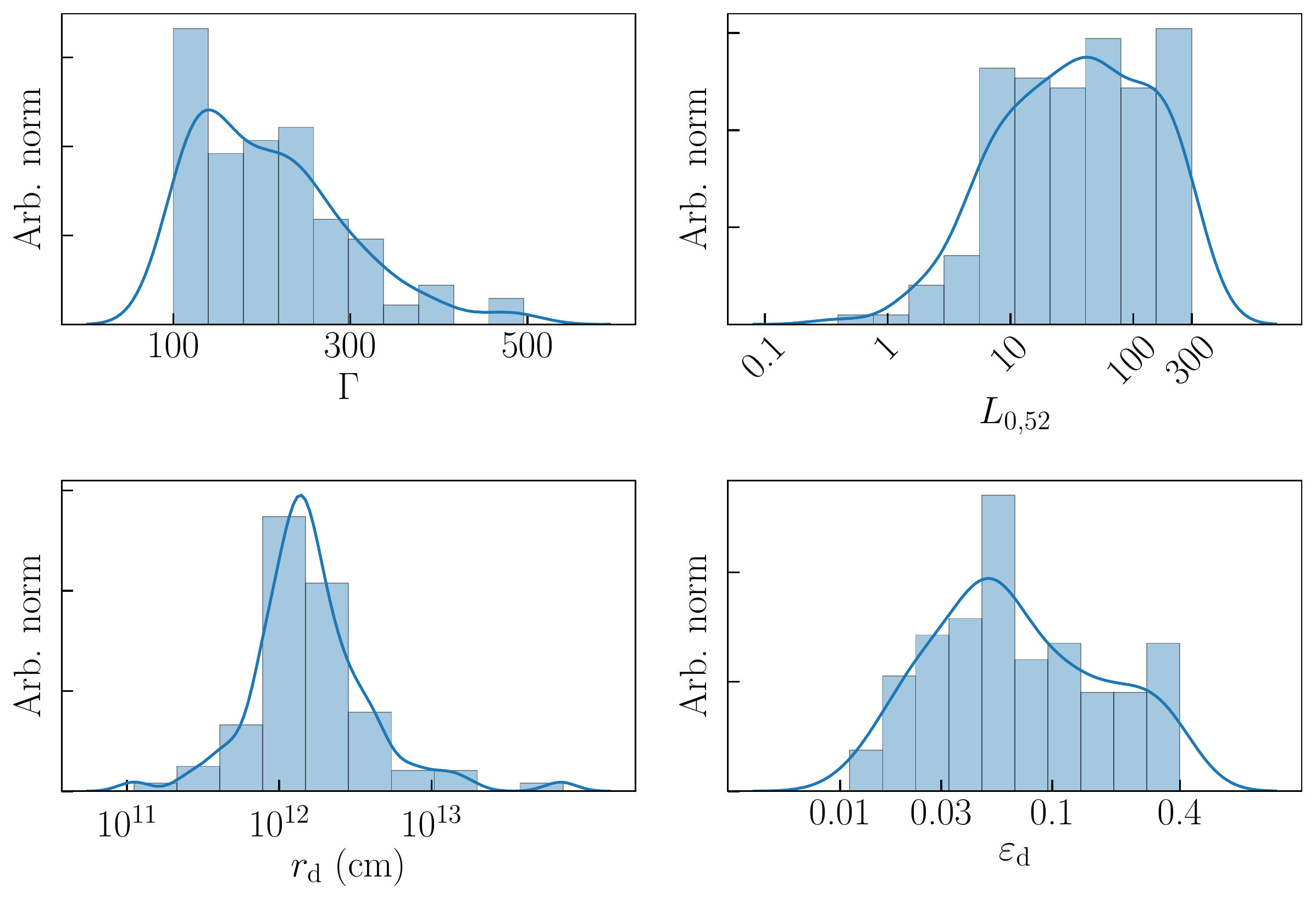}
\caption{Distribution of best-fitting parameter values, as well as the derived values of $\rd$, of the 171 fits which passed the GOF and SNR cut. The corresponding Gaussian kernel density estimates are also shown, providing an un-binned representation of the distributions. The introduction of kernel density estimates scales the histograms so that they represent probability distributions, and the y-axes are normalised accordingly.}
\label{fig:globParDistHist}
\end{figure*}

In Fig.~\ref{fig:globParDistHist} we present the global parameter distributions for the free model parameters of our accepted fits, as well as the dissipation radius, $\rd$. We have included Gaussian kernel density estimates in the histograms to aid with visual interpretation. The histograms span the entire parameter space, except for $\rd$, which has allowed values in the range $3\times 10^{8} -4\times 10^{15} $ cm. 
We include $\rd$ since it may give information about the dissipation scenario and/or the GRB's relation with the progenitor.
$\Gamma$ is almost evenly spread in the range $100-250$, with a small peak at $\Gamma \sim 120$. $\Lum$ has the majority of its fits in the range $10-100$. The $\ed$ distribution peaks at $\ed \sim 0.05$, with 82 per cent of fits having $\ed < 0.2$. In table ~\ref{table:fitResults} we present all the best-fitting parameter values from successful fits. The full table containing all successfully fitted bursts and bins is available as supplementary online material. 

\begin{table}
\centering
\caption{Best-fitting parameter values for accepted fits with DREAM1.2. The 1 $\sigma$ asymmetric uncertainties are indicated by super- and subscripts. A missing uncertainty means that the uncertainty is unconstrained at the 1 $\sigma$ level. The complete table is available as online supplementary material.}
\label{table:fitResults}
\setlength\tabcolsep{3pt}
\begin{tabular}{lccccc}
\toprule
     Burst & 
     \multicolumn{1}{p{1.4cm}}{\centering Time bin \\ (s)}& 
     $\varepsilon_{\mathrm{d}}$ &
     $L_{0,52}$ & 
     $\Gamma$ &   
     \multicolumn{1}{p{1.3cm}}{\centering $\rd$ \\ ($10^{12}$ cm)} \\ \hline 
\setlength\extrarowheight{3pt} \\[-7pt]
090424 &                                            0.0-0.1 &  $0.075_{-0.040}^{+0.062}$ &     $9.6_{-3.9}^{+5.6}$ &        $140^{+53}$ &                                      $1.2 \pm 2.5$ \\[2pt]
  090424 &                                            0.1-0.2 &  $0.105_{-0.048}^{+0.023}$ &   $17.2_{-3.1}^{+10.4}$ &  $151_{-10}^{+55}$ &                                      $1.7 \pm 1.3$ \\[2pt]
  090424 &                                            0.2-0.4 &  $0.030_{-0.008}^{+0.008}$ &    $55.3_{-4.7}^{+9.0}$ &  $246_{-12}^{+26}$ &                                      $1.2 \pm 0.3$ \\[2pt]
  090424 &                                            0.4-0.5 &  $0.025_{-0.001}^{+0.003}$ &         $107_{-4}^{+5}$ &    $288_{-7}^{+7}$ &                                      $1.5 \pm 0.1$ \\[2pt]
  090424 &                                            0.5-0.6 &  $0.028_{-0.004}^{+0.009}$ &       $164_{-20}^{+11}$ &  $314_{-24}^{+12}$ &                                      $1.8 \pm 0.3$ \\
\end{tabular}
\end{table}

\subsection{Correlations}
\label{section:correlations}
It is of interest to search for correlations between the different best-fitting parameters of the model, as well as between our model parameters and the observed flux, isotropic equivalent rest frame luminosity, $L_{\mathrm{iso,z}}$, and redshift. We also examine correlations between our model parameters and the corresponding Band function parameters. Fluxes and luminosities were calculated with $k$-correction \citep{2002astro.ph.10394H}, using a cutoff power law in the stellar rest frame\footnote{This is the rest frame of the central engine, where the redshift has been accounted for.} energy range $1$~keV - $10 $~MeV. When performing the correlation analysis we consider only fits with well constrained parameter errors, \ie ~where the statistical errors on all parameters obtained during fitting are contained inside the parameter space. This leaves us with a subset of 114 bins for the correlation analyses involving $\Epz$ and 118 for the other correlations.

To search for correlations, we use a hierarchical Bayesian model for fitting a straight line to data with errors in both the dependent and independent variables, described in \cite{2007ApJ...665.1489K}. We use this technique to find estimates of the linear relation parameters, as well as estimates for the population linear correlation coefficient, $\rho$. The errors on derived parameters have been obtained with standard error propagation.

\begin{figure*}
\includegraphics[scale=0.42]{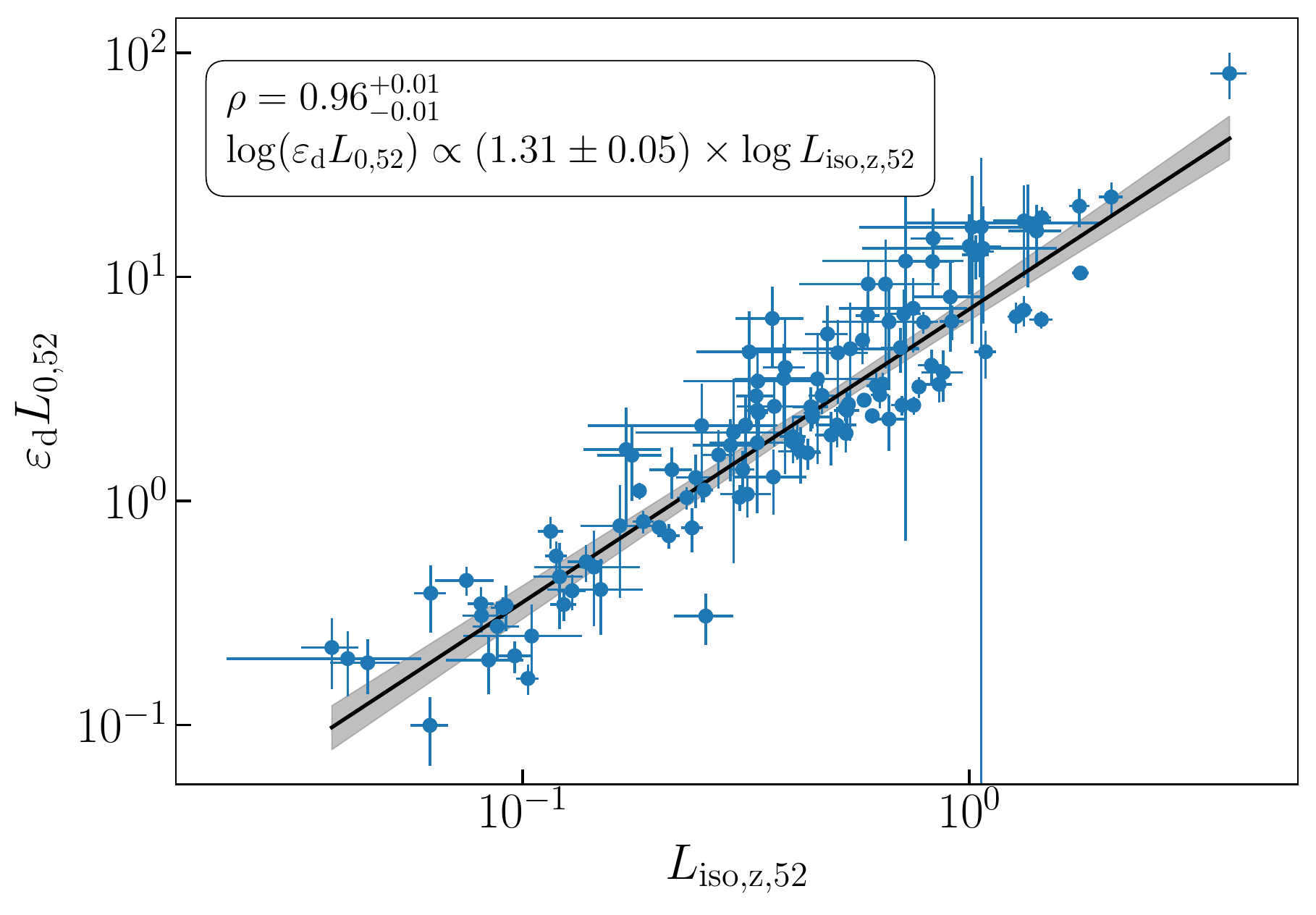}
\includegraphics[scale=0.42]{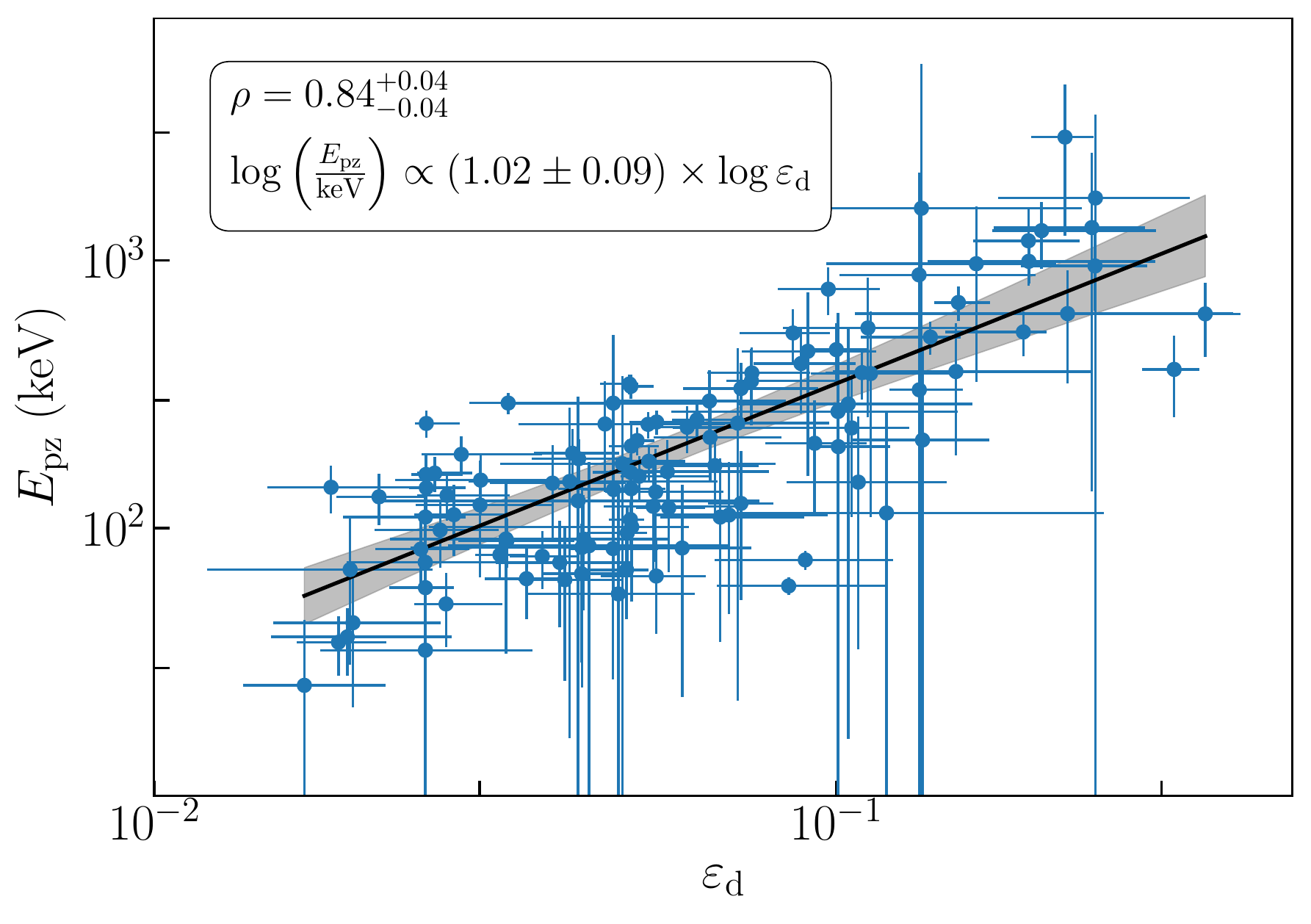}
\includegraphics[scale=0.42]{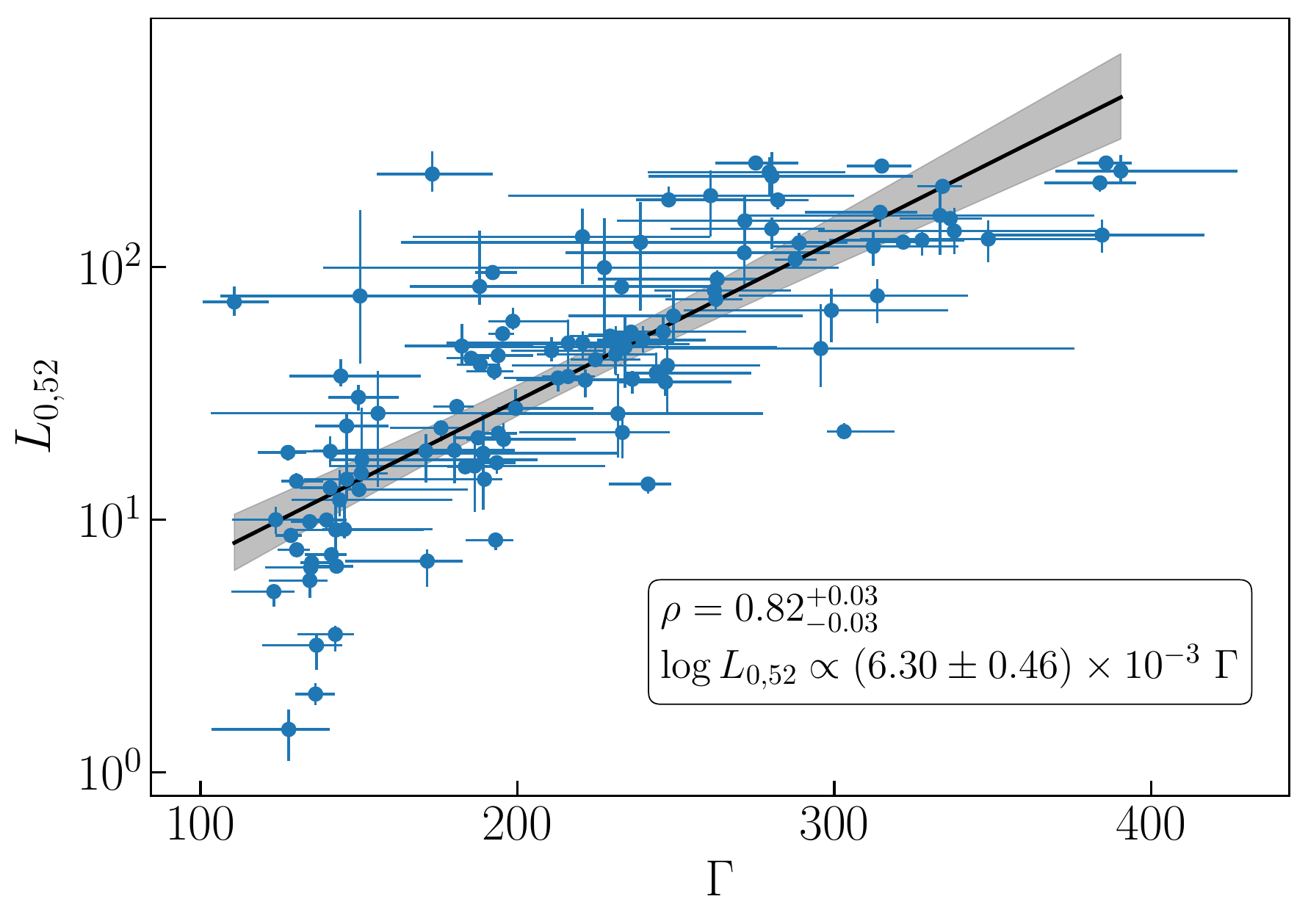}
\caption{Correlation between $\log (\ed L_{0,52}) - \log L_{\mathrm{iso,z}}$, $\log \Lum - \Gamma$ and $\log \Epz - \log \ed$, in the left, right and bottom panels, respectively. The legends show the estimates of the linear correlation coefficients, including the 68.3 per cent credible regions, as well as the linear regression slope estimate. 
In the top left panel, $L_{\mathrm{iso,z,52}} = L_{\mathrm{iso,z}} / (10^{52}$ erg s$^{-1}$).
The grey bands denote the 95 per cent pointwise confidence intervals of the regression line.}
\label{fig:correlations} 
\end{figure*}

As expected, we find a very strong correlation between the logarithm of the model luminosity, $L_{0,52}$, and the logarithm of the observed luminosity, $L_{\mathrm{iso,z}}$. If we instead consider $ \log (\ed \Lum) - \log L_{\mathrm{iso,z}}$, the correlation becomes even tighter. This correlation serves as a sanity check.
Additionally, we find a correlation between the model parameters $\Gamma$ and $\log \Lum$, as well as between $\log \ed - \log \Epz$, where $\Epz$ is the Band function peak energy in the stellar rest frame.
All correlations are presented in Fig.~\ref{fig:correlations}. The estimated linear correlation coefficients, including their 1 $\sigma$ credible intervals, as well as the regression slopes are also shown in the figure.

\begin{figure*}
\includegraphics[scale=0.35]{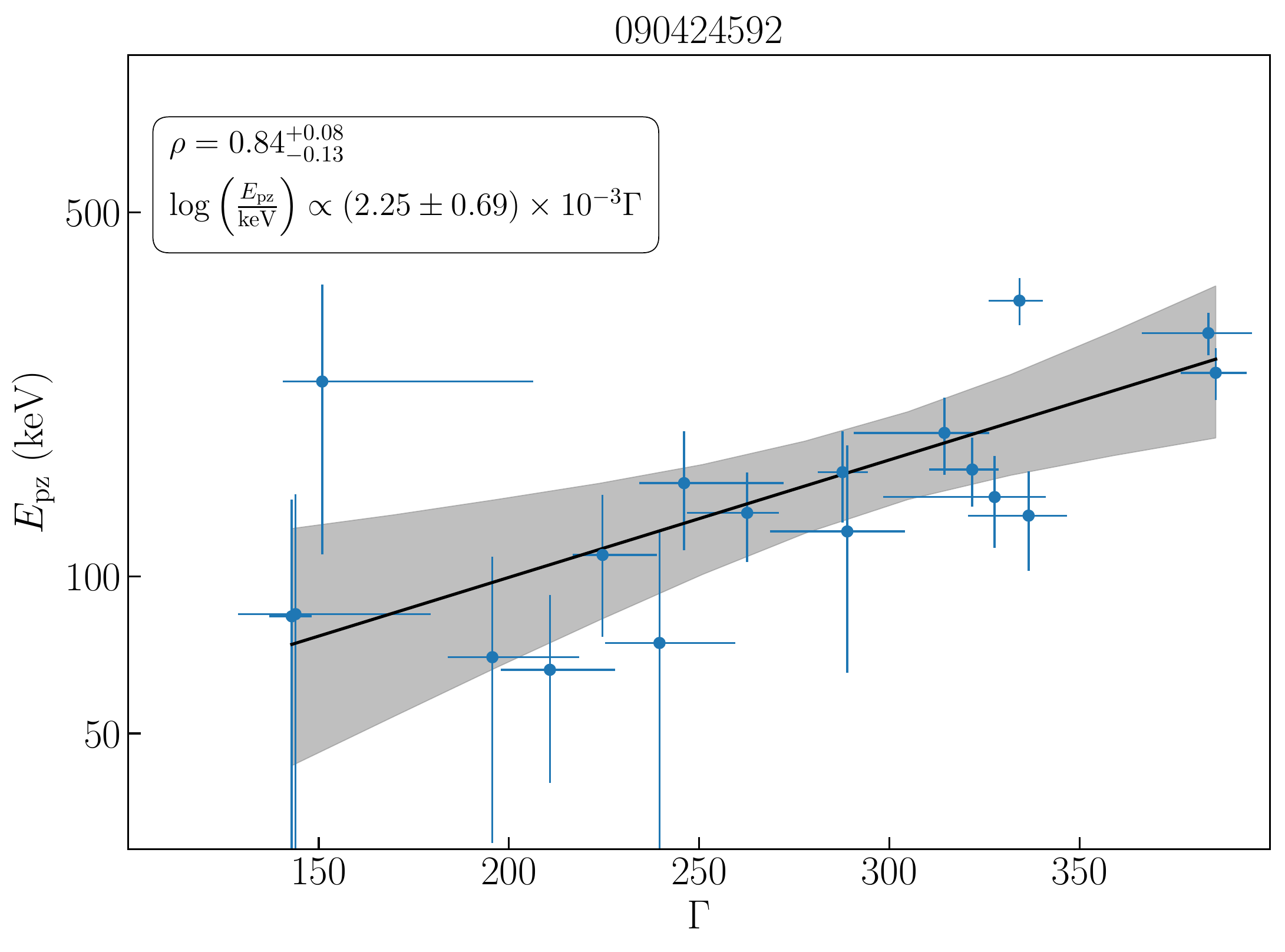}
\includegraphics[scale=0.35]{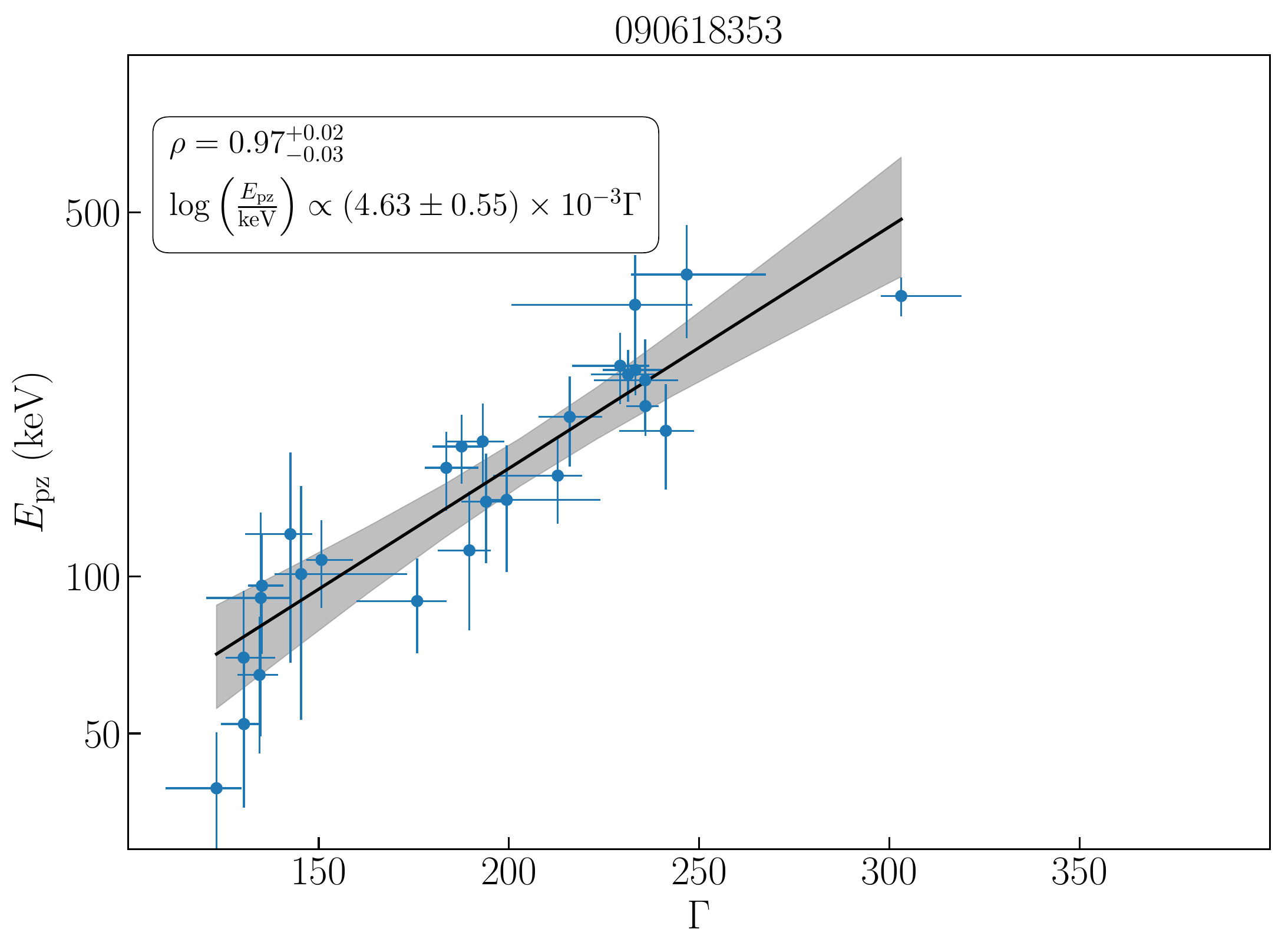}
\caption{Correlation between model parameter $\Gamma$ and the Band function $\log \Epz$ in GRB~090424 and GRB~090618. These are the only two bursts in the sample where this correlation is significant. The legends show the estimates of the linear correlation coefficients, including the 68.3 per cent credible regions, as well as the linear regression slope estimates. The grey bands denote the 95 per cent pointwise confidence intervals on the regression lines.}
\label{fig:GammaEpz}
\end{figure*}

Finally, we also find a correlation between $\log \Epz$ and $\Gamma$, which is not present across the entire sample, but instead only exists within some bursts. We have only investigated this correlation in bursts with at least 5 accepted time bins. In Fig.~\ref{fig:GammaEpz} we present the two bursts in which the correlation is significant. The correlation is considered significant if the 95 per cent highest posterior density region of the marginal posterior for the correlation coefficient does not include 0.
The correlations are discussed further in section \ref{subsection:discuss_corr}. Notably, there are no correlations with the Band function $\alpha$ or $\beta$. The implications of this are discussed in section \ref{subsection:BandTableComp}.

\subsection{Accepted versus rejected fits}
\label{goodvsbad}
To determine why only 171 out of 634 fits are accepted, we consider the parameter distributions of best-fitting parameters from our model, the physical quantities flux, redshift and $L_{\mathrm{iso,z}}$, as well as parameter values of the corresponding Band function fits. 

Examining the parameter distributions of rejected fits, we find that 302 out of 463 rejected fits have $\Lum = 300$. The remaining 161 spectra, which have not reached the upper bound in $\Lum$, exhibit a greater variation in their parameter values. Additionally, out of these 302 spectra, 260 also have $\ed = 0.4$. Furthermore, manual inspection of spectra shows that the fits that hit the upper boundary in $\Lum$ and $\ed$ are significantly worse than those that do not.

\begin{figure}
\includegraphics[scale=0.38]{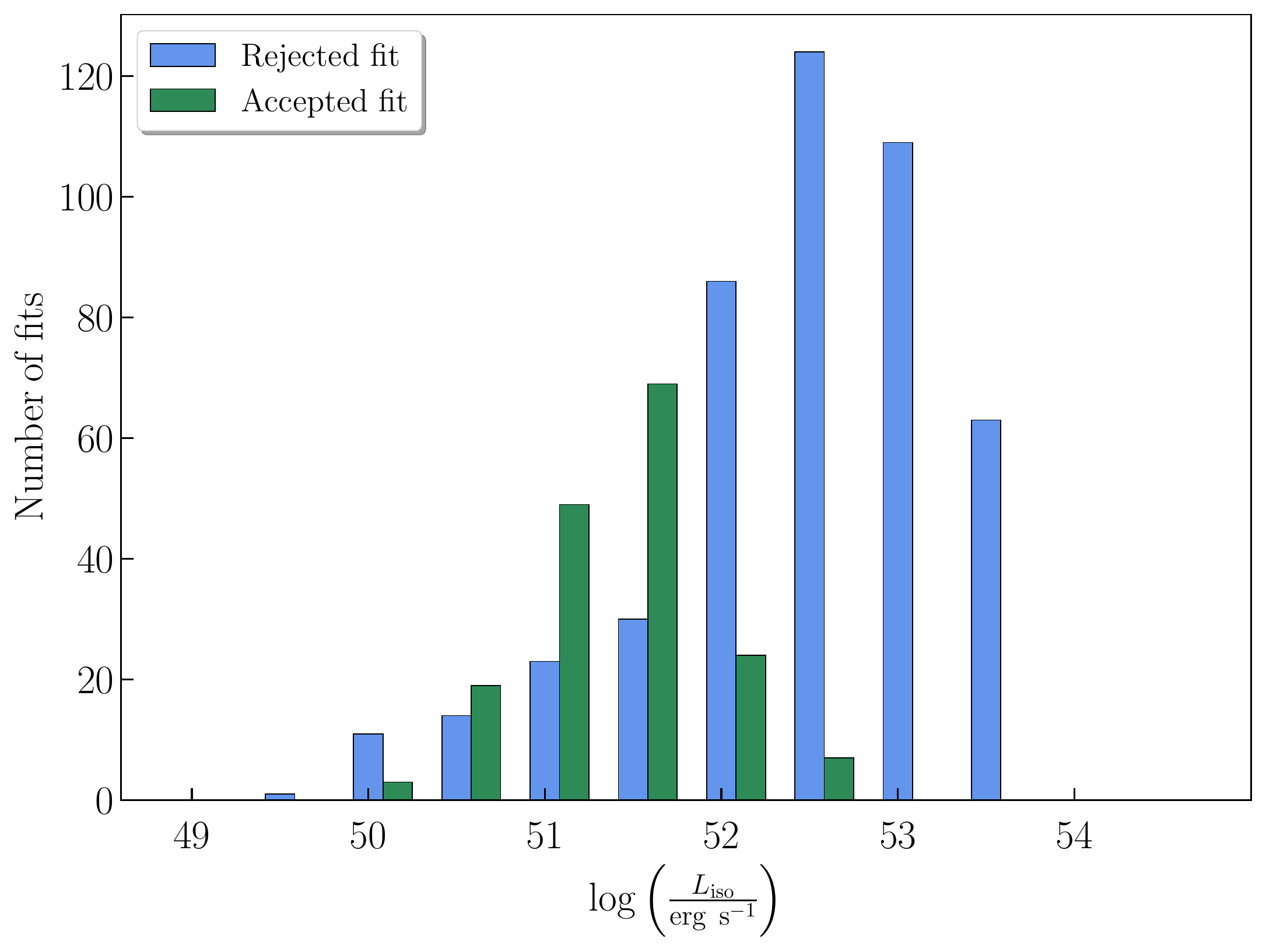}
\caption{Histograms of accepted and rejected fits in terms of their isotropic equivalent luminosity, $L_{\mathrm{iso,z}}$. The histograms are uniform in logarithmic space with two bins per decade, and cover the range of $10^{48} - 10^{55}$ erg s$^{-1}$. The bars have been separated for illustrative purposes. Blue bars show the number of rejected fits and green bars show the number of accepted fits, with the acceptance criteria being the GOF test described in section \ref{subsection:gof}. $L_{\mathrm{iso,z}}$ was obtained from fits with a cut-off power law.}
\label{physparamgoodvsbad}
\end{figure}

\begin{figure*}
\includegraphics[scale=0.35, trim=0cm 0cm 0cm 0cm]{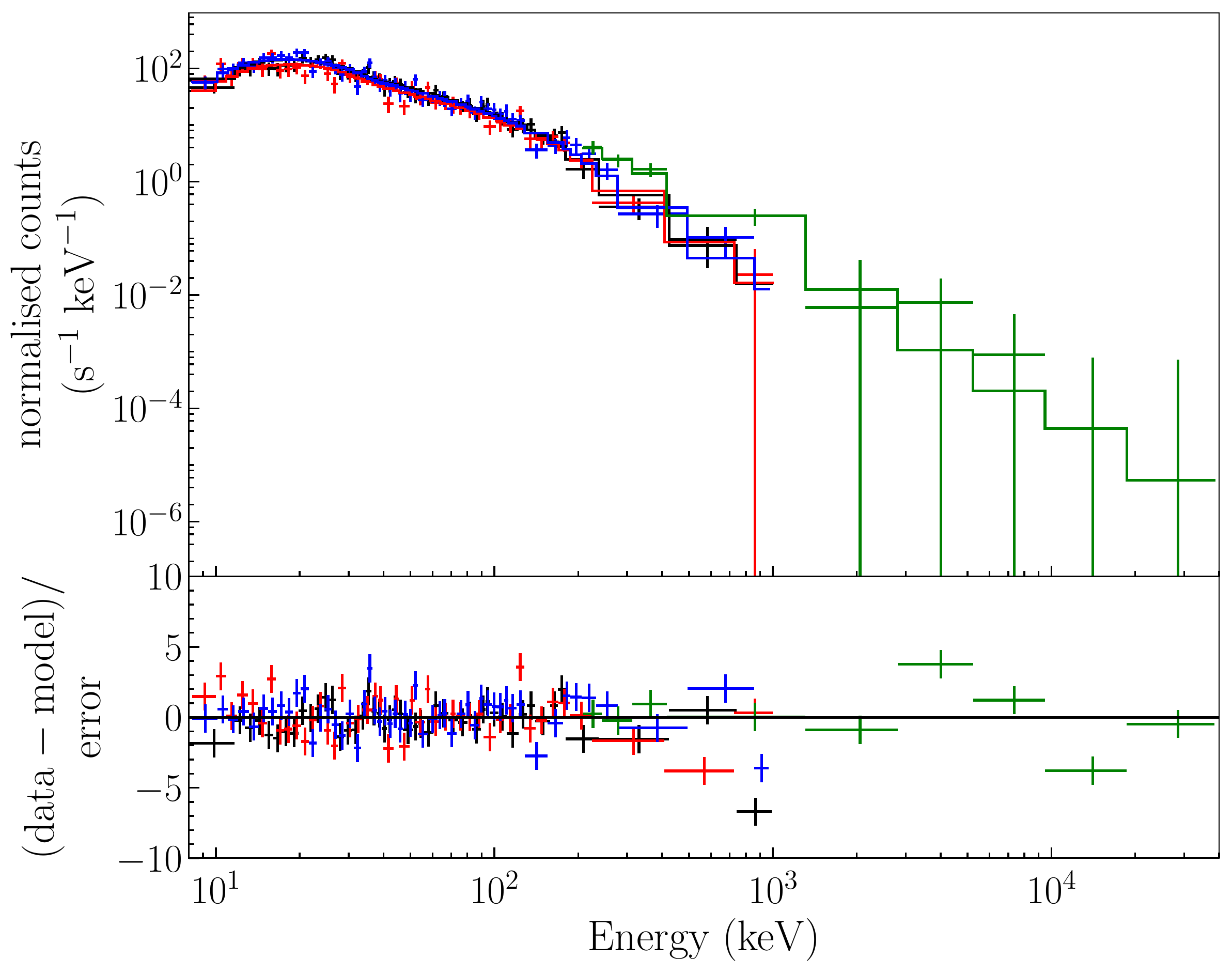}
\includegraphics[scale=0.35, trim=0cm 0cm 0cm 0cm]{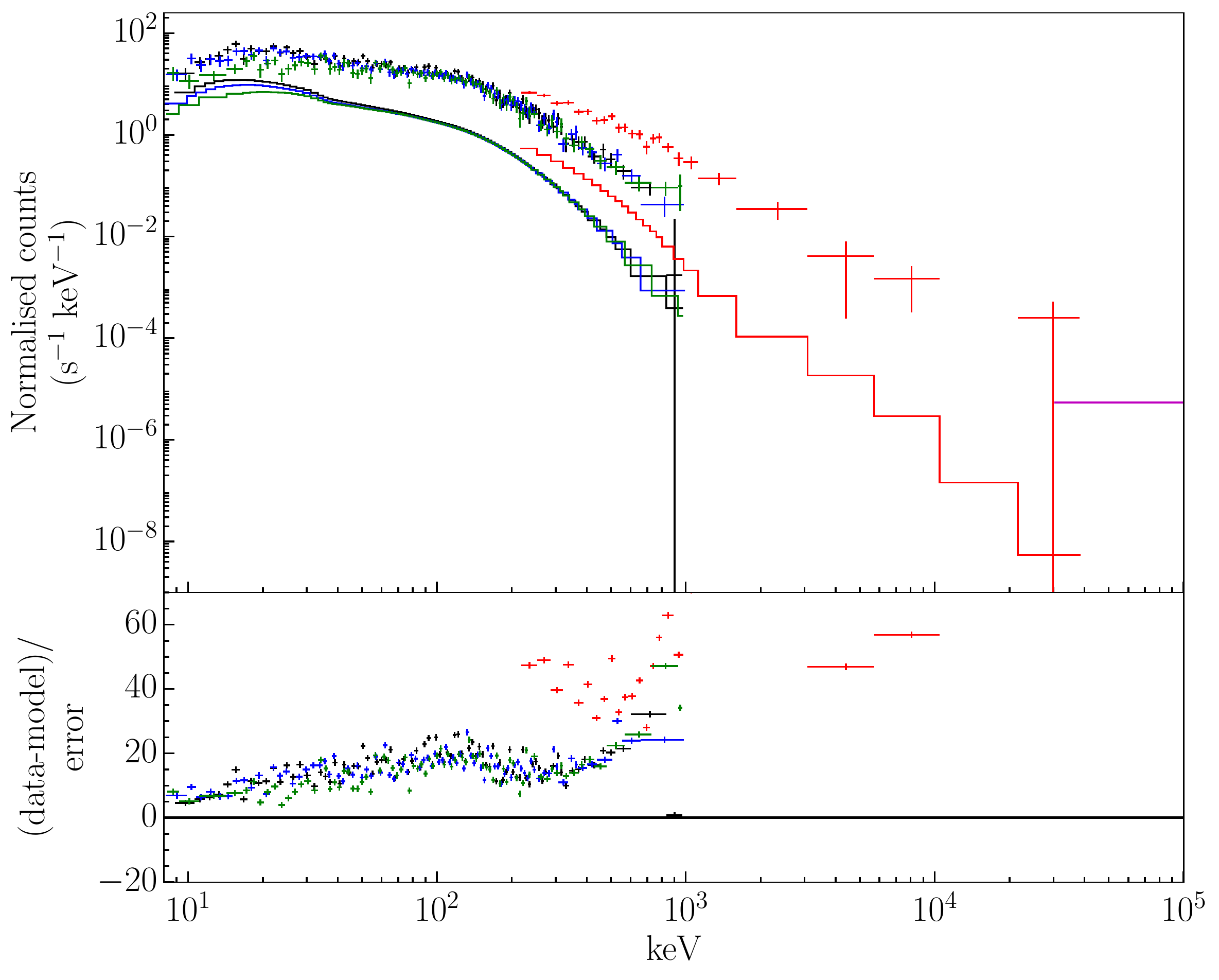}
\caption{Examples of accepted and rejected fits using DREAM1.2. The figure shows the fit results for GRB~091127 and GRB~130518A at $1.2-1.4$~s and $26.3-27.2$~s, in the left and right panel, respectively. For GRB~091127 data from NaI 6,7 and 9, as well as BGO 1, are shown in black, red, blue and green, respectively. For GRB~130518A data from NaI 3, 6 and 7, as well as BGO 0 and the LLE data, are shown in black, red, blue, green and magenta, respectively. The data have been re-binned to errors of 3 $\sigma$ for visual clarity. The residuals are produced using the {\scriptsize XSPEC}, `delchi' command (for the statistic used here this does not correspond to contributions to the statistic). The best-fitting parameters are $\ed = 0.07^{+0.02}_{-0.01}$, $L_{0,52} = 41^{+4}_{-4}$, $\Gamma = 188^{+9}_{-11}$, for GRB~091127, and $\ed = 0.40$, $L_{0,52} = 300$, $\Gamma = 150$, for GRB~130518A. Note that the parameter values found for GRB~130518A represent a model which does not fit the data.}
\label{fig:examplefits}
\end{figure*}

When analysing correlations between physical quantities and whether a fit is successful or not, we find that bursts with higher redshifts have a lower fraction of successful fits, and similarly for the flux. Even tighter is the correlation between a high $L_{\mathrm{iso,z}}$ and a high fraction of rejected fits. The latter relation is shown in Fig. \ref{physparamgoodvsbad}. These results, together with the pegged values of $\Lum$ in most rejected fits, demonstrate that it is the model failing to reach the required normalisation, \ie ~the predicted flux, which is the cause for the majority of the rejected fits. 
There is no discernible pattern in the best-fitting Band $\alpha$ or $\beta$ parameters for rejected fits. This is an important result which we discuss further in section \ref{subsection:BandTableComp}. However, the spectra with the 26 highest value of $E_{\mathrm{p}}$ are not successfully fitted with DREAM1.2, and there is a decrease in the fraction of accepted spectra from $E_{\mathrm{p}} \gtrsim 300$~keV. This suggests that DREAM1.2 works better to describe spectra with lower values of $E_{\mathrm{p}}$.
In Fig.~\ref{fig:examplefits} we show examples of typical accepted and rejected fits applying DREAM1.2 to our sample. This clearly illustrates the normalisation problem for the rejected fits.

\subsection{Time-evolution within bursts}
For a physical model the time evolution may serve as a consistency check (since we expect physical parameters to behave in some systematic way) as well as a way to assess the model. Many of the analysed bursts do not have a sufficient number of bins with acceptable fits for a study of the time evolution. 11 bursts in our sample have more than 4 surviving bins. Only these are considered for the analysis of the temporal evolution of best-fitting parameters. See table~\ref{appendix:table:burstsumtable} for a summary of the number of accepted time bins for each burst.

For these bursts, there are some tentative systematic parameter evolutions. As expected, we find that $L_{0,52}$ approximately follows the light curves. As mentioned in A15, we observe a decreasing value of $\Gamma$ for some bursts, whereas there is no apparent pattern in others. The dissipation radius, $\rd$, generally increases during a burst. The level of dissipation, $\ed$, appears to have no consistent temporal behaviour.

\begin{figure}
\includegraphics[scale=0.4]{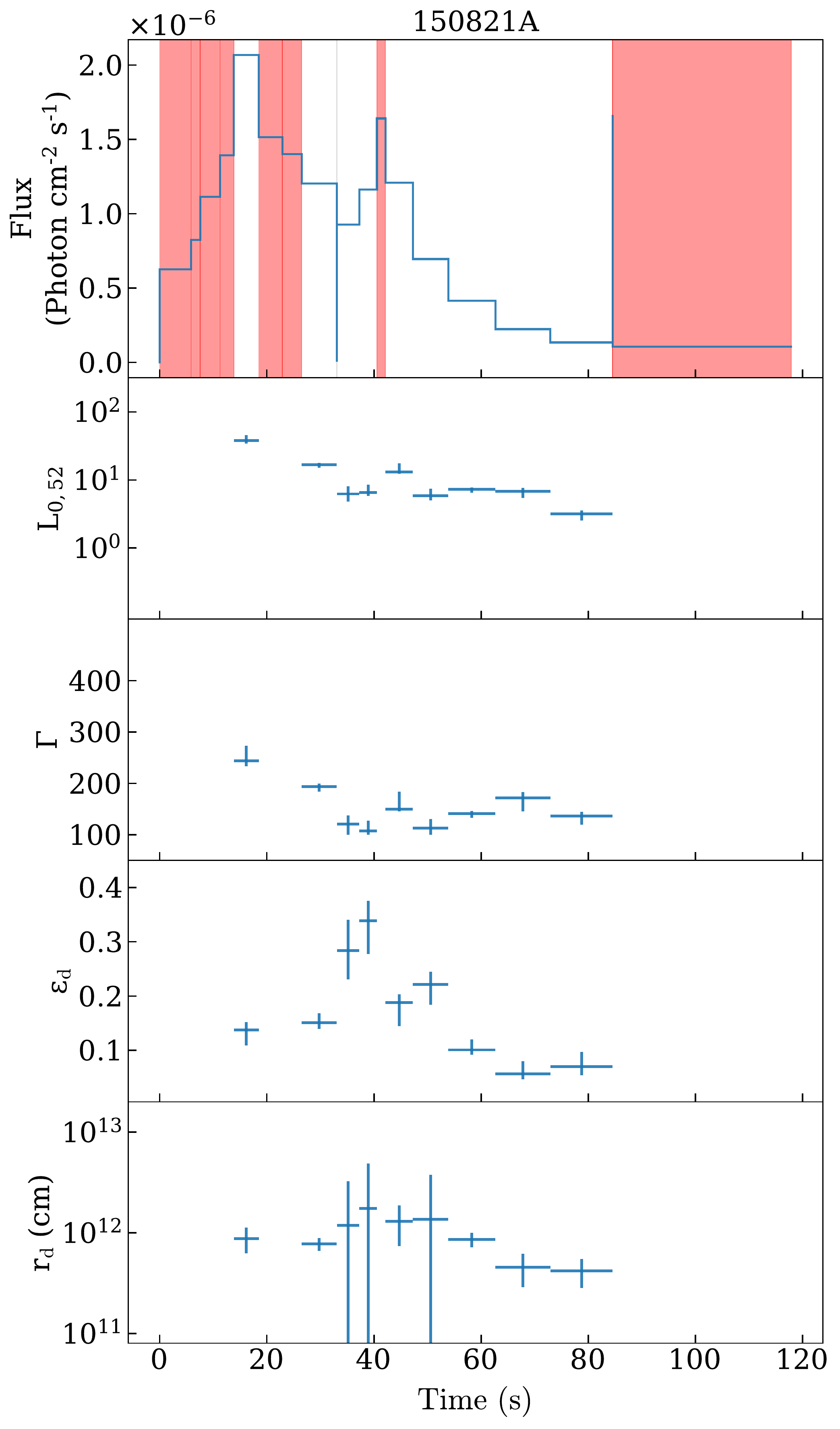}
\caption{Time evolution of the best-fitting parameters of GRB~150821A, using DREAM1.2. The first panel shows the flux light curve, binned according to the Bayesian blocks. Only parameter values from accepted fits are presented. The red regions in the light curve show rejected time bins. Parameter values are plotted in the centre of the respective time bins. Time bins with grey shading and no presented parameter values indicate that the bin is not analysed due to a low SNR.}
\label{timeevol_plot}
\end{figure}

In Fig.~\ref{timeevol_plot}, we provide an example of the time evolution of $L_{0,52}$, $\Gamma$, $\ed$ and $\rd$ for the accepted fits of GRB~150821A, plotted together with the flux light curve. $L_{0,52}$ correlates with the light curve, as seen in most bursts. 
In this particular burst $\Gamma$ also tends to follow the light curve, and $\ed$ appears to anti-correlate with it. Finally, $\rd$ remains almost constant throughout the burst duration. We provide equivalent plots for all bursts with more than 4 accepted time bins in appendix~\ref{appendix:D:fluxLCandParamEvol}. We discuss the implications of the temporal evolutions in section~\ref{subsection:discuss_phys_implications}. 

In the flux light curve of Fig.~\ref{timeevol_plot}, we note that there are a few very thin time intervals with unusually large or small fluxes. This is a consequence of our Bayesian blocks binning and these fluxes are not reliable. However, due to their very short durations, these bins have not passed the SNR cut and are thus not part of the analysis.

\section{Discussion}
\label{section:discussion}
We begin by relating to the discussion of A15 and point out some of the key differences between the two works.
Secondly we consider the rejected fits, to see where the model fails and discuss how this relates to the underlying assumptions. We also discuss the relation to the Band function, correlations as well as other physical models and previous works.

\subsection{Comparison with A15}
\label{subsec:A15}

As shown in section~\ref{section:Themodel}, there are several differences between the model presented in this work and that in A15. Although the models are produced using the same numerical code, the improvements we have made since A15 have changed the model predictions. The single largest improvement is the introduction of adiabatic cooling (see section~\ref{subsection:adcool}). 
One of the conclusions in A15 was that $\tau$ is an important parameter which sets the distance between the two peaks in the spectrum and decides if the spectrum appears single or double peaked. In this work we have instead chosen to keep $\tau$ fixed and argue that it has low impact for most fits. There are two reasons for these differences. Most importantly, the introduction of the adiabatic cooling has increased the degeneracy between $\tau$ and $\Lum$. As a result, the model generally prefers higher values of $\tau$ and becomes increasingly insensitive to high values of $\tau$.
Secondly, low values of $\tau$ predominately lead to double peaked spectra, but we find that our sample mainly consist of single peaked spectra. This is not surprising given the small number of GRBs with strong double peaks reported in the literature \citep{2011ApJ...727L..33G,2012ApJ...757L..31A}. 
We found that most spectra in our sample prefer $\tau \gtrsim 15$, allowing us to keep it fixed without changing the number of accepted fits significantly. Individual GRBs may still prefer different values of $\tau$ and it may be warranted to keep $\tau$ as a free parameter in some studies.

\subsection{Implications and analysis of rejected fits}
\label{subsection:discuss_badfits}

The fact that 463 time bins are not well described by our model shows that the simplest version of the internal shock scenario below the photosphere with negligible magnetisation employed here does not fully account for GRB prompt emission.
As shown in section~\ref{goodvsbad}, the main reason for fits being rejected is that the model cannot reproduce the observed flux. 
Below we discuss how this result is affected by our model assumptions and the limitations on the parameter space.

\begin{figure}
\includegraphics[scale=0.4]{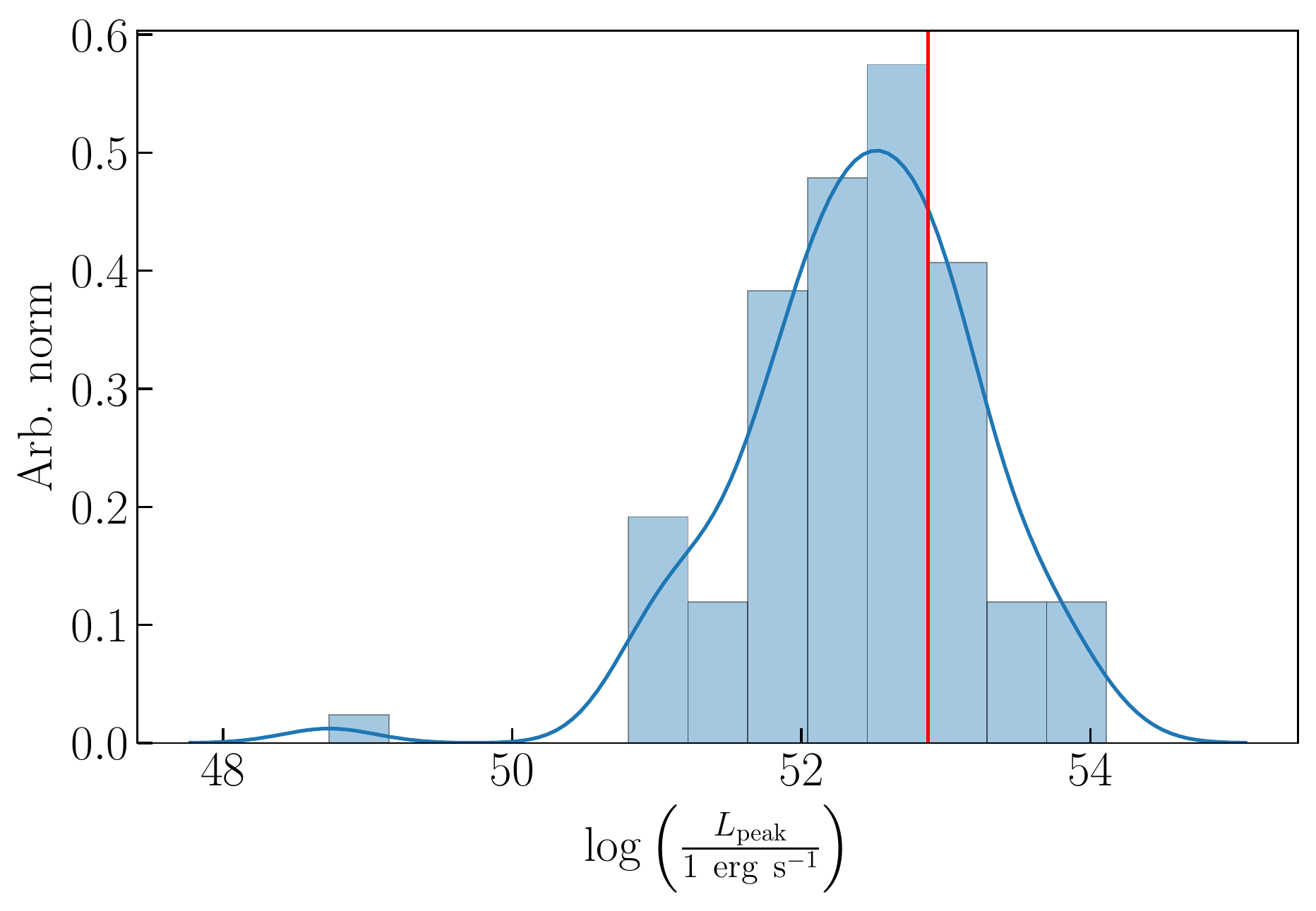}
\caption{$L_{\mathrm{peak}}$ distribution from \protect\cite{2010PASJ...62.1495Y}, showing the $k$-corrected isotropic equivalent luminosity in the 1 second peak interval spectrum, $L_{\mathrm{peak}}$, from 101 bursts. The red line indicates the maximum $L_{\mathrm{iso,z}}$ for which we have an accepted fit with DREAM1.2 ($7.5\cdot 10^{52}$ erg s$^{-1}$). The blue line shows a Gaussian kernel density estimate of the distribution.}
\label{fig:LisoDREAMmax}
\end{figure}

We start by considering these results in relation to the luminosity distribution of GRBs. In Fig.~\ref{fig:LisoDREAMmax} we compare the maximal $L_\mathrm{{iso,z}}$ of an accepted fit ($7.5\cdot 10^{52}$ erg s$^{-1}$), to the $L_{\mathrm{peak}}$ distribution from \cite{2010PASJ...62.1495Y}, where $L_{\mathrm{peak}}$ is the value of $L_\mathrm{{iso,z}}$ for the 1 second peak spectrum of each burst. 27/101 bursts in this sample have an observed $L_{\mathrm{peak}} > L_{\mathrm{iso,z,max}}$. Thus, to the degree that the sample in \cite{2010PASJ...62.1495Y} is representative of the GRB population, the current version of our model is incapable of describing the brightest $\sim 30$ per cent of GRBs, only considering its low predicted observed luminosity. We note that this is only a first-order estimate. It does not imply that the remaining 70 per cent of all GRBs can be described by the model, only that this fraction is not immediately ruled out by the luminosity.

One possibility to accommodate the problem of insufficient predicted flux would be to expand the parameter space, \eg ~to higher values of $\Lum$. As noted in section~\ref{tablemodel}, the boundary of $\Lum =300$ that we have imposed is not absolute, but nearing the limit of what is physically reasonable. To investigate this we constructed a model including the parameter value $\Lum=1000$. This model does yield additional accepted fits (219 in total), but is not sufficient to fully solve the issue, with many spectra still having higher fluxes than the model can account for. 
The highest $L_\mathrm{{iso,z}}$ we can produce with DREAM1.2 is 
$\max(\Lum) \cdot \max(\ed) \cdot \ee \cdot a \approx 4 \cdot 10^{53}$, where $a$ is the adiabatic cooling factor presented in section~\ref{subsection:adcool}.
As shown in Fig.~\ref{physparamgoodvsbad}, the fraction of accepted fits goes down significantly well below this value. This is because the spectral shape changes and the model becomes less flexible when $\Lum$ and $\ed$ are forced to high values, and the fits are instead discarded in the GOF test. 
Larger values of $\ed$ are possible in principle, but we are not able to test this with our current set-up without ignoring the possibility of a re-acceleration phase. Additionally, as shown in Fig.~\ref{fig:globParDistHist}, most fits prefer low values of $\ed$, with only $18$ per cent of fits having $\ed > 0.2$, suggesting that higher values of $\ed$ may not be required.

We note that the distribution of $\Gamma$ presented in Fig.~\ref{fig:globParDistHist} peaks at the lower end of the parameter range. However, there are very few failed fits that actually hit the the lower boundary, making it unlikely that lower values of $\Gamma$ will resolve the flux issue. The current range of this parameter is also in good agreement with other observations (see section~\ref{tablemodel}).
Changing the parameters that were kept fixed in this study may also help resolve the issue. Most interesting are $\eb$ and $\tau$. In particular, a higher magnetisation will increase the number of photons provided by synchrotron emission. We have performed fits with $\eb = 0.1$ and found that it does not result in a significantly larger number of successful fits. A study including other values of $\eb$ needs to be conducted before any conclusions regarding the magnetisations can be drawn. In the case of $\tau$, we note that lower values lead to more photons and less adiabatic cooling, which results in higher fluxes. However, we have investigated this by using a model where $\tau$ is allowed to vary in the range $1-500$ and found that it does not significantly change the number of accepted fits. It is possible that dissipation in the optically thin region would provide additional accepted fits, but this is not a part of the subphotospheric dissipation scenario that we are testing. 

Another option is to consider an alternative dissipation scenario. This would allow us to increase the predicted flux while still considering subphotospheric dissipation in an outflow with low magnetisation. We therefore examine our model assumptions, particularly the internal shock scenario, which implicitly sets the nozzle radius as $r_0 = r_{\mathrm{d}}/\Gamma^2$. This in turn sets the temperature of the initial BB, which, given a specific luminosity, sets the number of photons in the outflow. One way to accommodate the flux issue may therefore be to remove the assumption $r_0 = r_{\mathrm{d}}/\Gamma^2$, instead making $r_0$ a free model parameter. 
This will be investigate in a future work.
Alternatively, by invoking additional dissipation deeper in the outflow, in the Wien or Planck zone, it would be possible to increase the number of seed black body photons \citep{2013ApJ...764..157B}.

Another possible origin for rejected fits is that there are additional emission components present. In our model we have assumed a one-zone emission and no synchrotron emission. 
There are cases of GRBs where, \eg ,~an additional power-law component is used to fit the data, such as in GRB~090902B, \citep{Rydeetal_2010A_ApJ}. This particular burst is in our sample, but only contains rejected fits, due to the model under-predicting the flux. This is discussed further in section \ref{subsection:discuss_phys_implications}.

\subsection{Relation to the Band function}
\label{subsection:BandTableComp}
\begin{figure*}
\includegraphics[scale=0.4]{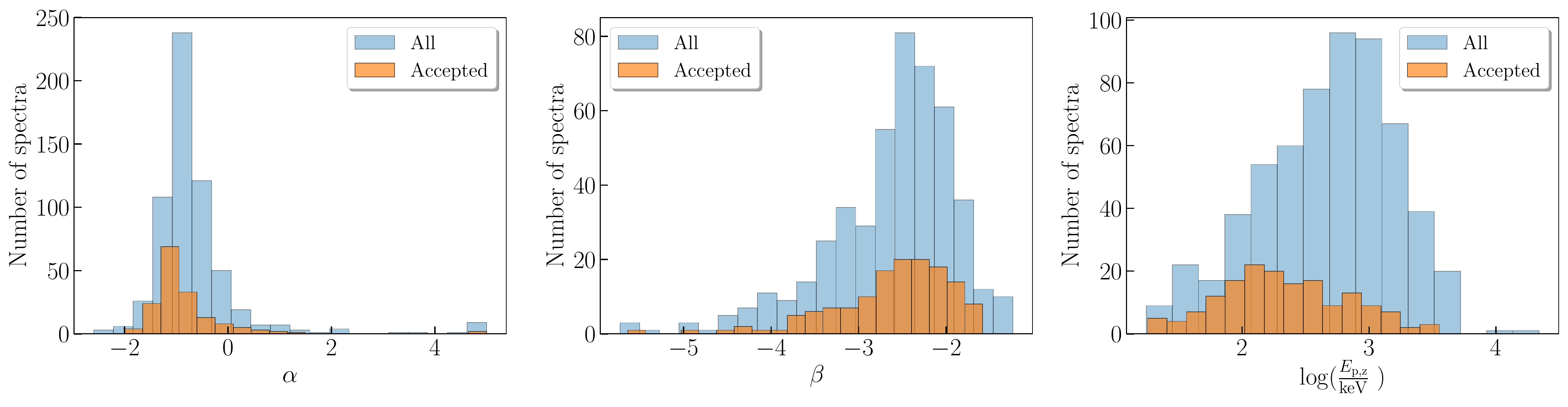}
\caption{Distribution of best-fitting Band function parameter values. The orange histograms show the 171 fits which made the GOF and SNR cut using the DREAM model. The blue histograms show the parameter distribution of all analysed 634 spectra. For visualisation purposes, a small number of fits have been omitted from the histograms, lying outside the plotted ranges. For the accepted sample, there are 7, 32 and 8 fits not shown in this figure, with $\alpha < -3$, $\beta < -6$ and $\Epz < 10$, respectively.}
\label{fig:BandParHist}
\end{figure*}

As noted in section \ref{section:correlations}, there is no relation between the Band function $\alpha$ and $\beta$ parameters and whether the DREAM model provides an acceptable fit to the spectra. A plausible explanation for this is that there is spectral complexity not captured by the Band function. In Fig.~\ref{fig:BandParHist}, we show the distribution of best-fitting Band function parameters for the distribution of accepted spectra, as well as for all analysed spectra. We note that the accepted fits covers essentially the same ranges as the full distribution.

\begin{figure}
\includegraphics[scale=0.3]{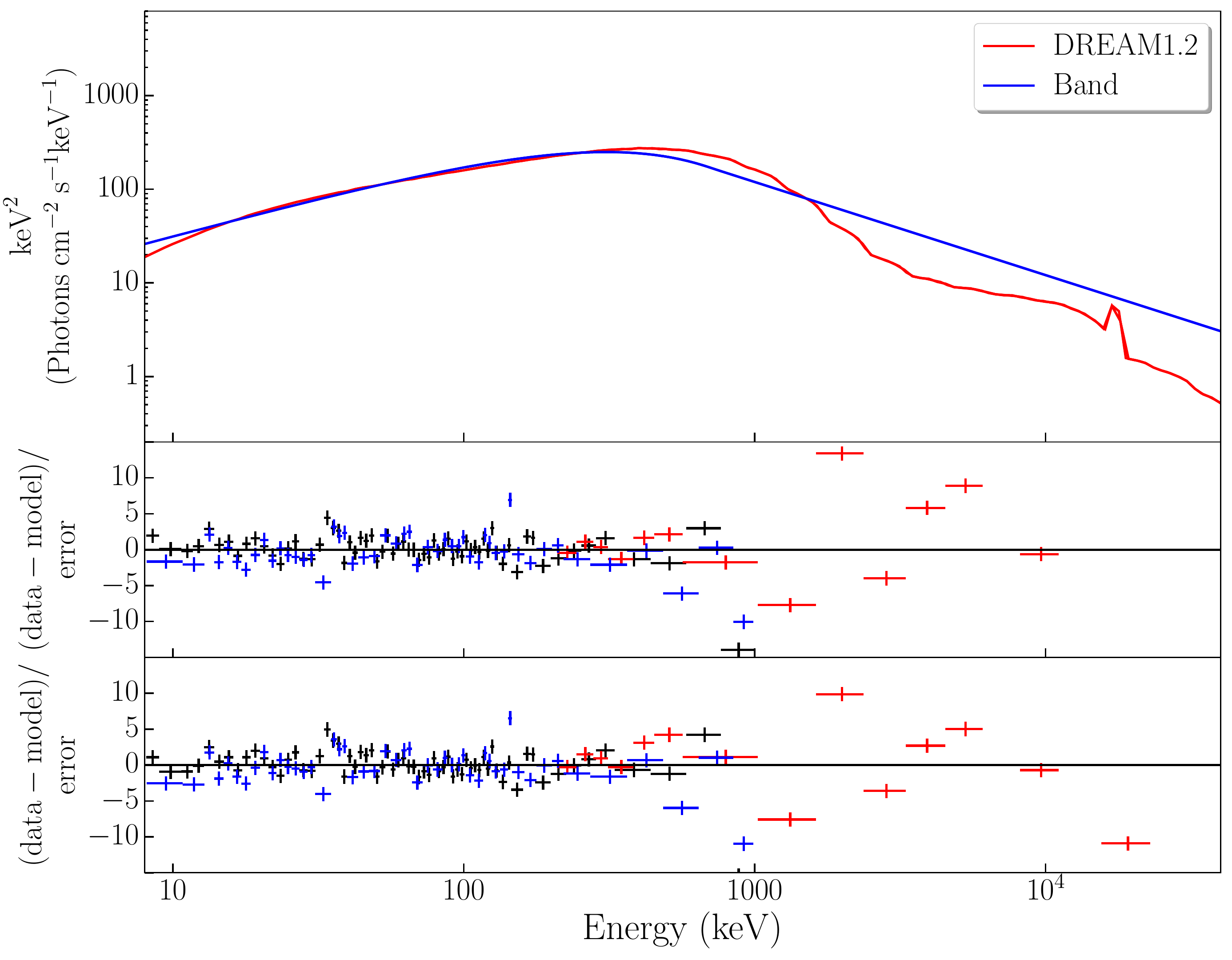}
\caption{The top panel shows best-fitting models for GRB~150821A in the time interval $42.1-47.3$~s using DREAM1.2 and the Band function. For clarity we have removed the data points. The middle and lower plot windows show the residuals of the fits with DREAM1.2 and the Band function, respectively.}
\label{BandTableComp}
\end{figure}

By possibly ignoring features in the spectrum, incomplete, or even misleading, spectral information might be retrieved when fitting with the Band function, leading to erroneous conclusions. For example, as pointed out in A15, the possible existence of two breaks in the spectrum could imply that we need a more nuanced notion of $E_{\mathrm{p}}$, meaning that a single peak energy might not be a relevant characteristic for all spectra.
To illustrate these issues we show an example in Fig.~\ref{BandTableComp} of fits to the spectrum of GRB~150821A in the time interval $21.1-47.3$~s using DREAM1.2 and the Band function, plotted together. Particularly noteworthy is that $\alpha = -1.2$ is not what is usually associated with the low-energy slope of photospheric models. The key point is that it does not need to be, because, as shown in Fig.~\ref{BandTableComp}, values of $\alpha$ may be averages of more complex spectral slopes.

Comparing values of the Band function $\alpha$ with the asymptotic low-energy slope of the thermal radiation is only meaningful if 
a) the asymptotic behaviour of the BB is contained in the fitted energy range
and b) if the data actually represents a power law. Neither of these conditions are necessarily true. We show here, as well as in A15, that a) the fitted energy range is above the asymptotic slope of the seed BB and b) that the data may be equally well described by a non-power law behaviour at low energies. 

We stress that Fig.~\ref{BandTableComp} does not show that the presented spectrum must be curved at low energies.
However, the forward folding and measurement uncertainties make it hard to assess any model by comparing it to fits with another model. We emphasise that physical models must instead be assessed based on direct comparison to data, as well as their physical predictions. These concerns, as well as the shortcomings of the Band function for describing the underlying data, are also discussed in detail in, \eg , \cite{2014ApJ...784...17B,2015MNRAS.451.1511B} and \cite{2017ifs..confE..74B}.

Given that photospheric emission is usually associated with very hard values of $\alpha$, it is an important result that the model can describe spectra with $\alpha$ as soft as -2 (see Fig.~\ref{fig:BandParHist}). Our finding that the model can describe the full range of Band parameters is with the approximation of no geometric broadening. The effect of geometric broadening is in general a softening of the low-energy spectral slope (\cf~Fig.~\ref{fig:adcool:cool}).

\subsection{Correlations}
\label{subsection:discuss_corr}
The correlations between the model parameters of DREAM1.2 and the Band function have no physical interpretation. Instead, the $\ed - \Epz$ and $\Gamma-\Epz$ correlations can be understood in terms of the position of spectral peaks. In DREAM1.2, most spectra are slightly double peaked, which means that $\Epz$ corresponds to some average of the two peaks. For high $\ed$, $\Epz$ will be found closer to the high-energy peak (\cf ~section~\ref{tablemodel} and \ref{subsection:BandTableComp}) and thus we see a positive correlation between these parameters.
As can be seen in Fig.~\ref{paramVarPlot}, $\Gamma$ yields more complex changes to the spectral shape. This leads to, not surprisingly, that the $\Gamma-\Epz$ correlation is less prevalent in the sample than the $\ed-\Epz$ correlation.

Regarding the $\log \Lum-\Gamma$ correlation, we note that an $L_{\mathrm{iso}} - \Gamma$ correlation has previously been reported by \cite{2012ApJ...751...49L} and \cite{2018A&A...609A.112G}. 
\cite{2012ApJ...751...49L} argue that the positive correlation is consistent with the idea of a hyper-accreting black hole central engine, where the jet is powered by a neutrino-driven dominated accretion flow. 
In these studies $L_{\mathrm{iso}}$ is the time-integrated luminosity. \cite{2012ApJ...751...49L} calculate the Lorentz factor used to find the correlation with the
afterglow onset method \citep{1999ApJ...520..641S}, whereas \cite{2018A&A...609A.112G} compare several different methods.
The correlation we present here is self-consistent and we find it independently of previous works, which used completely different methods to calculate $\Gamma$. Additionally, our model also allows for an alternative origin of the correlation, as being a result of the radiative transfer process in the outflow. This interpretation is independent of assumptions about the central engine, other than the assumptions from the fireball model.

In appendix~\ref{appendix:A} we demonstrate that the $\log L_{0,52}-\Gamma$ relation is not caused by degeneracies in the model, or by intrinsic correlations.
Although $\Gamma$ and $L_{0,52}$ are related in the model through their mutual effect on $\rd$, there are no assumptions on the value of $\rd$, and the two parameters are not a priori functionally correlated or dependent. 
However, as shown in appendix~\ref{appendix:A}, when excluding spectra with low SNR we remove spectra with simultaneously high $\Gamma$ and low $\Lum$. 
This indicates that the correlation may be partially due to selection effects. The impact of selection effects on GRB parameter correlations have been studied before. \cite{2013A&A...557A.100H} consider the $E_{\mathrm{peak}} - E_{\mathrm{iso}}$ correlation ($E_{\mathrm{peak}}$ and $E_{\mathrm{iso}}$ being the peak energy of the prompt spectrum and the isotropic equivalent energy, respectively), and find that there is a bias against detecting GRBs with large $E_{\mathrm{peak}}$ and low $E_{\mathrm{iso}}$ due to their lower SNR. 
\cite{2018A&A...609A.112G} also consider the impact of selection bias and find that the $\Lum-\Gamma$ correlation they observe is not a result of selection effects.
Similarly, it seems like the selection effect we find in appendix~\ref{appendix:A} cannot fully account for the observed $\log L_{0,52}-\Gamma$ correlation. However, a more detailed statistical analysis where selection effects are specifically modelled for will be carried out in future work.

\subsection{Implications for the physical properties of jets and progenitors}
\label{subsection:discuss_phys_implications}
Even though most fits are rejected (463 out of 634 fits), we have shown that the problem occurs predominately for bright bursts and that our sample is biased towards bright bursts. The majority of all bursts have luminosities which are consistent with the model (see section~\ref{subsection:discuss_badfits}). Additionally, there are 6 bursts in our sample where more than 60 per cent of the analysed time bins are accepted (see table~\ref{appendix:table:burstsumtable}). This justifies considering the implications of the best-fitting parameters of the accepted fits.

From Fig.~\ref{fig:globParDistHist} we see that $\rd$ is generally distributed around $10^{12}$ cm. 
All values of $\rd$ from successful fits are above the $\sim 10^{11}$ cm radius typically associated with Wolf-Rayet-like (WR) stars, usually assumed to be the progenitors of long GRBs. As pointed out in section~\ref{subsec:Results:bestfitpars}, values of $\rd$ found from the fits inhibit only a fraction of the allowed parameter space, with values as low as $3 \times 10^{8}$ cm being allowed. Hence, the relatively narrow interval in which we find $\rd$ indicates that the dissipation is not caused by interaction with the progenitor star itself. These radii could instead be associated with internal or external shocks, as well as interactions of the outflow with a baryonic shell lifted from the WR star, as found by \cite{2014ApJ...796...81G} and \cite{2014ApJ...791...46T}.

Furthermore, we find that most values of $\ed$ lie around a few percent, which is consistent with what is typically assumed for internal shocks \citep{1995Ap&SS.231..441M,1997ApJ...490...92K}. Although we use an internal shock scenario for our dissipation process, there is nothing in our numerical code which forces the efficiency to this level, which makes the result intriguing. 
Regarding the nozzle radius, $r_0$, we find, from the relation $r_0 = \rd/\Gamma^2$, that $r_0$ typically takes on values $r_0 \sim 10^{7}-10^{8}$ cm.
This result lies in proximity of the usually assumed value of $r_0 \sim 10^{7}$ cm, which is obtained when a variability time-scale of milliseconds is assumed for the central engine. Similar results ($10^{6.5} \lesssim r_0 \lesssim 10^{9}$) have been found also by \cite{2015ApJ...813..127P}. Note that, as for $\ed$, there is nothing in the fitting procedure or code which forces these values. 
As mentioned in A15, the time evolution of $\Gamma$ can be interpreted to correlate with central engine activity and could in principle be used to constrain accretion scenarios. Having $\Gamma$ decrease during a burst implies a central engine luminosity that decreases faster than the mass loading, or even an increasing mass loading.

When studying the temporal evolution of our model, we note that most GRBs have accepted and rejected fits interspersed, \cf ~Fig.~\ref{timeevol_plot} and corresponding figures in appendix~\ref{appendix:D:fluxLCandParamEvol}. We do no not interpret this as subphotospheric dissipation being turned on and off throughout the GRB. This instead points to the fact that our model is a limited implementation of the physical scenario considered, having only three free parameters. Although additional free parameters or expanded parameter space are not expected to be able to fully solve the issue of the under-predicted flux (see section~\ref{subsection:discuss_badfits}), it can likely improve fits for bursts with moderate luminosity.

\subsection{Other physical models and previous work}
\label{discuss_origin_implications}
In this section we consider previous work on subphotospheric dissipation, as well as other physical models, particularly those which have previously been fitted to bursts in our sample. 

GRB~090902B is one of the best examples of a burst with photospheric emission and it has been the focus of several papers.
\cite{Rydeetal_2010A_ApJ} describe GRB~090902B by a two-component scenario by modelling the photospheric emission as a multicolour BB and adding an additional power-law component for the non-thermal emission. Both the thermal and non-thermal emission are modelled empirically.
\cite{2012MNRAS.420..468P} also describe GRB~090902B by a two-component scenario. They use an empirical multicolour BB to model the photospheric emission component. Using the numerical code of \cite{2005ApJ...628..857P}, this multicolour BB is then injected as the seed photon population for optically thin dissipation to produce the non-thermal power law component.

In contrast, in this work we tried fitting GRB~090902B with a single component model from the photosphere. We were unable to produce any successful fits using this model, due to insufficient predicted flux. We also tested fitting this burst with a two component scenario, where the thermal component was produced by DREAM1.2, and the non-thermal component was represented by a power law. Also in this case the thermal component obtained an insufficient flux. The conclusion from our fits to GRB090902B is therefore that the dissipation scenario we assume in DREAM1.2 is not the correct one for this burst. However, we point out that a photospheric origin of the main component in GRB090902B is still probable, but a different dissipation scenario is needed. For instance, GRB090902B  was modelled by \cite{2016ApJ...831..175V}, assuming a completely different scenario for subphotospheric dissipation. One of the key differences between the model used by \cite{2016ApJ...831..175V} and our model is that they consider a jet which starts to dissipate at $r_d \sim 3 \times 10^{10}$ cm, which is below their saturation radius. This is in contrast to our dissipation radius which, by construction, is always above the saturation radius. Allowing photon production from thermal and non-thermal mechanisms deeper in the outflow yields a higher photon production, particularly around the Wien peak, which makes it possible to obtain a higher predicted flux without resorting to higher fireball luminosities.

Another burst in our sample which has been fitted with physical models is GRB~080916C. \cite{2009ApJ...700L..65Z} suggested a multi-component description of this burst by showing that there is an additional energy source above the photosphere. This burst was also described using the ICMART model of \cite{2011ApJ...726...90Z}. In our analysis, GRB~080916C could not be fitted due to insufficient predicted flux, and a comparison of model predictions cannot be made.

There are other physical models which have been fitted to or compared to data, such as the external shock model of \cite{2016ApJ...822...63B} for GRB~141028A, and the synchrotron model of \cite{2016ApJ...816...72Z} for GRB~130606B. 
Out of these, only GRB~141028A is in our sample. However, only 2 of the 15 time-resolved spectra are accepted, while the others are rejected, mainly due to the high flux. It is  clear that the scenario of localised subphotospheric dissipation cannot describe this GRB as a whole, and a comparison with \cite{2016ApJ...822...63B} is not possible. Studies of larger overlapping samples of GRBs using physical models are needed in order to make more thorough comparisons with our work. 

The fact that all spectral slopes present in the Band function parameter space can be fitted with our model is a clear indication of our model's strength and its relevance. The fact that the current version of the model clearly does not suffice to describe all observed GRBs is an important conclusion of this work. This, together with the indications that there are possible alterations of the dissipation scenario which may resolve some of the identified key issues, warrant further studies.

\section{Summary and Conclusion}
\label{section:conclusions}
We have considered a model for subphotospheric dissipation as the origin of GRB prompt emission and fitted it to \fermi GRB data. The dissipation is localised and we assume internal shocks as the dissipation mechanism. Additionally, the model does not take into account geometric effects, a fuzzy photosphere or jet hydrodynamics. We consider the scenario where there are no significant magnetic fields presents.
A table model, DREAM1.2, was created from simulations using the numerical code of \cite{2005ApJ...628..857P}.
Using DREAM1.2 we performed a time-resolved analysis of 36 bursts, using Bayesian blocks as a binning method. We analysed a total of 634 time-resolved spectra. Out of these we find that 171 spectra are well described by our model, passing the GOF test and having no parameters on the boundaries of the parameter space. This corresponds to an acceptance-rate of about 27 per cent, with 10 bursts having at least 50 per cent accepted fits.

From the successful fits we conclude that:
\begin{itemize}
\item The model can fit all spectral slopes found in our sample in terms of Band function parameters. The spectra are dominated by Comptonisation and have negligible contributions from synchrotron radiation.
\item Most values of $\ed$ are in good agreement with the level of efficiency expected from the internal shock scenario at $\ed \sim 1-10$ per cent. It should be noted that there is nothing in the underlying numerical code which ties $\ed$ to these values, despite assuming internal shocks.
\item We find a correlation between the model parameters $L_{0,52}$ and $\Gamma$. This correlation is a possible equivalence of the $L_{\mathrm{iso}} - \Gamma$ correlation found by \cite{2012ApJ...751...49L} and \cite{2018A&A...609A.112G}. Simulations show that the correlation is not a model artefact. However, we find that it may be partially due to selection effects.
\item The fact that there is very little correlation between our model parameters and the Band parameters emphasises the importance of using physical models to analyse GRB data.
\end{itemize}
From the rejected fits we conclude that:
\begin{itemize}
\item With 73 per cent of spectra not being successfully described using DREAM1.2 it is clear that the scenario of photospheric emission with localised subphotospheric dissipation by internal shocks and without magnetic fields, does not suffice to describe the majority of bursts in the sample.
\item The main reason for fits being rejected is that the model under-predicts the observed flux. 
The inclusion of additional parameter space of the model, or the inclusion of significant magnetic fields may yield additional accepted fits. However, we do not expect this to be able to completely resolve the flux issue.
Alternative scenarios of subphotospheric dissipation or additional emission components could instead be invoked to describe the brightest GRBs.
\item The analysed sample is biased toward bright GRBs. Comparing the maximum $L_{\mathrm{iso,z}}$ of time bins successfully described by DREAM1.2 to the $L_{\mathrm{peak,iso}}$ distribution of \cite{2010PASJ...62.1495Y} yields the estimate that $\sim 1/3$ of the GRB population are too luminous to be described using DREAM1.2.
\end{itemize}

We conclude that our model cannot describe the full population of GRBs, but that it appears to be capable of describing a significant part of the population. Thus, if all GRB prompt emission is of photospheric origin, the scenario we have considered in this work cannot be the only dissipation scenario.
In order to further constrain the dissipation scenario, a possible next step is to include \swift data at lower energies, outside the \fermi GBM energy range. Additionally, the parameter space of the model can be explored further, including the effect of significant synchrotron emission or optically thin dissipation.

\section*{Acknowledgements}
We would like to thank Daniel Mortlock for his advice on statistical matters. This work was supported by the Göran Gustafsson Stiftelse, the Swedish National Space Board and the Knut \& Alice Wallenberg Foundation. Parts of the simulations were performed using resources at PDC Centre for High Performance Computing (PDC-HPC). Part of the simulations were performed on resources provided by the Swedish National Infrastructure for Computing (SNIC) at the National Supercomputer Centre (NSC). 
This research made use of Astropy, a community-developed core Python package for Astronomy, \citep{2013A&A...558A..33A}, matplotlib, \citep{Hunter:2007}, Scipy \citep{Scipy2018} and Pandas \citep{mckinney-proc-scipy-2010}. FR is supported by the Göran Gustafsson Foundation for Research in Natural Sciences and Medicine.




\bibliographystyle{mnras}
\bibliography{SUPER_bibliography}



\appendix

\section{Model validation and correlations}
\label{appendix:A}
In this appendix we present validation tests of the model. We also show how the quality and reliability of fits are dependent on the SNR and motivate the SNR cut presented in section~\ref{subsection:fitting:datasection}. 
We also perform simulations to confirm that the $\log\Lum-\Gamma$ correlation presented in section~\ref{section:correlations} is driven by the data, as opposed to being a model artefact.

When performing the validation tests we use the same MC scheme as outlined in section~\ref{subsection:gof}, with some important changes. Firstly we need to provide a response and background to the fake data. We use the background and response of GRB~090618 in the time interval 0-1.5 seconds after the trigger, letting it represent typical conditions for our bursts. Secondly we need a redshift, which will greatly impact the SNR. When simulating a spectrum at a given redshift the SNR will primarily be set by the model parameter $\Lum$. Note that this means that spectra simulated with different parameters at a given redshift are expected to yield vastly different SNR. For real data the observed SNR will of course depend on the redshift of the burst, the angles of the GBM detectors, the exposure time, and the background, apart from the intrinsic strength of the signal.

\begin{figure}
\includegraphics[scale=0.50]{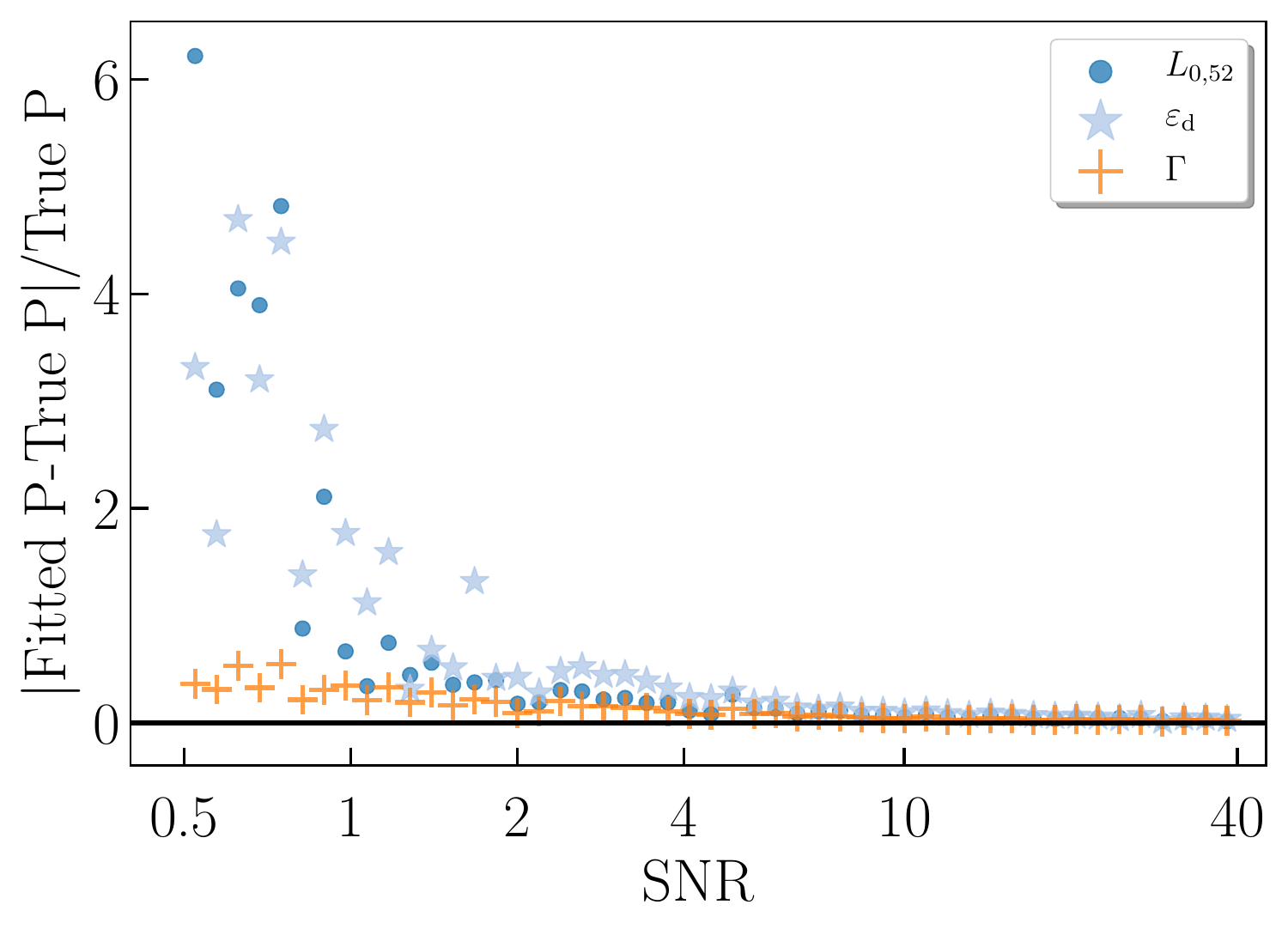}
\caption{Relative error as a function of the SNR. The SNR is spaced logarithmically  in 50 steps in the range $0.5 - 40$. It is the average of the relative error in each SNR interval which is plotted. The horizontal black line denotes 0 relative error. The colours and marker symbols indicate the different parameters as given in the legend.}
\label{appendix:A:RelErr_SNR}
\end{figure}

We start by simulating spectra using the {\scriptsize XSPEC} \textit{fakeit} command, drawing the parameter values from the entire parameter ranges, and fitting the model to the simulated data. 
To examine the quality of the fits we consider the relative errors of all parameters, defined as $P_{\mathrm{Err}} = |(P_{\mathrm{sim}} - P_{\mathrm{fit}})|/P_{\mathrm{sim}}$, for a parameter $P$, as a function of SNR. We create 2000 simulated spectra, assuming $z=1$. This redshift allows us to consider the errors under conditions representative of our data sample. 
In Fig.~\ref{appendix:A:RelErr_SNR} we show the average of the relative error within discrete SNR intervals in the SNR region $0.5 - 40$. As expected, the relative error is 0 at large SNR, but increases when the SNR is low.  The fact that the relative error becomes very large for $\Lum$ and $\ed$ at SNR$\lesssim 2$, whereas it remain relatively low for $\Gamma$, is due to the fact that $\Gamma$ can only vary over half an order of magnitude, whereas $\Lum$ and $\ed$ vary over several orders of magnitude. We note that there is a significant decrease in the relative errors between SNR $2 - 4$ and that the errors approach 0 asymptotically for a larger signal. 
We stress that the exact values of the errors presented in Fig.~\ref{appendix:A:RelErr_SNR} will vary depending on the redshift, as well as on the background and response considered. 

Most spectra in the real data sample with very low SNR are discarded due to fits ending up on the parameter boundaries, while others fail the GOF test presented in section~\ref{subsection:gof}. 
However, even after these two cuts, there would remain fits to spectra with SNR so low that the fits do not provide meaningful information. In order to remove these spectra we apply an SNR cut of SNR$>4$ in the analysis. This cut is a trade-off between ensuring reliable fits and having enough spectra to analyse. As seen in Fig.~\ref{appendix:A:RelErr_SNR}, the average error due to the imperfect signal at SNR=4 and at a redshift of $1$ is, $0.08$, $0.07$ and $0.23$ for $\Lum$, $\Gamma$ and $\ed$, respectively.

\begin{figure}
\includegraphics[scale=0.43]{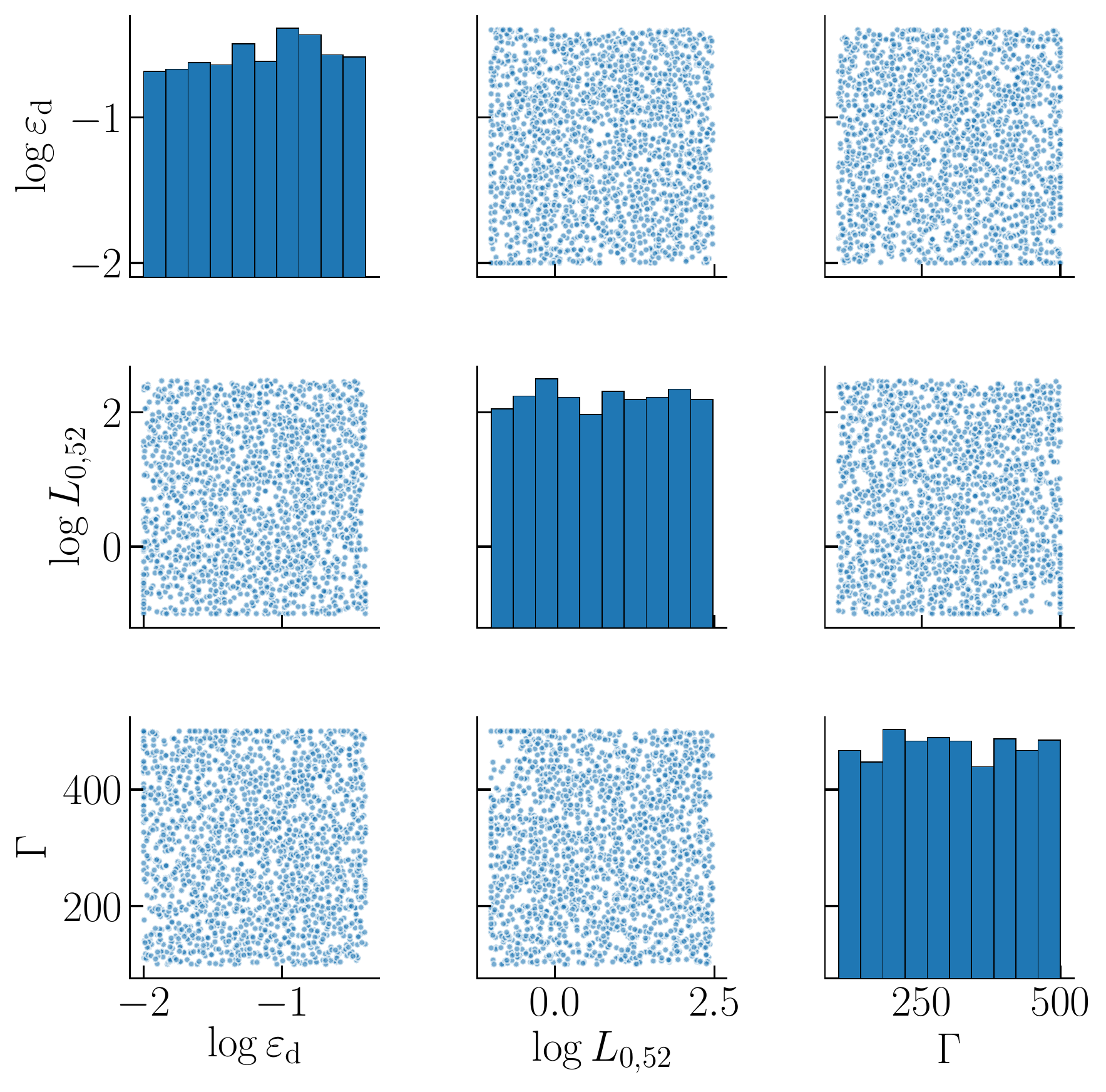}
\caption{The diagonal elements show the histogram of parameter values found when fitting our model to 2000 faked spectra, assuming a redshift of $z=0.1$. The scatter plots show the pairwise relation between all model parameters.}
\label{appendix:A:CorrCheck1}
\end{figure}

Here we consider the correlations presented in section~\ref{section:correlations}. In Fig.~\ref{appendix:A:CorrCheck1} we show the resulting parameter values from fits to 2000 faked spectra where $z=0.1$ was assumed. $\Lum$ and $\ed$ have been drawn from uniform logarithmic distributions over the entire range of values covered by our model, and $\Gamma$ was similarly drawn from a uniform linear distribution spanning all available parameter values. 
It is clear from this figure that there are no intrinsic model dependencies which forces the $\log \Lum - \Gamma$ correlation found in section~\ref{section:correlations}. The correlation is thus not a model artefact. 

\begin{figure}
\includegraphics[scale=0.43]{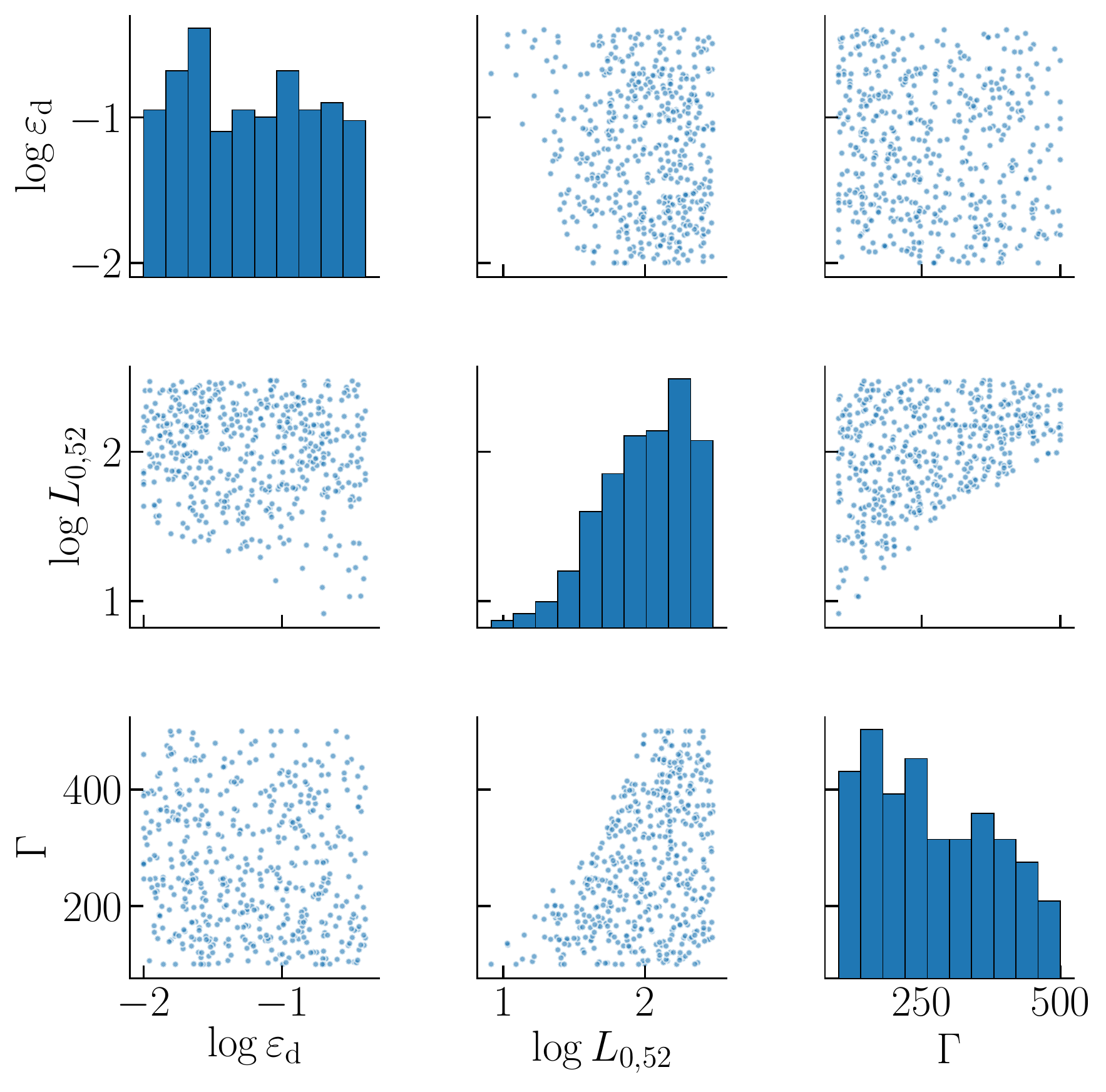}
\caption{The same as Fig.~\ref{appendix:A:CorrCheck1}, but assuming a redshift of $z=1$. The scatter plots show the pairwise relation between all model parameters. The data has been pruned so that only spectra with an SNR>4 are present in the plot. Note that the empty regions are indicative of some possible selection biases in the real data. However, the figure also shows that the $\log \Lum - \Gamma$ correlation is not an artefact of the model.}
\label{appendix:A:CorrCheck2}
\end{figure}

\begin{figure}
\includegraphics[scale=0.43]{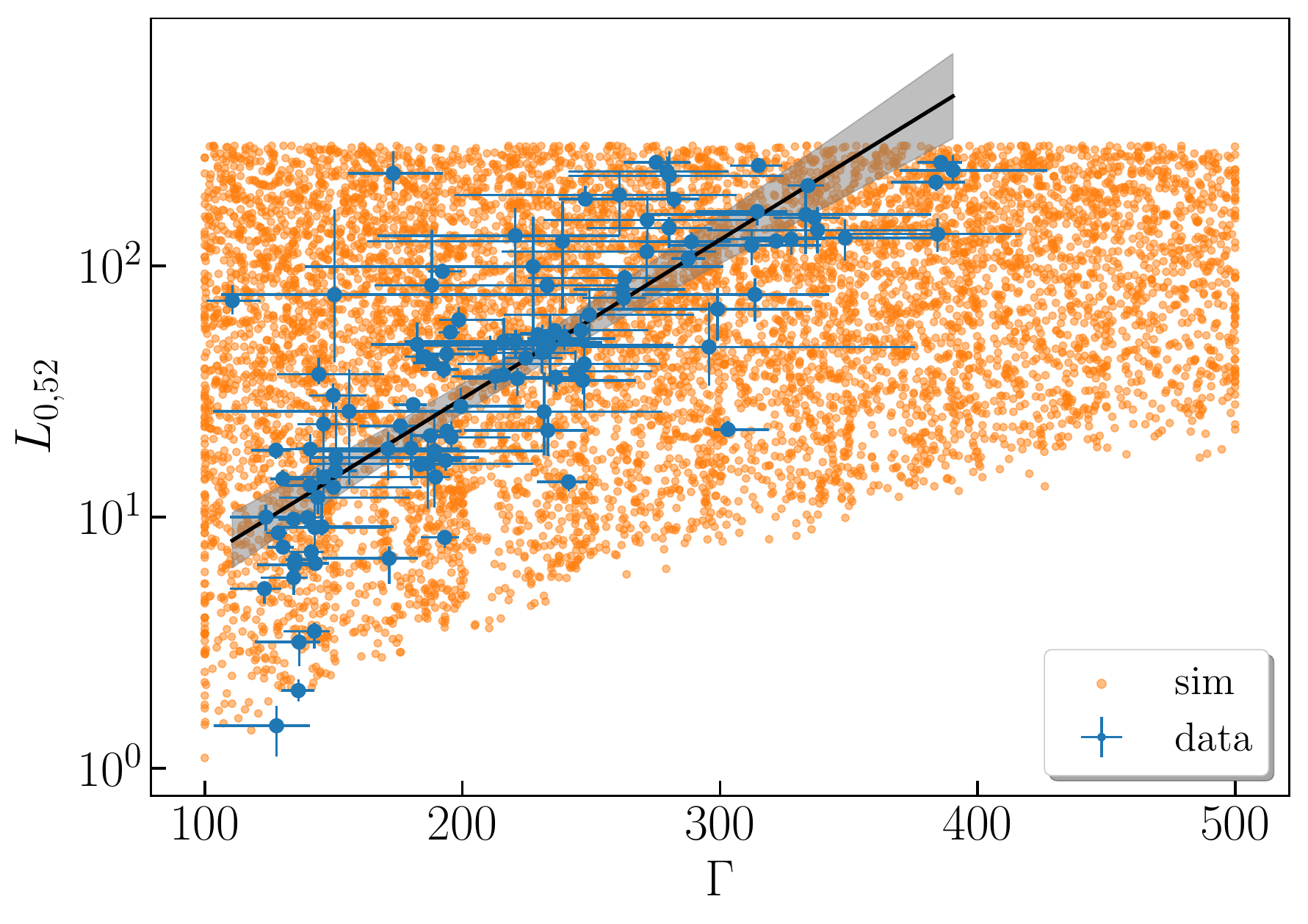}
\caption{Best-fitting $\Lum$ and $\Gamma$ values (orange points) from fits to spectra simulated from redshifts found in the data of the $\log \Lum - \Gamma$ correlation presented in Fig.~\ref{fig:correlations}. Only spectra with SNR>4 are present in the plot. The blue points, as well as the black line and grey confidence band,  are taken from Fig.~\ref{fig:correlations}.}
\label{appendix:A:CorrCheck3}
\end{figure}

However, the simulated data presented in this figure represent data of a far better quality than what is available in our real data sample. Hence, we also consider the same plot using the 2000 previously simulated spectra assuming $z=1$. We make a cut at SNR$>4$ to remove spectra with a poor signal, as we do in the analysis of real data. The results are presented in Fig.~\ref{appendix:A:CorrCheck2}. The empty regions in this plot show that we expect some parts of the parameter space to be devoid of fits at certain redshifts. This is a result of the model's physical nature and is not surprising. However, it suggests that the nature of the $\Lum-\Gamma$ correlation may be partially due to selection bias. 
To investigate this further, we simulate 200 spectra for each data point in the $\Lum-\Gamma$ correlation in Fig.~\ref{fig:correlations}, at the redshift of the respective data point. This gives us a simulated data set with as SNR cut-off representative of the real data. In Fig.~\ref{appendix:A:CorrCheck3} we show the best-fitting $\Lum$ and $\Gamma$ values from these simulated spectra, plotted together with the $\Lum-\Gamma$ correlation from Fig.~\ref{fig:correlations}. We see that the data points in the observed correlation generally do not lie along the boundary to the empty region. This indicates that the $\log \Lum - \Gamma$ correlation found in section~\ref{section:correlations} is unlikely to be caused primarily by selection bias.

\section{Evaluation of grid resolution}
\label{appendix:B}
\begin{figure}
\includegraphics[scale=0.43]{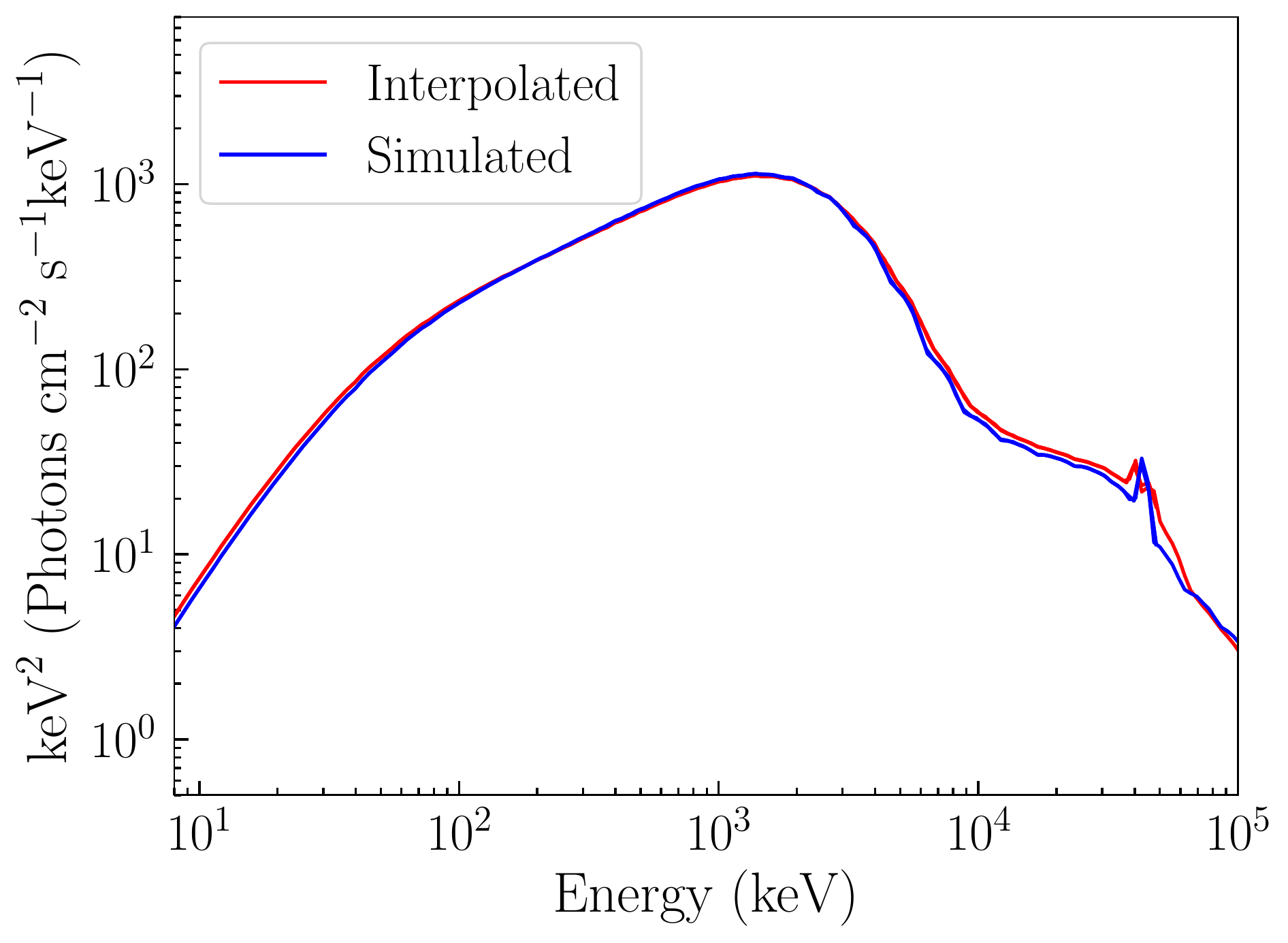}
\caption{An example of how an interpolated spectrum may compare to a spectrum simulated with the same parameter values. The blue line corresponds to the simulated, true, spectrum of our model, whereas the red line shows the interpolated spectrum from DREAM1.2. The parameter values for the two spectra are $\varepsilon_{\mathrm{d}} = 0.125$, $L_{0,52} = 75 $ and  $\Gamma = 375$.}
\label{interpolComp}
\end{figure}

When fitting with a tabulated model we naturally introduce 
uncertainties due to the finite resolution of the grid. Specifically, the interpolated spectra between grid points will differ slightly from spectra simulated at the same parameter values. 
In Fig.~\ref{interpolComp} we provide an example of an interpolated spectrum plotted together with the corresponding spectrum simulated using our numerical code. The spectra in the figure are generated for the parameter values $\varepsilon_{\mathrm{d}} = 0.125$, $L_{0,52} = 75 $ and  $\Gamma = 375$, to show what we may expect from a spectrum that has been maximally interpolated in every parameter. We would naturally expect smaller discrepancies the closer we are to a grid point. 

To quantify these uncertainties we simulate a subgrid, consisting of the holes in the parameter grid used to construct DREAM1.2. Each grid point in the subgrid was chosen to have maximal separation between its grid points and the grid points of DREAM1.2. We thus obtain the subgrid spanned by the parameters 
\begin{align*}
\Lum = &0.25, 0.75, 2.5, 7.5, 25, 75, 150, 250 \\
\Gamma = &125, 175, 225, 275, 325, 375, 425, 475 \\
\ed = &0.0175, 0.0375, 0.0625, 0.0875, 0.125, \\ 
&0.175, 0.225, 0.275, 0.325, 0.375,
\end{align*}
yielding 640 grid points in total.

We simulate spectra from all sets of parameter values of the subgrid and fit DREAM1.2 to them. The spectra were simulated without noise and background in order to isolate the effects of the limited resolution. 
We construct the relative error of a parameter as presented in appendix~\ref{appendix:A}.
In Fig.~\ref{interpolResHistograms} we present the resulting relative error distributions for our three model parameters. The median values and the two-sided 1 $\sigma$ errors 
for the different parameters are $0.08^{+0.08}_{-0.04}$, $0.05^{+0.05}_{-0.03}$ and $0.08^{+0.12}_{-0.05}$ for $\Lum$, $\Gamma$ and $\ed$, respectively. We stress that these distributions represent the points in the model grid where the errors pertaining to evaluating the model between grid points are the largest. These uncertainties are smaller than the statistical errors on the best-fitting parameters, which are found to be $0.12_{-0.06}^{+0.16}$, $0.09_{-0.06}^{+0.43}$ and $0.20_{-0.10}^{+0.21}$ for $\Lum$, $\Gamma$ and $\ed$, respectively. We conclude that the uncertainty in parameter estimates are not dominated by effects of the limited resolution and that the grid of our model is adequately spaced.

\begin{figure*}
\includegraphics[scale=0.43]{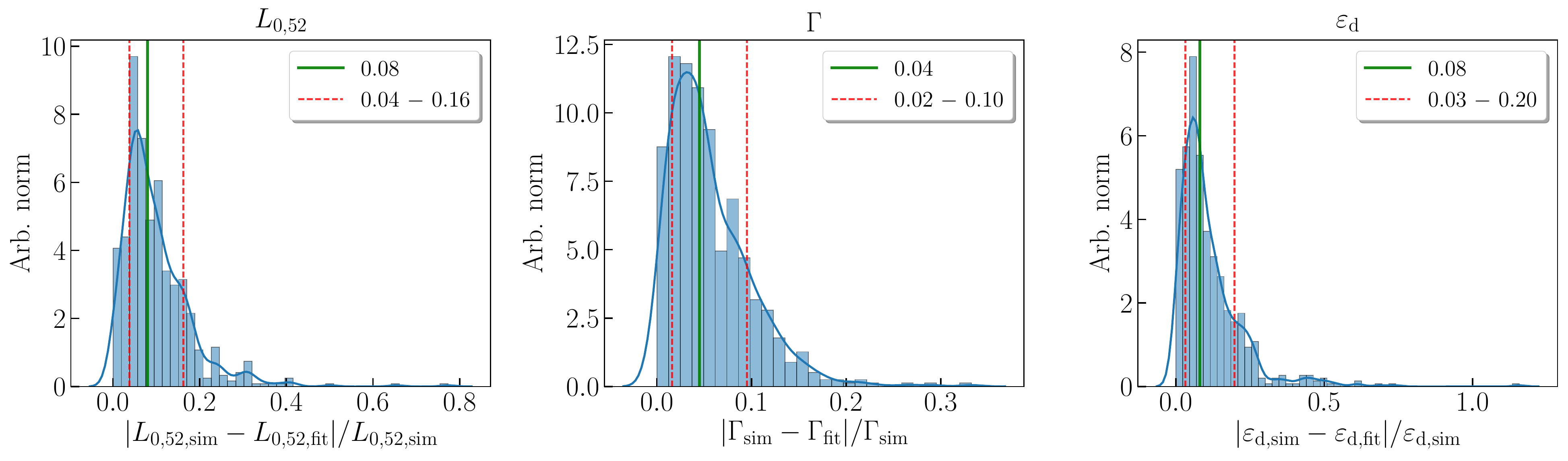}
\caption{Histogram of relative errors (values from the fits versus values from which the spectra were simulated) in each fitted parameter. The blue lines show Gaussian kernel density estimates of the distributions. The green lines correspond to the median value and the red dashed lines show the $31.7$ per cent quantiles of the distribution, representing the asymmetric 1 $\sigma$ confidence intervals. The legend presents the numerical values of the confidence interval bounds.}
\label{interpolResHistograms}
\end{figure*}

Finally, {\scriptsize XSPEC} provides an option of whether to use linear or logarithmic interpolation in the table model. When examining the effects of interpolation we do so using one model with linear interpolation and one model with logarithmic interpolation and compare the results. These tests indicate that we obtain a a slightly better precision using linear interpolation, although the discrepancy is not prevalent in the entire parameter space. We have thus chosen to use the linear interpolation for our model.

\section{Impact of LAT-LLE data}
\label{appendix:C:LLEdata}
\begin{figure*}
\includegraphics[scale=0.4]{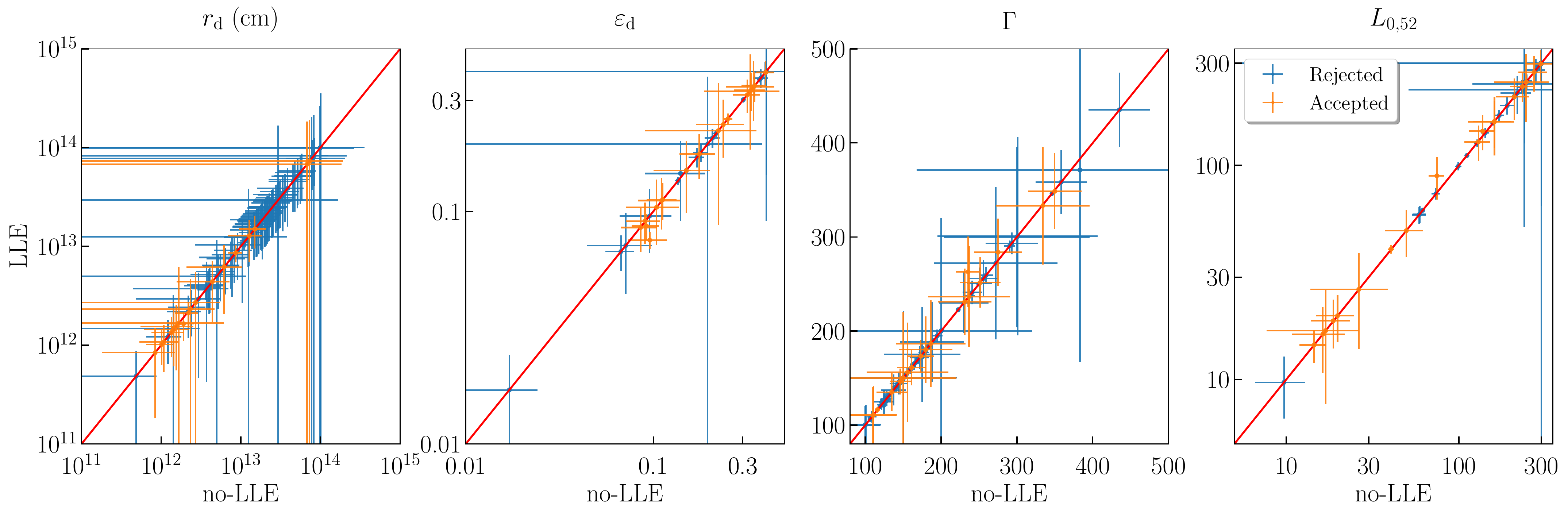}
\caption{Best-fitting parameter values of all $230$ spectra that have LLE data, where SNR$>4$. The y-axis and x-axis show parameter values from the analysis with and without LLE, respectively. $211$ out of the $230$ fits hit the boundary in either $\ed$ or $\Lum$, and are thus rejected (note that it is hard to see most points since they are almost all plotted on top of each other). Orange and blue points represent fits corresponding to accepted and rejected fits, respectively. The red line shows the 1:1 correspondence between parameter values from the respective fits.}
\label{fig:LLEplot}
\end{figure*}

Here we discuss the impact of LAT-LLE data on our analysis. This is relevant since DREAM1.2 has its largest uncertainties at high energies (see section~\ref{subsection:adcool}), and we do not want these uncertainties to bias our fits. To test this we performed all analysis also without the LLE data. The results are summarised in Fig.~\ref{fig:LLEplot}. We find that fitting without the LLE data yields that only 3 spectra change their best-fitting parameter values by more than 1 per cent. Additionally, no additional spectra are accepted or rejected in the GOF test described in section~\ref{subsection:gof}.

The low impact of the LLE data in this study is mainly a result of two effects. Firstly, when performing the binning described in section~\ref{binning_section}, the SNR of the LLE data are often very low. For comparison, the median SNR of the LLE spectra is 2.3, whereas it is 20.5 for the corresponding NaI spectra. When the difference in SNR is large and there is no strong tension with the model at high energies, the fits are predominately driven by the larger number of counts at low energies.

The second reason why the LLE data do not significantly affect the results is that the vast majority of spectra with bright LLE data are already rejected due to their high flux. Specifically, there are only 21 analysed spectra with LLE data where $\ed$ and $\Lum$ do not hit the boundary of the parameter space. Out of these, only 6 spectra have SNR$>1$ in the LLE data. Thus, it is not surprising that the general impact of the LLE data is negligible in this study. Therefore, we conclude that the relatively larger model uncertainties at high energies have negligible to no effect on the results.

\section{Flux light curves and parameter evolutions}
\label{appendix:D:fluxLCandParamEvol}
\begin{figure*}
\includegraphics[scale=0.40]{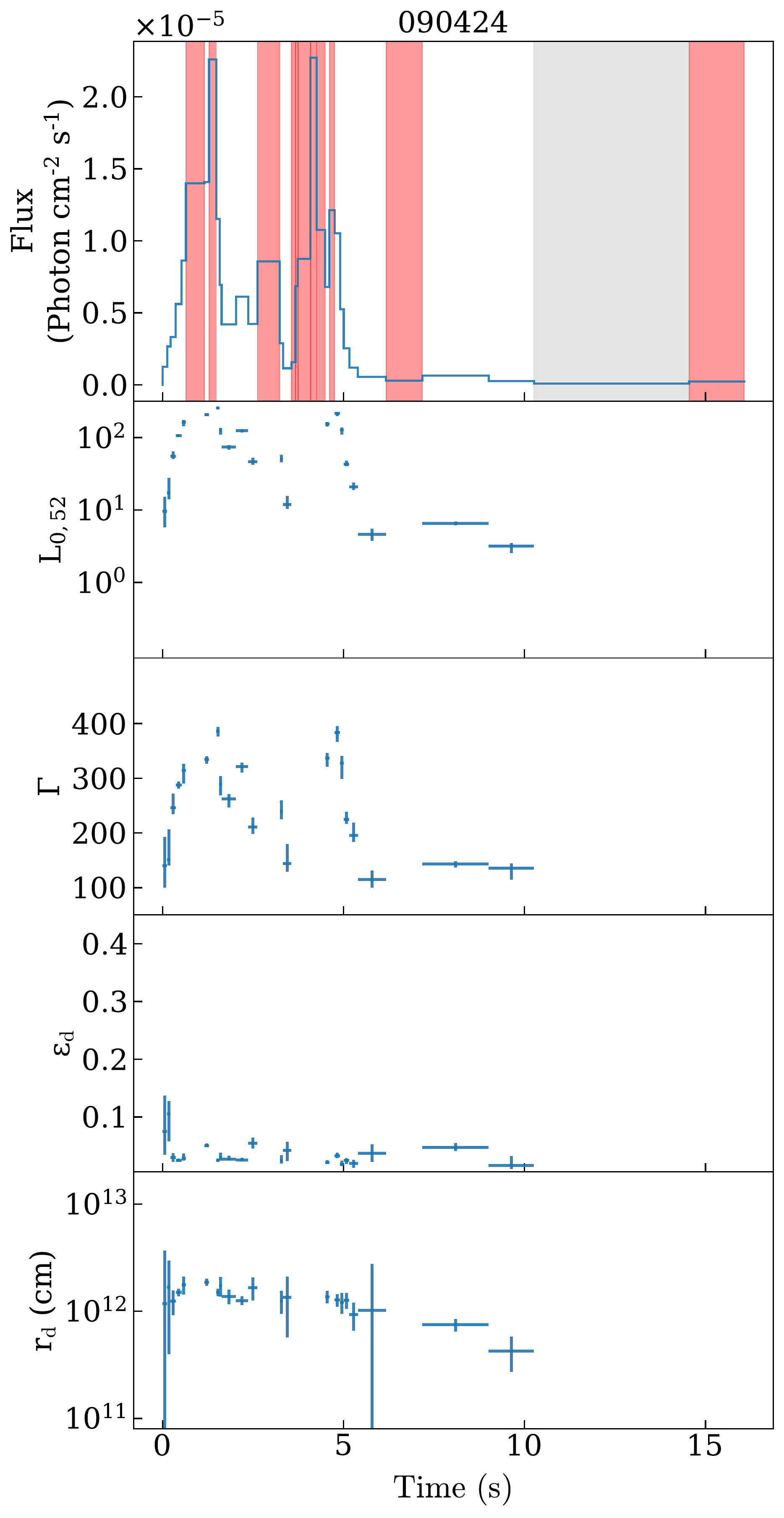}
\includegraphics[scale=0.40]{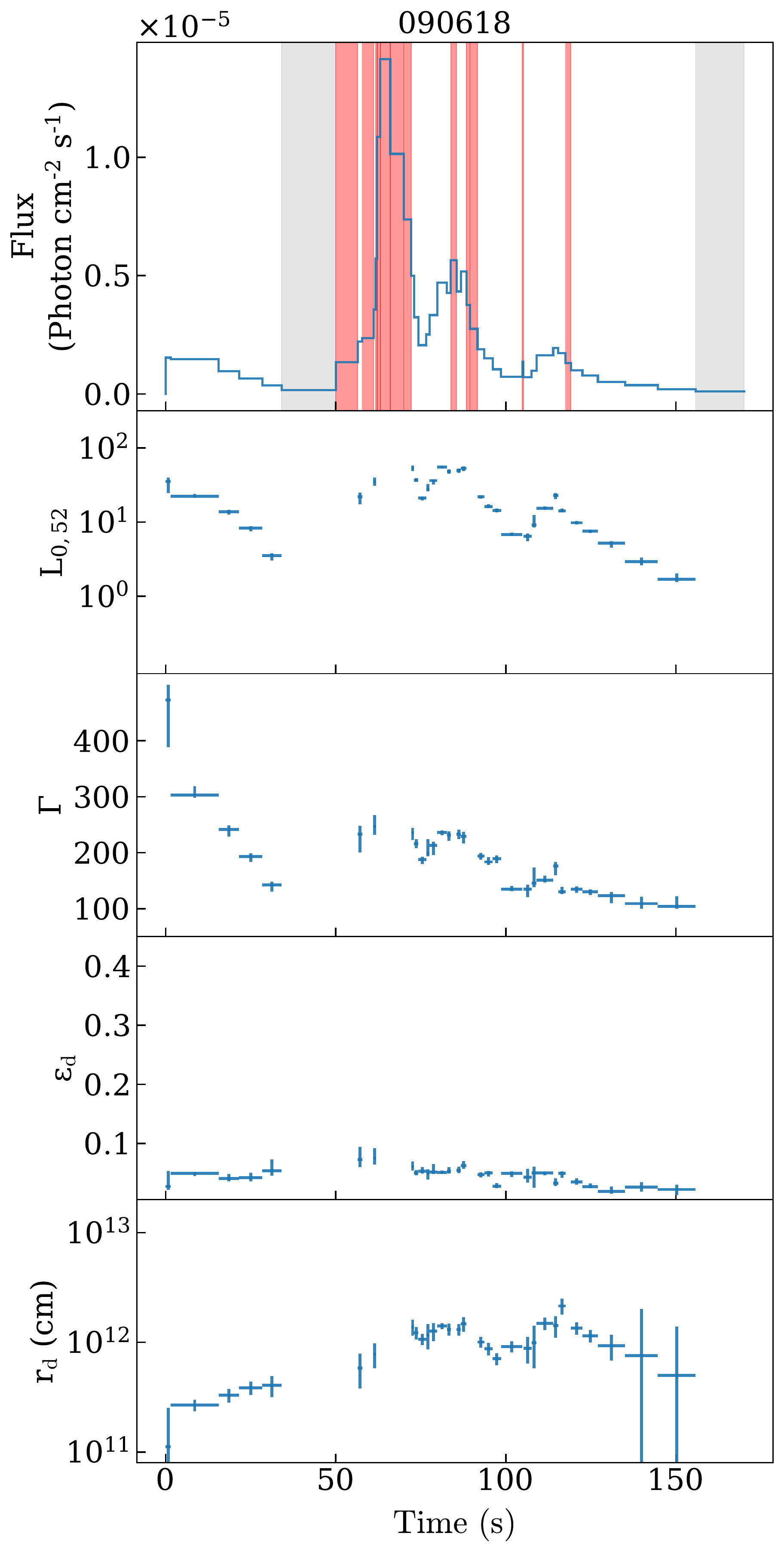}
\caption{Time evolution of the best-fitting parameters using DREAM1.2. The first panel shows the flux light curve, binned according to the Bayesian blocks. Only parameter values from accepted fits are presented. The red regions in the light curve show rejected time bins. Parameter values are plotted in the centre of the respective time bins. Time bins with grey shading and no presented parameter values indicate that the bin is not analysed due to a low SNR.}
\label{fig:timeevol_plots_appendix}
\label{interpolResHistograms}
\end{figure*}

In Fig.~\ref{fig:timeevol_plots_appendix} we provide plots equivalent to Fig.~\ref{timeevol_plot}, for all bursts in our sample with at least 4 accepted time bins.

\begin{figure*}
\includegraphics[scale=0.40]{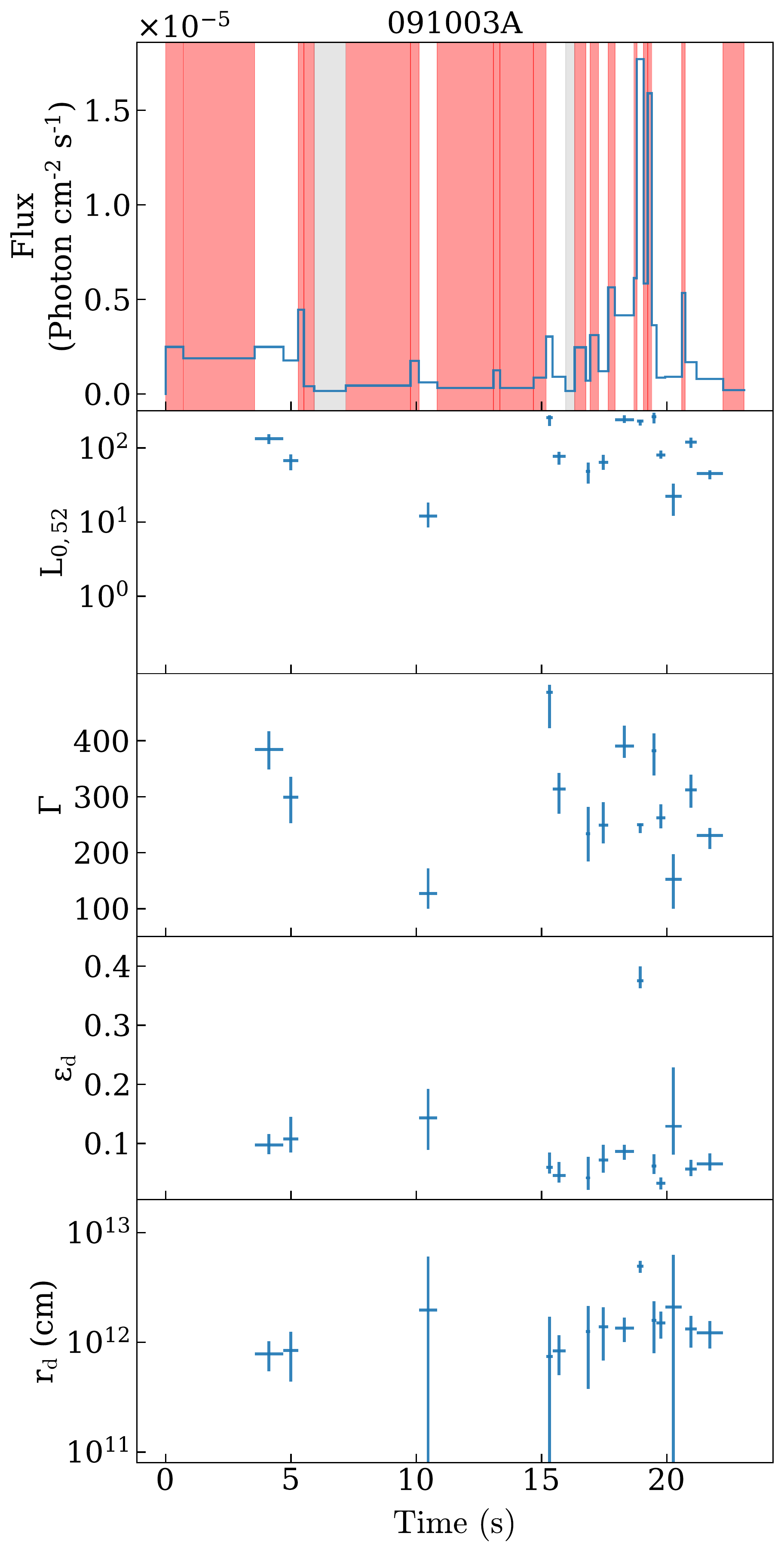}
\includegraphics[scale=0.40]{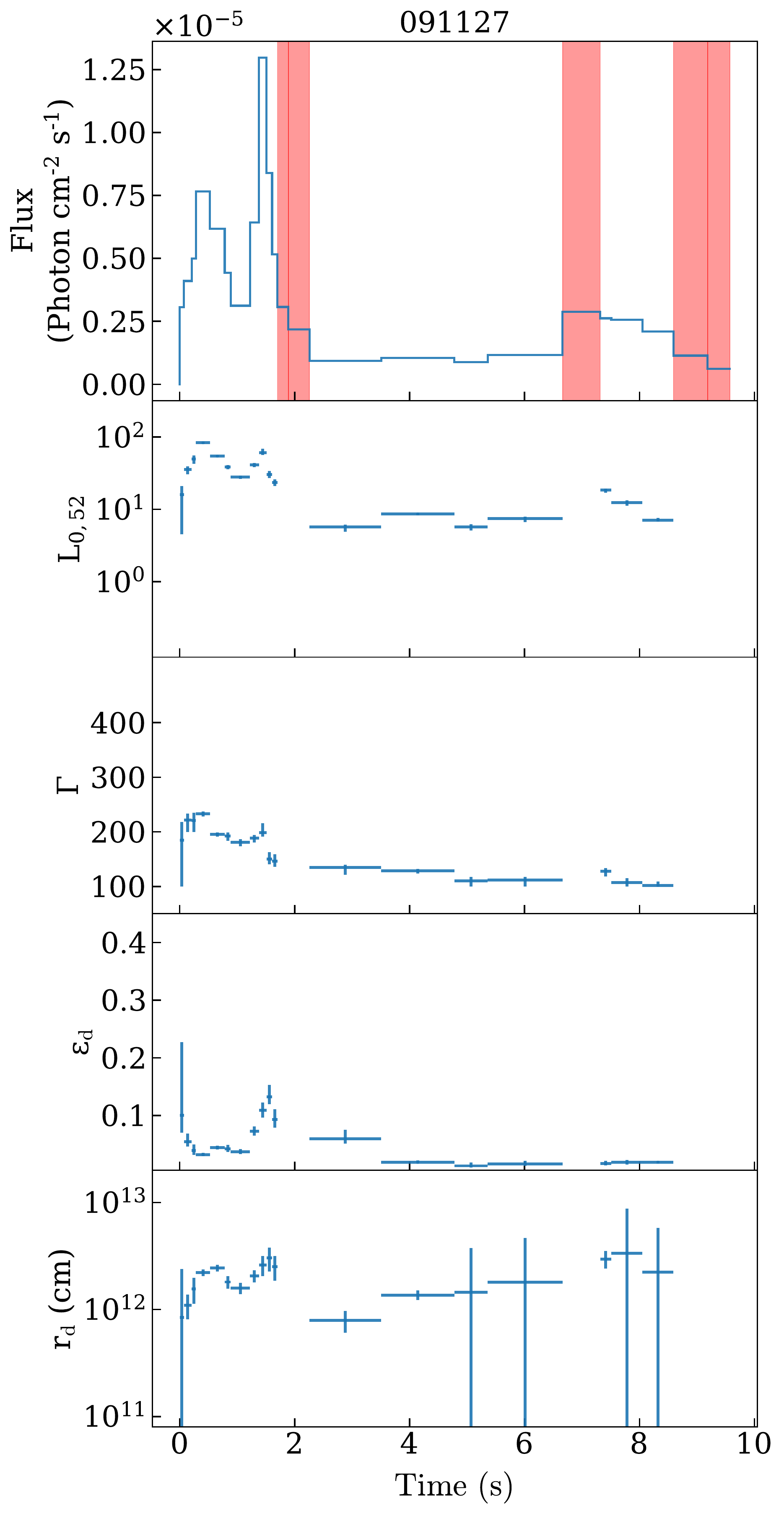}
\end{figure*}
\begin{figure*}
\includegraphics[scale=0.40]{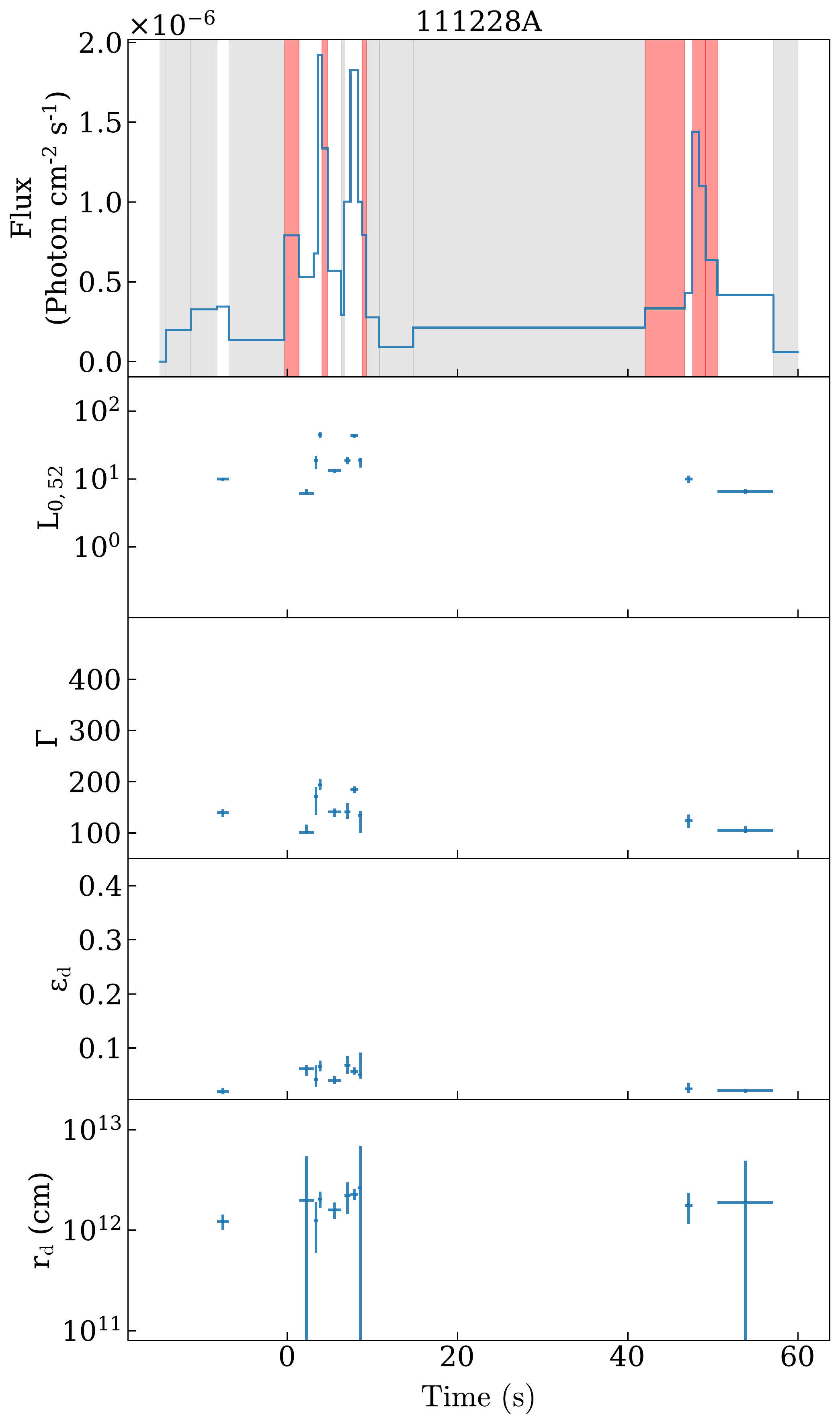}
\includegraphics[scale=0.40]{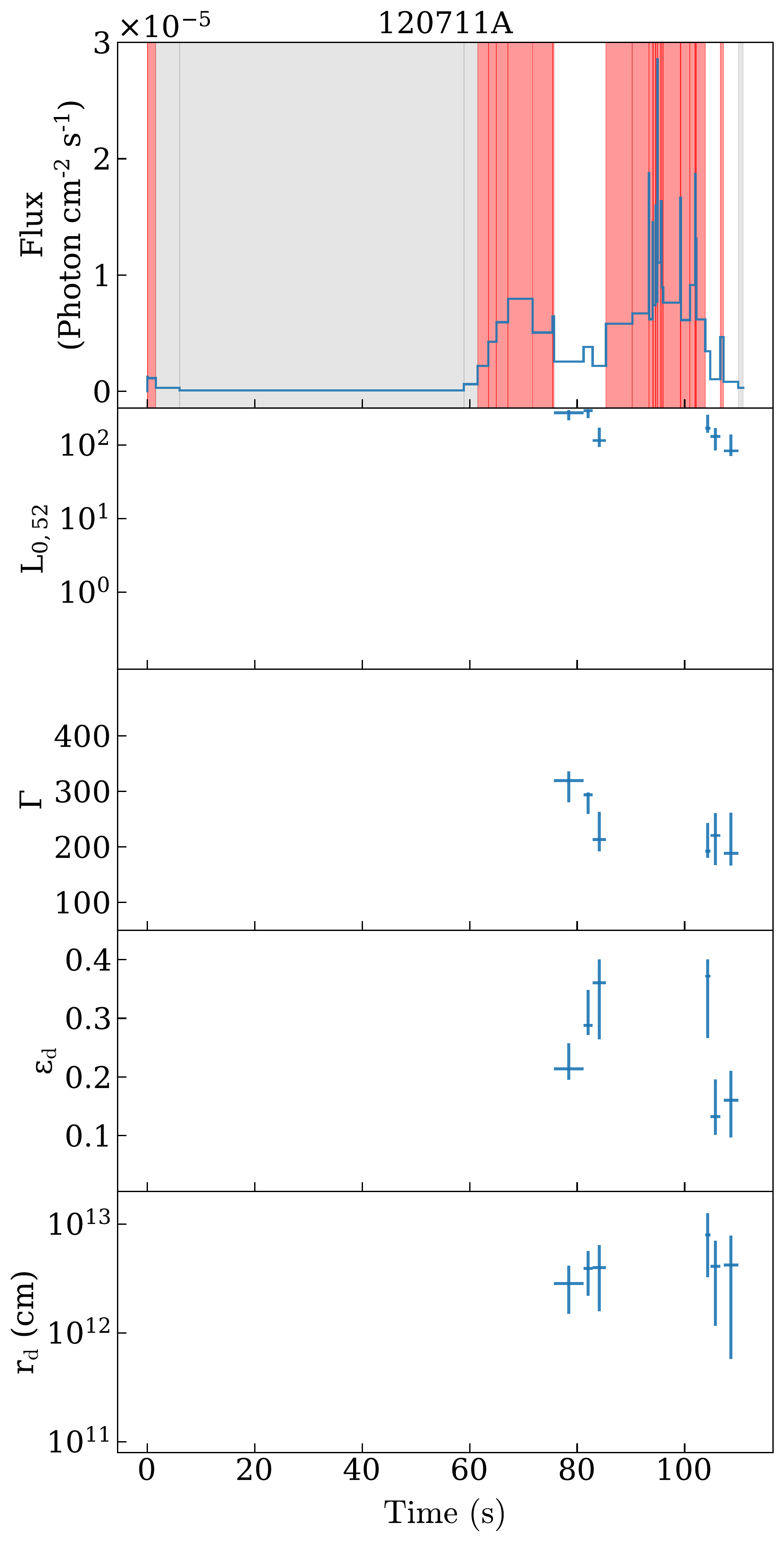}
\end{figure*}

\begin{figure*}
\includegraphics[scale=0.40]{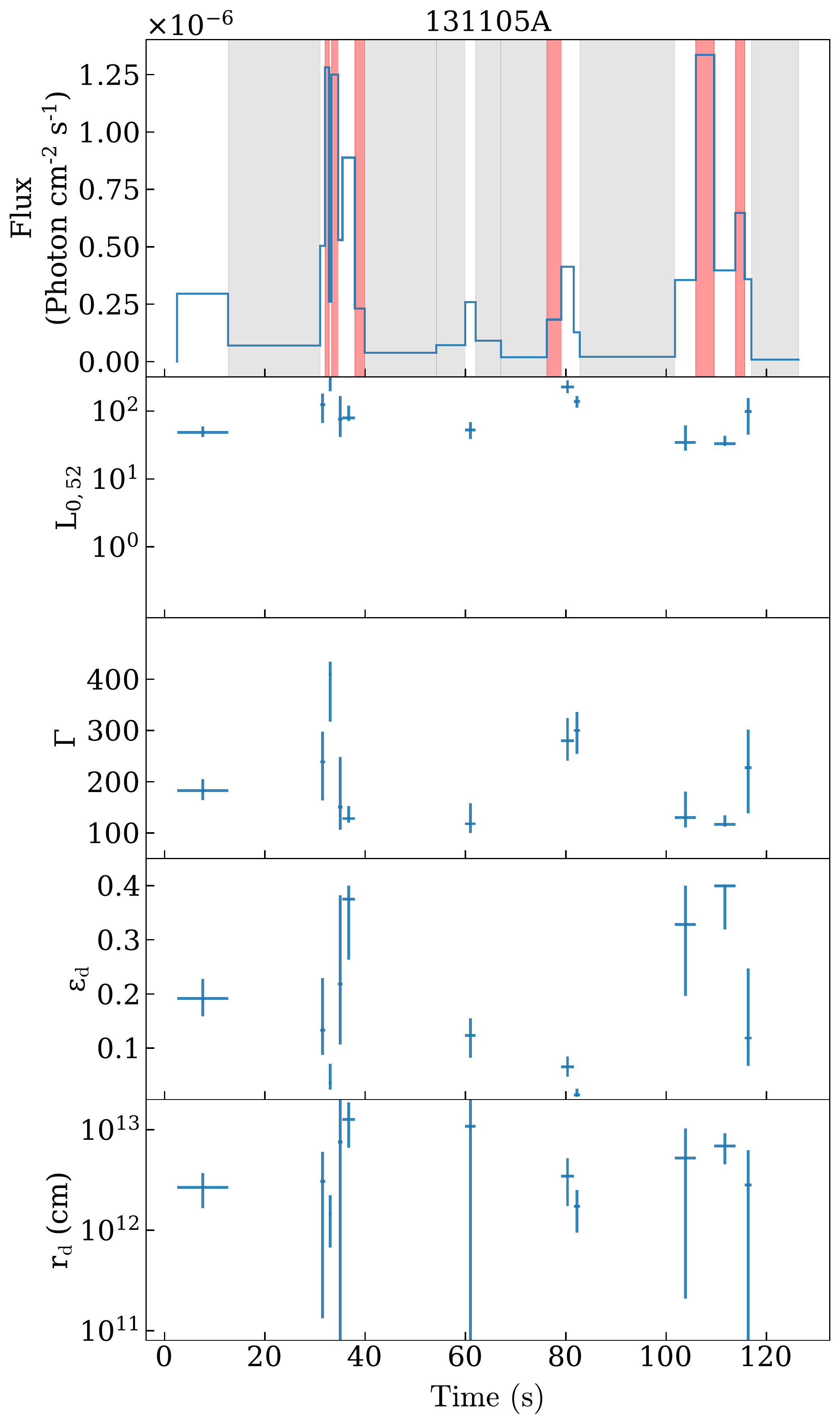}
\includegraphics[scale=0.40]{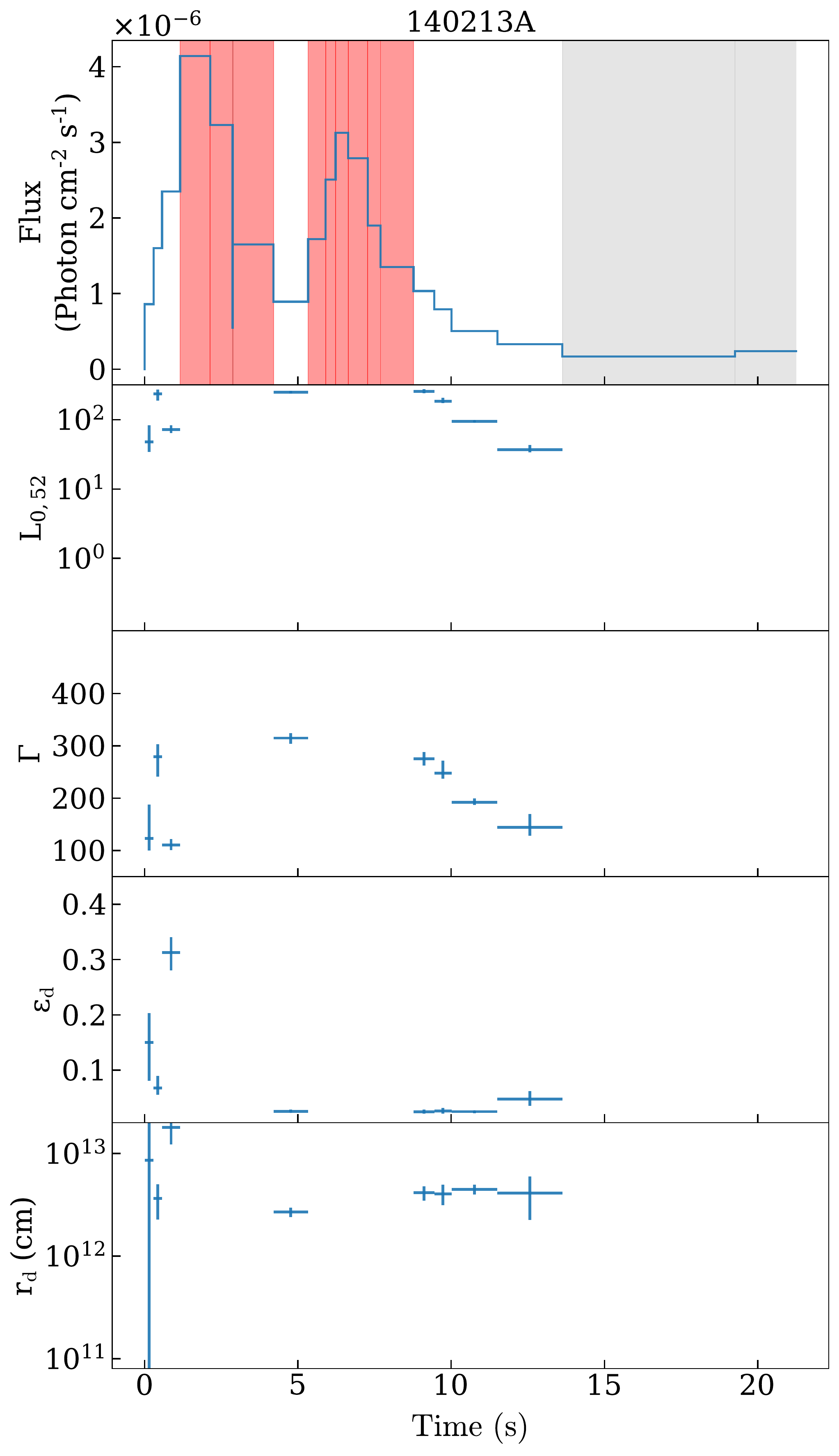}
\end{figure*}
\begin{figure*}
\includegraphics[scale=0.40]{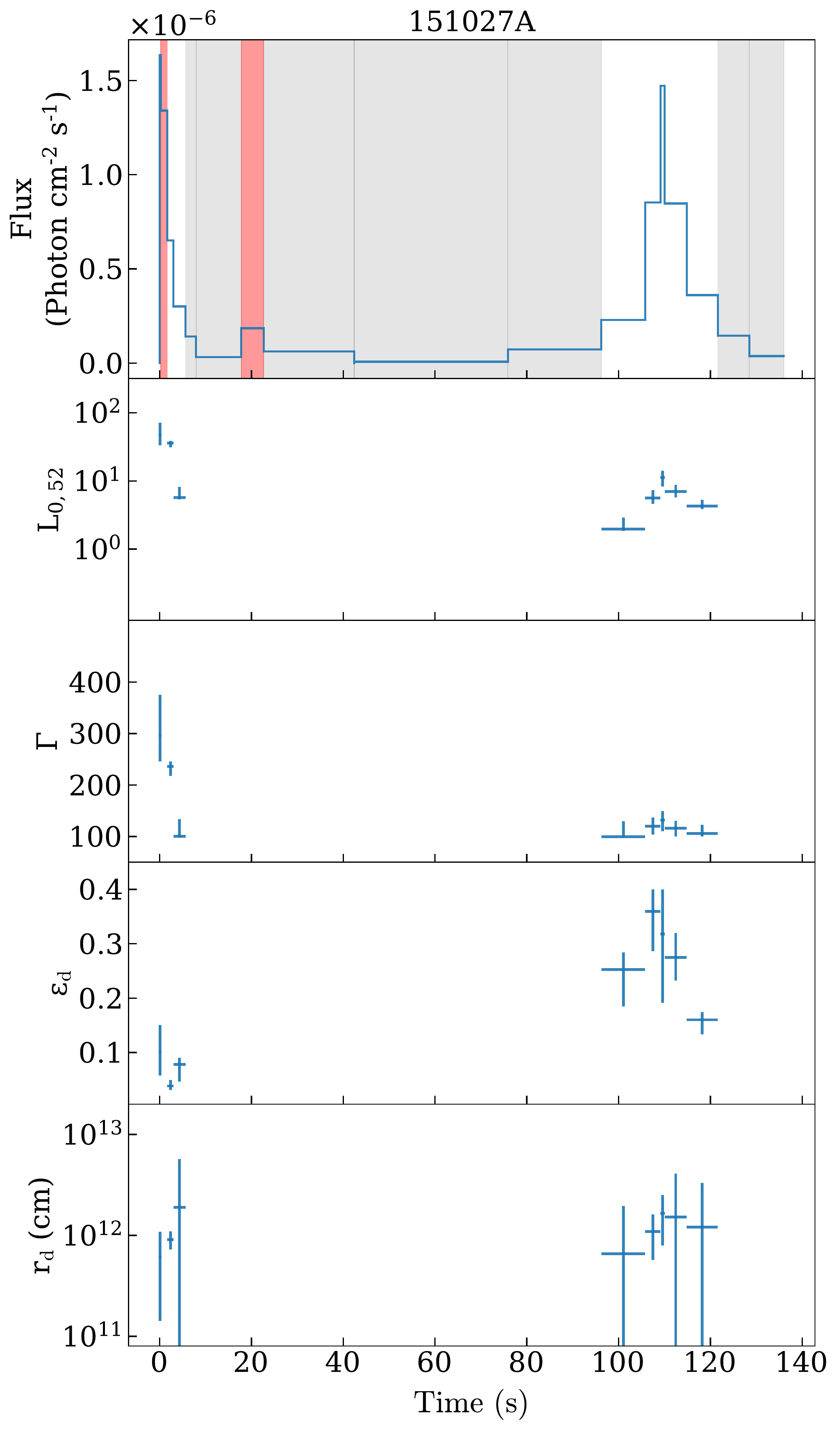}
\includegraphics[scale=0.40]{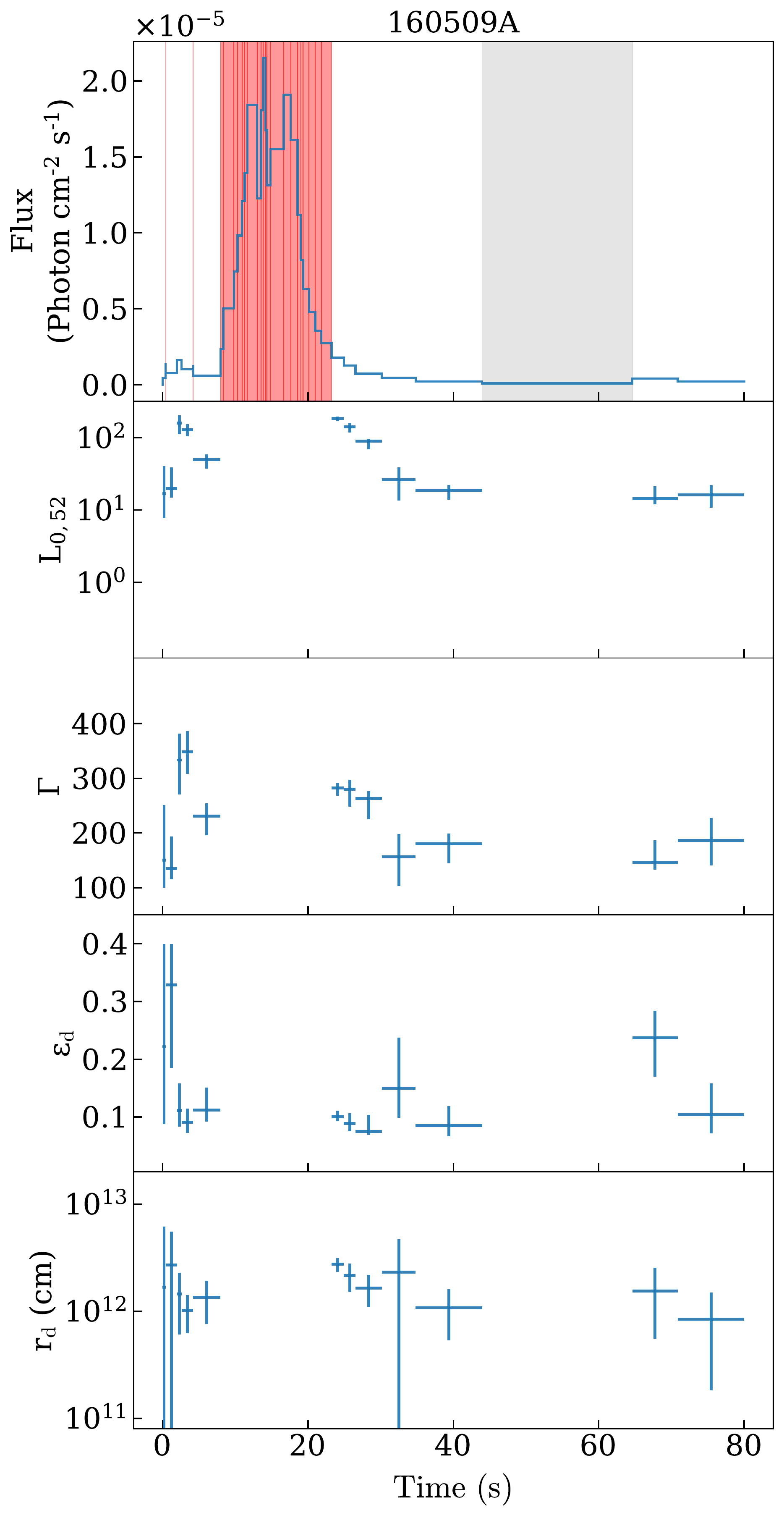}
\end{figure*}

\section{Summary of fitted spectra for each burst}
\label{appendix:E:burstsumtable}
This section contains table \ref{appendix:table:burstsumtable}, which summarises the number of time bins created by our Bayesian blocks binning. It also shows the number of bins that remain after the GOF test described in~\ref{subsection:gof}.

\begin{table}
\centering
\caption{Summary of the number of analysed and accepted spectra for each burst. The first column denotes the number of time bins which passes the SNR cut for each burst. The second column shows how many of these spectra passed the GOF test described in section~\ref{subsection:gof}.}
\label{appendix:table:burstsumtable}
\begin{tabular}{lcccc}
\toprule
   burst & 
   \multicolumn{1}{p{2cm}}{\centering Analysed \\ spectra} &
   \multicolumn{1}{p{2cm}}{\centering Accepted \\ spectra}\\
\midrule
080916C &               15 &                0 \\
  081121 &                9 &                0 \\
  081221 &               18 &                0 \\
 090323A &               23 &                0 \\
  090424 &               32 &               21 \\
  090516 &                0 &                0 \\
  090618 &               42 &               30 \\
 090902B &               61 &                1 \\
 090926A &               28 &                1 \\
 090926B &                4 &                0 \\
 091003A &               32 &               14 \\
  091127 &               23 &               18 \\
 100414A &               21 &                2 \\
 100728A &               13 &                1 \\
 100814A &                4 &                1 \\
 100906A &               14 &                3 \\
 110731A &               12 &                0 \\
 111228A &               17 &               10 \\
 120119A &               14 &                2 \\
 120711A &               34 &                6 \\
 120716A &                6 &                0 \\
 130215A &                3 &                2 \\
 130420A &                2 &                1 \\
 130518A &               23 &                2 \\
 130925A &               20 &                2 \\
 131105A &               17 &               11 \\
 140206A &               11 &                0 \\
 140213A &               17 &                8 \\
 140423A &                0 &                0 \\
 140512A &                7 &                4 \\
 141028A &               13 &                1 \\
 150314A &               16 &                0 \\
 150403A &               20 &                1 \\
 150821A &               18 &                9 \\
 151027A &               10 &                8 \\
 160509A &               35 &               12 \\
\hline
\\[-7pt] total & 634 & 171 \\
\bottomrule
\end{tabular}
\end{table}

\bsp	
\label{lastpage}
\end{document}